\documentclass[11pt,a4paper]{article}
\usepackage[a4paper]{geometry}
\usepackage{amssymb,latexsym,amsmath,amsfonts,amsthm}
\usepackage{comment}
\usepackage{overpic}
\usepackage{mathrsfs,enumerate}
\usepackage{amsmath}
\usepackage{amsthm}
\usepackage{amssymb}
\usepackage{amsfonts}
\usepackage{epsf}
\usepackage{indentfirst}

\DeclareMathOperator{\diag}{diag} \DeclareMathOperator{\Res}{Res}
\DeclareMathOperator{\Bal}{Bal} \DeclareMathOperator{\Airy}{Airy}
\DeclareMathOperator{\Bessel}{Bessel}
\DeclareMathOperator{\PV}{PV}
\newcommand{\er}{\mathbb{R}}
\newcommand{\cee}{\mathbb{C}}
\newcommand{\lam}{\lambda}
\newcommand{\Lam}{\Lambda}

\renewcommand{\Re}{\mathrm{Re}\,}

\renewcommand{\vec}{\mathbf}

\newcommand{\supp}{\mathrm{supp}}
\newcommand{\vecv}{\mathbf{v}}
\newtheorem{thm}{Theorem}[section]

\newtheorem{lem}[thm]{Lemma}
\newtheorem{prop}[thm]{Proposition}

\numberwithin{equation}{section}
\newcommand{\re}{\text{\normalfont Re}}
\newcommand{\im}{\text{\normalfont Im}}

\newtheorem{remark}{\indent \textrm{Remark}}
\newcommand{\al}{\alpha}

\newcommand{\eq}{\begin{equation}}
\newcommand{\nq}{\end{equation}}
\newcommand{\eqa}{\begin{eqnarray}}
\newcommand{\nqa}{\end{eqnarray}}

\begin{document}

\title{Non-intersecting squared Bessel paths with one positive starting and ending point}
\author{Steven Delvaux\footnotemark[1], Arno B. J. Kuijlaars\footnotemark[1],
Pablo Rom\'{a}n\footnotemark[1]  \footnotemark[2] and Lun Zhang\footnotemark[1]}
\maketitle
\renewcommand{\thefootnote}{\fnsymbol{footnote}}
\footnotetext[1]{Department of Mathematics, Katholieke Universiteit
Leuven, Celestijnenlaan 200B, B-3001 Leuven, Belgium. E-mail:
\{steven.delvaux,arno.kuijlaars,pablo.roman,lun.zhang\}@wis.kuleuven.be.
}

\footnotetext[2]{CIEM, FaMAF, Universidad Nacional de C\'ordoba, Medina Allende
s/n Ciudad Universitaria, C\'ordoba, Argentina.  E-mail: roman@famaf.unc.edu.ar.}

\begin{abstract}
We consider a model of $n$ non-intersecting squared Bessel processes with one
starting point $a>0$ at time $t=0$ and one ending point $b>0$ at time $t=T$.
After proper scaling, the paths fill out a region in the $tx$-plane. Depending
on the value of the product $ab$ the region may come to the hard edge at $0$,
or not. We formulate a vector equilibrium problem for this model, which is
defined for three measures, with upper constraints on the first and third
measures and an external field on the second measure. It is shown that the
limiting mean distribution of the paths at time $t$ is given by the second
component of the vector that minimizes this vector equilibrium problem. The
proof is based on a steepest descent analysis for a $4 \times 4$ matrix valued
Riemann-Hilbert problem which characterizes the correlation kernel of the paths
at time $t$. We also discuss the precise locations of the phase transitions.
\end{abstract}

%---------------------------------------------------------------------
\section{Introduction}

%----------------------------------------------------------------------
%\subsection{Non-intersecting squared Bessel processes}
Let $X(t)$ be a squared Bessel process with parameter $\al>-1$,
i.e., it is a diffusion process (a strong Markov process  with
continuous sample paths) on $\mathbb{R}^+$ with transition
probability density given by (cf.~\cite{AndreiNPaavo,
KonigOconnell})
\begin{align}
p_t^{\al}(x,y)&=\frac{1}{2t}\left(\frac{y}{x}\right)^{\al/2}e^{-\frac{x+y}{2t}}
I_{\al}\left(\frac{\sqrt{xy}}{t}\right), &\qquad x,y>0,\label{p_t alpha} \\
p_t^{\al}(0,y)&=\frac{y^{\al}}{(2t)^{\al+1}\Gamma(\al+1)}e^{-\frac{y}{2t}},
&\qquad y>0,\label{p_t alpha 0}
\end{align}
where
\begin{equation}
I_{\al}(z)=\sum_{k=0}^{\infty}\frac{(z/2)^{2k+\al}}{k!\Gamma(k+\al+1)}
\end{equation}
is the modified Bessel function of the first kind of order $\al$; see
\cite[p.375]{HB92}. If $d=2(\alpha+1)$ is an integer, the squared Bessel
process can be viewed as the square of the distance to the origin of a
$d$-dimensional Brownian motion. Some applications of this diffusion process
can be found, for example, in \cite{KarShr}.

We consider a model of $n$ independent copies $X_j(t)$, $j=1,\ldots,n$,
conditioned such that for some $T > 0$,
\begin{itemize}
  \item $X_j(0)=a_j, X_j(T)=b_j$ for $j=1, \ldots, n$ for some given values
  $0\leq a_1<a_2<\cdots<a_n$, $0 \leq b_1<b_2<\cdots<b_n$,
  \item the paths do not intersect each other for $0<t<T$.
\end{itemize}

It follows from
the seminal paper of Karlin and McGregor \cite{KM} that the positions of the
non-intersecting squared Bessel paths at any given time $t\in(0,T)$ are a
determinantal point process. This
means that there exists a correlation kernel $K_n$ such that for each positive
integer $m$ the $m$-point correlation function takes the determinantal form
\begin{equation}
\det[K_n(x_j,x_k)]_{j,k=1,\ldots,m}.
\end{equation}

Indeed, we have by \cite{KM} that the positions
of the paths at time $t$ have a joint probability density
\begin{equation}\label{pdf different a b}
\mathcal {P}_{n,t}(x_1,\ldots,x_n)=\frac{1}{Z_{n,t}}
\det[f_j(x_k)]_{j,k=1,\ldots,n} \det[g_j(x_k)]_{j,k=1,\ldots,n},
\end{equation}
where $f_j(x)=p_t^\al(a_j,x)$, $g_j(x)=p_{T-t}^\alpha(x,b_j)$, $j=1,\ldots,n$,
and $Z_{n,t}$ is the normalization constant
\begin{equation}
    Z_{n,t} = \int_{(0,\infty)^n}\mathcal{P}_{n,t}(x_1,\ldots,x_n)dx_1 \ldots dx_n.
\end{equation}
Hence, \eqref{pdf different a b} is a biorthogonal ensemble \cite{Bor99}, which
is a special case of a determinantal point process. The correlation kernel is
given by
\begin{equation}\label{cor kernel}
K_n(x,y)=\sum_{j=1}^{n}\phi_j(x)\psi_j(y),
\end{equation}
where the functions $\phi_j$, $\psi_j$, $j=1,\ldots, n$ are such that
\begin{equation*}
\begin{aligned}
\textrm{span}(\phi_1,\ldots, \phi_n)=\textrm{span}(f_1,\ldots,f_n), \quad
\textrm{span}(\psi_1,\ldots,\psi_n)=\textrm{span}(g_1,\ldots,g_n),
\end{aligned}
\end{equation*}
and
\begin{equation}
\int_0^{\infty} \phi_i(x) \psi_k(x) dx=\delta_{i,k},\qquad i,k=1,\ldots,n.
\end{equation}

We are interested in the situation where all paths start at the same position
at $t=0$ and end at the same position at $t=T$. Different types of initial and
ending conditions are considered in \cite{Kat11,KT04,KT11}, among others. In
the confluent limit $a_j\to a\geq 0$ and $b_j \to b\geq 0$, the  biorthogonal
ensemble structure is preserved.

If $a=b=0$, i.e., all paths start and end at the origin, this system is known
as chiral or Laguerre ensemble in random matrix theory,
cf.~\cite{DF,KIK08,KonigOconnell,TW07}. The case where $a>0$ and $b=0$ was
considered in \cite{KMW09} and \cite{KMW10}. In that situation, all paths,
after proper scaling, initially stay away from the hard edge at $x=0$, but at a
certain critical time $t^*$ the lowest paths hit the hard edge and are stuck to
it from then on. The limiting mean density of the paths at time $t$ is
characterized by a non-real solution of an algebraic equation of order three;
see also \cite[Appendix]{KMW09} and \cite{KR} for an interpretation in terms of
a vector equilibrium problem with two measures. If $t\neq t^*$, the local
correlations obey the usual scaling limit from the random matrix theory,
leading to the sine, Airy, and Bessel kernels \cite{KMW09}. A new kernel was
established near the origin at the critical time $t^*$ in \cite{KMW10}. The
kernel admits a double integral representation which resembles the Pearcey
kernel \cite{BK07,BoKu10,BH98}.

\begin{figure}[t]
\centering
\begin{overpic}[width=.4\textwidth]{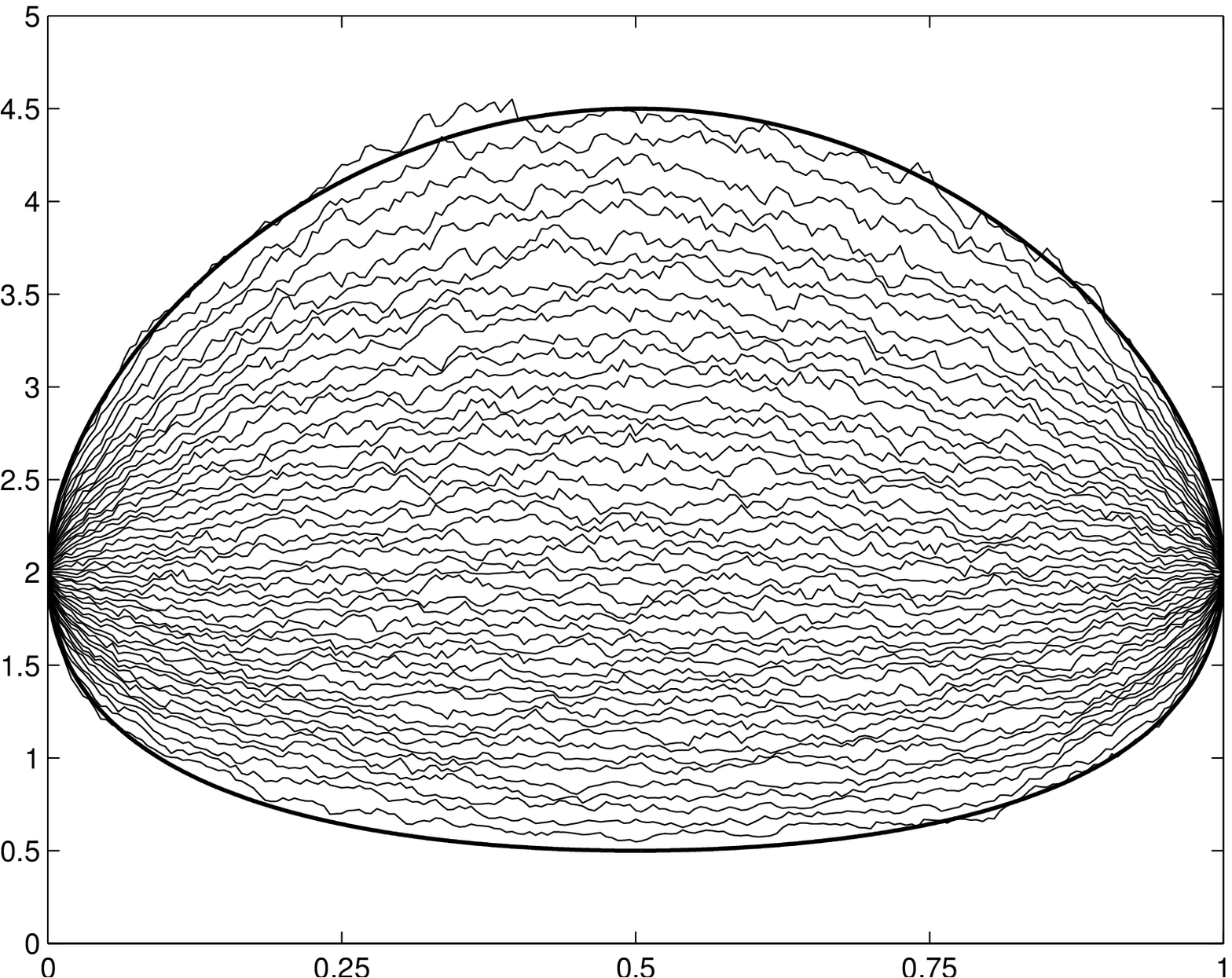}
\end{overpic}
\hspace{5mm}
\begin{overpic}[width=.4\textwidth]{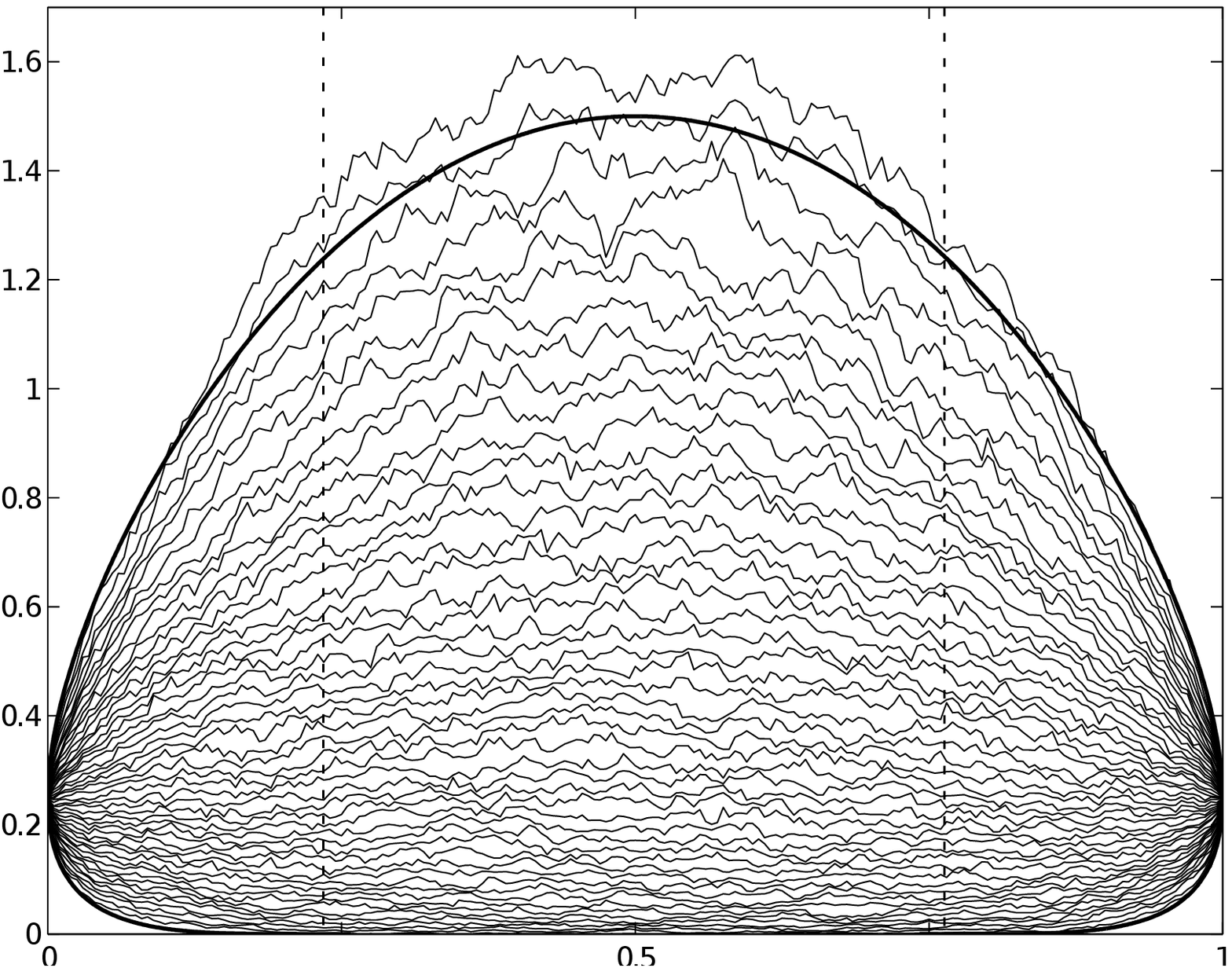}
%\put(8.5,-2){$0$}
\small{\put(25,-2){$t_1$} \put(74,-2){$t_2$}}
\end{overpic}
\caption{\label{fig:ab=4} 50 rescaled non-intersecting squared
Bessel paths with $a=b=2$ (left) and $a=b=1/4$ (right).}
\end{figure}

\begin{figure}[t]
\centering
\begin{overpic}[width=.4\textwidth]{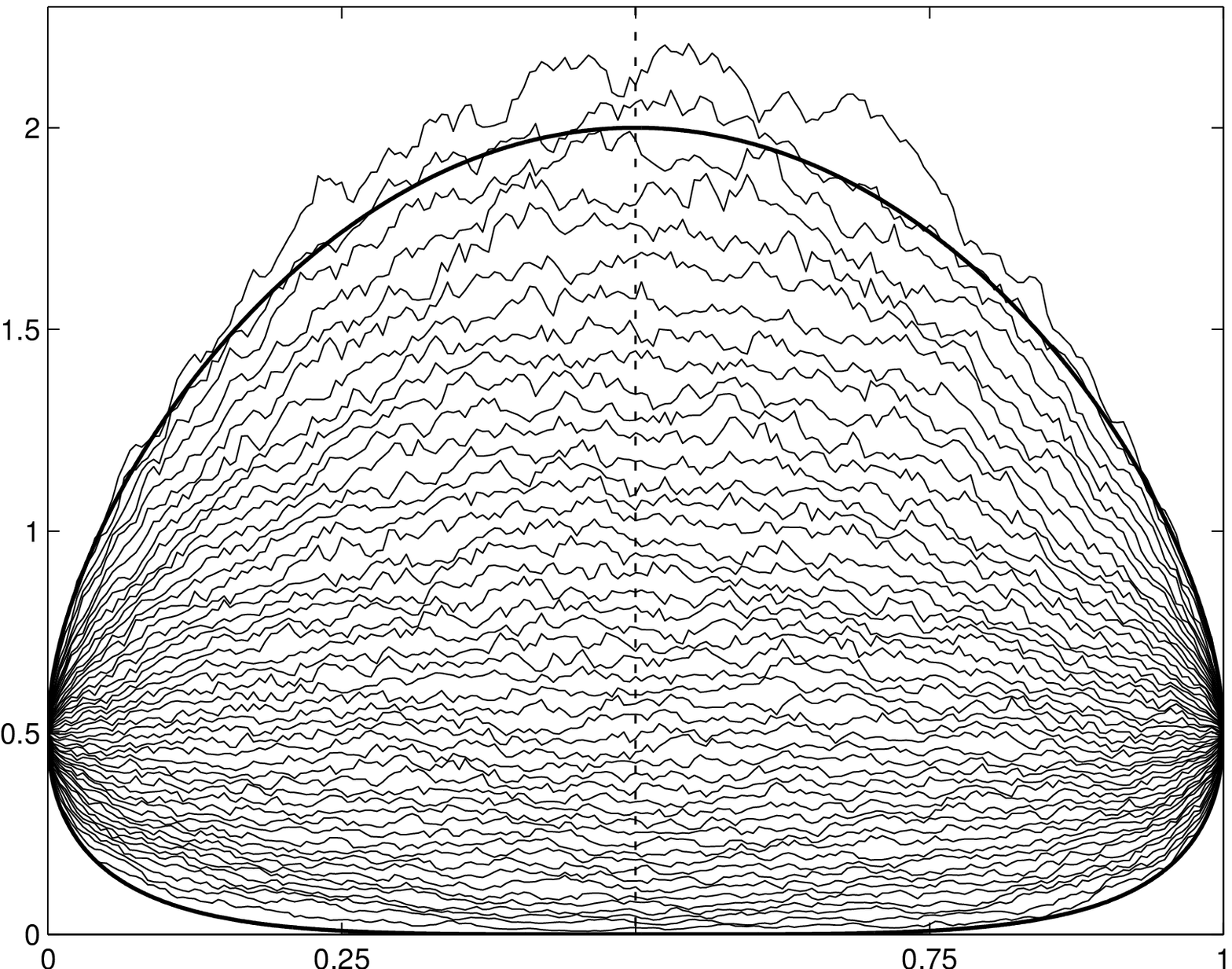}
\small{\put(50,-2){$t_c$}}
\end{overpic}
\hspace{5mm}
\begin{overpic}[width=.4\textwidth]{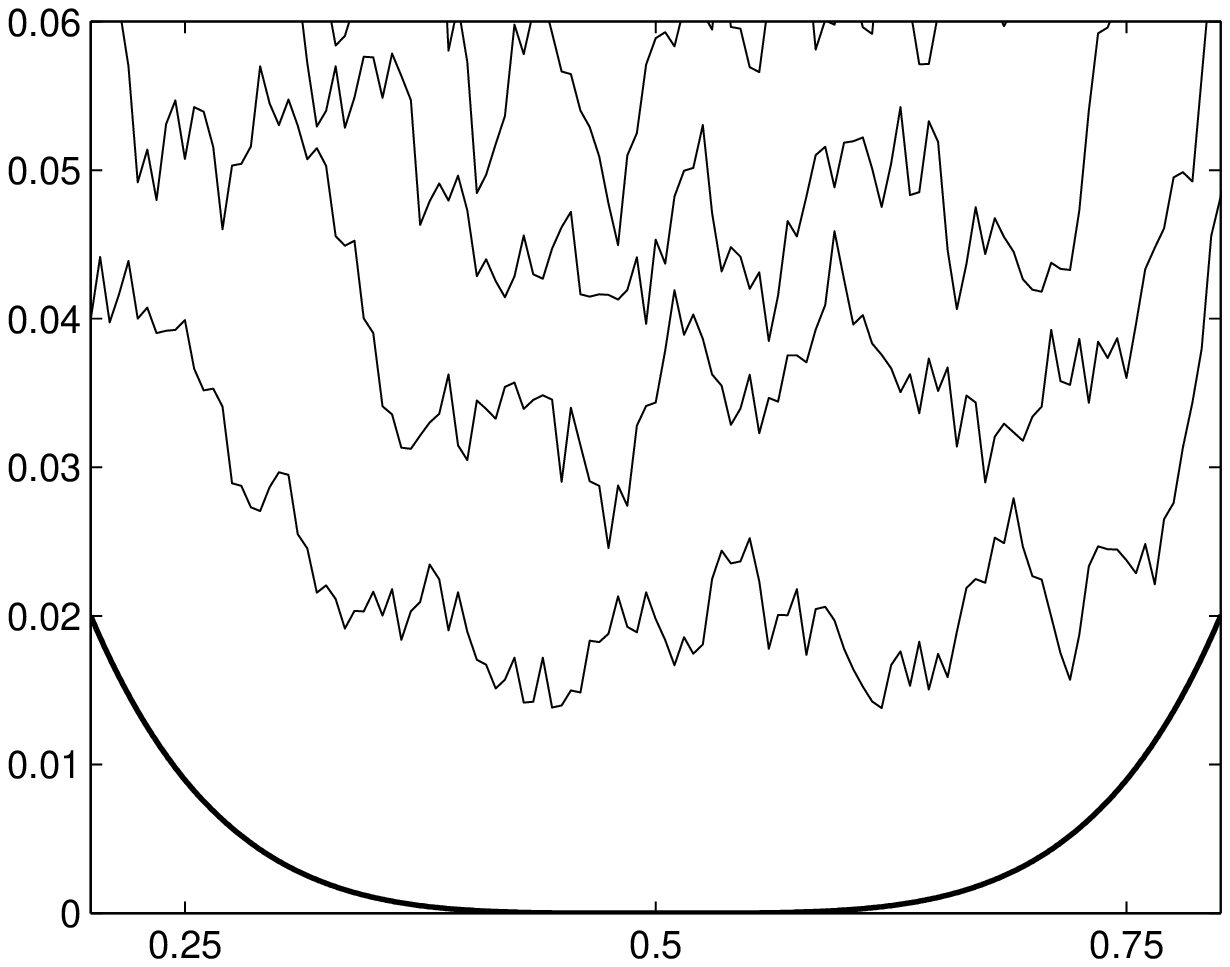}
\end{overpic}
\caption{\label{fig:ab=.25} 50 rescaled non-intersecting squared
Bessel paths with $a=b=1/2$ and $t_c=1/2$ (left) and around the
point $t_c$ (right).}
\end{figure}

In this paper, we  consider the case where both $a$ and $b$ are
positive. To obtain interesting results, we make a time
scaling as in \cite{KMW09} so that the time variable depends on the
number of paths $n$. That is, we replace the variables $t$ and $T$
by
\begin{equation}\label{time scaling}
    t \mapsto \frac{t}{2n}, \qquad T \mapsto \frac{1}{2n},
\end{equation}
so that $0<t<1$.
Now, letting $n\to\infty$, the paths fill out a
region in the $tx$-plane that looks like one of the regions shown in
Figures \ref{fig:ab=4} and \ref{fig:ab=.25}, depending on the
product $ab$. Suppose $a$ and $b$ stay far away from the hard edge
$x=0$, such that $ab>1/4$. Then we are in a situation as illustrated by
the left picture in Figure \ref{fig:ab=4}. All paths remain positive
for all time $t$. On the other hand, if $ab<1/4$, there are two
critical times $t_1$ and $t_2$, as depicted in the right picture of
Figure \ref{fig:ab=4}. All paths initially stay away from the hard
edge (a wall) and the lowest paths hit the hard edge at the time $t_1$. Since
all paths are required to end at a positive position $b$, the lowest
paths are stuck to the hard edge only for a while, and they leave
the wall at a second critical time $t_2$. The times $t_1$ and $t_2$ will come together if
$ab=1/4$, and we are then led to an intermediate situation shown in the
left picture in Figure \ref{fig:ab=.25}. The paths fill out
a region bounded by the solid line, which is tangent to the hard
edge at a multicritical time $t_c$; see also the right picture in
Figure \ref{fig:ab=.25} for a magnified view around $t_c$.

% \begin{figure}[t]
% \centering
% \begin{overpic}[width=.49\textwidth]{a=2b=2.eps}
% \end{overpic}
% \begin{overpic}[width=.49\textwidth]{a=.25b=.25.eps}
% \end{overpic}
% \caption{\label{fig:ab=4} 50 rescaled non-intersecting $\operatorname{BESQ}^2$ with
% $a=2$ and $b=2$ (left) and $a=1/4$ and $b=1/4$ (right).}
% \end{figure}

It is the aim of this paper to describe the region that is filled
in the large $n$ limit by the non-intersecting squared Bessel paths
with one positive starting point and one positive ending point. Our
results are stated in the next section.

%---------------------------------------------------------------------------
\section{Statement of results}

%-----------------------------------------------------------------------

%----------------------------------------------------------------
\subsection{A vector equilibrium problem}\label{Equilibrium problem}
The main ingredient of our analysis is a vector equilibrium problem
involving both an external field and two upper constraints.
To this end, we define, as usual (see \cite{SaffTotikBook}), the logarithmic energy of a
measure $\mu$  by
\begin{equation}
I(\mu)=\iint \log \frac{1}{|x-y|}d\mu(x)d\mu(y),
\end{equation}
and the mutual energy of two measures $\mu$, $\nu$ by
\begin{equation}
I(\mu,\nu)=\iint \log \frac{1}{|x-y|}d\mu(x)d\nu(y).
\end{equation}

Consider the energy functional
\begin{equation}\label{energy functional}
\begin{aligned}
E(\nu_1, \nu_2,\nu_3)
=\sum_{j=1}^3I(\nu_j)-\sum_{j=1}^{2}I(\nu_j,\nu_{j+1})+ \int
V(x)d\nu_2(x),
\end{aligned}
\end{equation}
where
\begin{equation}\label{V(x)}
V(x)=\frac{x}{t(1-t)} -\frac{2\sqrt{ax}}{t} -\frac{2\sqrt{bx}}{1-t}.
\end{equation}

The vector equilibrium problem is to minimize $E(\nu_1, \nu_2,\nu_3)$ among all
measures $\nu_1$, $\nu_2$ and $\nu_3$ satisfying the following conditions
(E1)--(E3):
\begin{enumerate}

\item[(E1)] $\nu_1$ is a measure on $\mathbb{R}^{-}:=(-\infty,0]$ with total mass
$1/2$ that satisfies the constraint
\begin{equation}
\nu_1\leq \rho_1,
\end{equation}
where $\rho_1$ is the measure on $\mathbb{R}^-$ with density
\begin{equation}\label{sigma a}
\frac{d\rho_1}{dx}=\frac{\sqrt{a}}{\pi t}|x|^{-1/2}, \qquad x\in\mathbb{R}^-.
\end{equation}

\item[(E2)] $\nu_2$ is a measure on $\mathbb{R}^{+}:=[0,\infty)$ with total mass $1$.

\item[(E3)] $\nu_3$ is a measure on $\mathbb{R}^{-}$ with total mass $1/2$
that satisfies the constraint
\begin{equation}
\nu_3\leq \rho_3,
\end{equation}
where $\rho_3$ is the measure on $\mathbb{R}^-$ with density
\begin{equation}\label{sigma b}
\frac{d \rho_3}{dx}=\frac{\sqrt{b}}{\pi(1-t)}|x|^{-1/2}, \qquad
x\in\mathbb{R}^-.
\end{equation}
\end{enumerate}

An electrostatic interpretation of the above equilibrium problem is the
following: Consider three types of charged particles. The particles of the
first and the third types are put on $\mathbb{R}^-$, both having total charge
$\frac{1}{2}$. The particles of the second type are put on $\mathbb{R}^+$ with
total charge 1. Particles of the same type repel each other. The particles of
the first and the second types interact with each other with an attraction that
is only half of the strength of the repulsion of particles of the same type. So
do the particles of the second and the third types. The particles of the first
and the third types do not interact directly. The particles of the second type
are influenced by the external field $V$. The measures $\rho_1$ and $\rho_3$ in
\eqref{sigma a} and \eqref{sigma b} play the role of upper constraints for the
first and the third measures. Namely, we minimize the equilibrium problem with
the extra condition that the particle densities of the first and the third
types do not exceed the densities of $\rho_1$ and $\rho_3$, respectively.

Vector equilibrium problems with mutual interaction can be traced back to
\cite{GR1,GR2}, see also \cite{NikiSoro1991}. The equilibrium problem with
constraints appeared before mainly in the asymptotic analysis of discrete
orthogonal polynomials, see e.g.~\cite{BKMM2007,DS1997}.

Our first result concerns the structure of the minimizer of the vector
equilibrium problem.
\begin{thm}\label{theorem:properties of minimizer} Let $a > 0$, $b > 0$ and $0 < t < 1$.
Then there is a unique minimizer $($$\mu_1$, $\mu_2$, $\mu_3$$)$ of the energy functional
\eqref{energy functional} under the conditions (E1), (E2), (E3). The minimizing measures
satisfy:
\begin{enumerate}[$($$a$$)$]
  \item The support of $\mu_1$ is the negative real axis and there
exists $r_1 \geq 0$ such that
  \begin{equation}
  \supp (\rho_1-\mu_1)=(-\infty, -r_1].
  \end{equation}
  The density of the measure $\rho_1-\mu_1$ vanishes like a square root at $-r_1$
  if $r_1>0$.

  \item The support of $\mu_2$ is bounded and there
  exist $0\leq p <q$ such that
  \begin{equation}
  \supp (\mu_2)=[p, q],
  \end{equation}
  and $\mu_2$ has a density which is real analytic in the interior of
  the support and vanishes like a square root at the endpoint $q$ and at $p$ if $p>0$.

  \item The support of $\mu_3$ is the negative real axis and there
  exists $ r_3\geq 0$ such that
  \begin{equation}
  \supp (\rho_3-\mu_3)=(-\infty, -r_3].
  \end{equation}
  The density of the measure $\rho_3-\mu_3$ vanishes as a square root at $-r_3$
  if $r_3>0$.
\end{enumerate}
\end{thm}
Note that $p,q,r_1$, and $r_3$ depend on $a, b$ and $t$.

The proof of Theorem \ref{theorem:properties of minimizer} will be given in
Section \ref{analysis of equ pro}. Throughout the whole paper, we will denote
\begin{align}\label{def of Delta}
\Delta_{1}:=(-\infty, -r_1),\quad \Delta_{2}:=(p,q), \quad
\textrm{and}\quad \Delta_{3}:=(-\infty,-r_3),
\end{align}
i.e., $\Delta_j$, $j=1,2,3$ is the interior of the sets $\supp(\rho_1-\mu_1)$,
$\supp (\mu_2)$ and $\supp(\rho_3-\mu_3)$, respectively.

\subsection{Classification into cases}
It turns out that at least one of $r_1$, $p$, $r_3$ is equal to $0$, and generically
no two consecutive ones are $0$. Generically we also have a $O(|x|^{-1/2})$ behavior
of the density at the hard edge $0$.   Then there are three generic cases.

\begin{description}
\item[Case I] $r_1 = r_3 =0$ and $p >0$. The densities of $\rho_1-\mu_1$
and $\rho_3-\mu_3$ blow up like $O(|x|^{-1/2})$ as $x \to 0-$.

\item[Case II] $p > 0$ and one of $r_1, r_3$ is zero and the other is positive.
Thus there are two subcases which play a symmetric role.
\begin{description}
\item[subcase IIa] $r_1>0$, $p > 0$ and $r_3=0$.
\item[subcase IIb] $r_1=0$, $p > 0$ and $r_3>0$.
\end{description}
\item[Case III] $r_1>0$, $p=0$ and $r_3>0$.
\end{description}

In Case I we have to add the condition that the densities of
$\rho_1-\mu_1$ and $\rho_3-\mu_3$ blow up like an inverse square
root at $0$ in order to be in a generic case. There is also a
non-generic case where one of these densities vanishes as a square
root. This is exactly what happens in the transition from Case I to
Case II. See Case V below.

The non-generic cases are:
\begin{description}
\item[Case IV] Exactly two consecutive values from $r_1, p, r_2$ are zero. Thus $p=0$ and
one of $r_1, r_3$ is also zero, while the other is positive.

\item[Case V] $r_1=r_3=0$ and $p > 0$. One of the densities of $\rho_j-\mu_j$ vanishes like a square root
$O(|x|^{1/2})$ as $x \to 0-$, while the other blows up like
$O(|x|^{-1/2})$ as $x \to 0-$.

\item[Case VI] All three are zero: $r_1= p = r_3=0$.
\end{description}

The Case I corresponds to the situation as shown by the left picture
in Figure \ref{fig:ab=4}. In this case $ab > 1/4$ and all paths stay away from the
hard edge at $x=0$ for all $t\in(0,1)$. Both constraints are never active.

The right picture of Figure \ref{fig:ab=4} which deals with $ab < 1/4$,
contains the three Cases II, III, IV. If $t<t_1$ or $t>t_2$, then we are in Case II, where only one
constraint is active. If $t_1<t<t_2$, then we are then in Case III, where
both constraints are active. The times $t_1$ and $t_2$ correspond to
Case IV, which represents a phase transition between soft edge and
hard edge.

The final Cases V and VI are contained in Figure~\ref{fig:ab=.25} for the
situation $ab= 1/4$ when the domain just touches the wall.
We are in Case V if $t\neq t_c$, and in case VI if $t=t_c$.

The Cases IV--VI are non-generic critical cases. They correspond to certain phase transitions
between the generic cases; see also Section \ref{subsec:phase transition} for a phase diagram
in the situation where $ab < 1/4$.

We shall only consider the generic cases in this paper. For
convenience (and without loss of generality), it is assumed that
\begin{equation}\label{assumption on r}
    r_3  \geq r_1
\end{equation}
in Cases II and III, which implies that
$\Delta_{3} \subset \Delta_{1}\subset \mathbb{R}^-$ holds in all three cases.

%----------------------------------------------------------------------
\subsection{Variational conditions}
Denote by
\begin{equation} \label{logarithmic potential}
    U^{\mu}(x)=\int\log\frac{1}{|x-y|}d\mu(y)
\end{equation}
the logarithmic potential of a measure $\mu$. The minimizer
$(\mu_1,\mu_2,\mu_3)$ is characterized by the following
Euler-Lagrange variational conditions:
\begin{prop}\label{prop:variation conditions}
The measures $\mu_1$, $\mu_2$ and $\mu_3$ satisfy for some constant
$l\in\mathbb{R}$,
\begin{align}
2U^{\mu_2}(x)&=U^{\mu_1}(x)+U^{\mu_3}(x)-V(x)+l, \qquad x\in
\supp(\mu_2), \label{var con mu2 1}
\\
2U^{\mu_2}(x)&>U^{\mu_1}(x)+U^{\mu_3}(x)-V(x)+l, \qquad x\in\mathbb{R}^+
\setminus \supp(\mu_2), \label{var con mu2 2}
\end{align}
where $V(x)$ is given in \eqref{V(x)};
\begin{align}
2U^{\mu_1}(x)&=U^{\mu_2}(x), \qquad x\in \supp(\rho_1-\mu_1),
\label{var con mu1 1}\\
2U^{\mu_1}(x)&<U^{\mu_2}(x), \qquad x\in \mathbb{C}\setminus
\supp(\rho_1-\mu_1)\label{var con mu1 2},
\end{align}
and
\begin{align}
2U^{\mu_3}(x)&=U^{\mu_2}(x), \qquad x\in
\supp(\rho_3-\mu_3),\label{var con mu3 1}
\\
2U^{\mu_3}(x)&<U^{\mu_2}(x), \qquad x\in \mathbb{C}\setminus
\supp(\rho_3-\mu_3).\label{var con mu3 2}
\end{align}
\end{prop}
Note that the inequalities in \eqref{var con mu2 2}, \eqref{var con
mu1 2} and \eqref{var con mu3 2} are all strict. The strictness of
the inequality in \eqref{var con mu2 2} follows from the strict
convexity of the external fields acting the second measure; see
Lemma~\ref{prop of nu2} below.

The proof of Proposition \ref{prop:variation conditions} will be
given in Section~\ref{analysis of equ pro}.

%WHY DO WE HAVE STRICT INEQUALITIES IN \eqref{var con mu2 2},
%\eqref{var con mu1 2} \eqref{var con mu3 2} ???

%-------------------------------------------------------------------

\subsection{Limiting mean distribution}
The following is the main theorem of this paper.
\begin{thm}\label{theorem:limiting mean distribution}
Let $a, b > 0$, $0 < t < 1$, and let $K_n$ denote the correlation
kernel for the positions of the non-intersecting squared Bessel
paths at time $t$ in the rescaled time variables \eqref{time
scaling}. Let $(\mu_1,\mu_2,\mu_3)$ be the minimizer of the vector
equilibrium problem stated in Section \ref{Equilibrium problem}.
Suppose that we are in one of the generic cases I, II, or III,
described above, then
\begin{equation} \label{Knlimit}
\lim_{n\to\infty}\frac{1}{n} K_n(x,x) = \frac{d\mu_2}{dx}(x),\qquad
x\in\mathbb{R}^+.
\end{equation}
\end{thm}
The limit \eqref{Knlimit} expresses that the density of the measure $\mu_2$ is
the limiting mean density of the non-intersecting squared Bessel paths
at time $t$ as $n\to \infty$.

It is an unusual feature that it is the second measure which plays the main
role here. In recent papers \cite{DuitK2009,DKM} concerning
eigenvalue distribution of a two-matrix model in random matrix
theory, a vector equilibrium problem with three measures appears as
well. In that situation, however, the first component of the
minimizer is relevant to the limiting mean distribution.

We prove Theorem \ref{theorem:limiting mean distribution} by means
of a steepest descent analysis of a relevant Riemann-Hilbert
problem. From this analysis, we also obtain the universality results
for the local path correlations. For Cases I, II, III, we obtain the
sine kernel (in the bulk), Airy kernel (at the soft edges) and
Bessel kernel of order $\alpha$ (at the hard edge $0$), which are
the typical limiting kernels in random matrix theory, see e.g.\
\cite{Deift99book,For,Mehta}. We will not discuss this any further
but instead refer to the papers
\cite{AptBleKuij2005,BK04,BK07,DuitK2009} for a more detailed
analysis in similar situations.

We expect the conclusion of Theorem \ref{theorem:limiting mean distribution} to
remain valid in the Cases IV--VI. The local path correlations at the origin in
Case IV should be the same as those derived in \cite{KMW10}, since they both
correspond to phase transitions between a soft and a hard edge. The Cases V--VI
represent new critical phenomena. It is conjectured that they are related to
the inhomogeneous Painlev\'{e} II equation, which bears some connections with
the `critical separation' case for non-intersecting Brownian motion with two
starting points and two ending points \cite{AFV2,DelKui1, DKZ, DG}. These results
will be reported in a future publication.

Theorem \ref{theorem:limiting mean distribution} will be proved in
Section \ref{section:proof:limitdistribution}.

%--------------------------------------------------------------------------
\subsection{About the proof of Theorem \ref{theorem:limiting mean distribution}}
The proof of Theorem \ref{theorem:limiting mean distribution} relies
on the Riemann-Hilbert (RH) problem for multiple orthogonal
polynomials of mixed type and its connection to the correlation
kernel \eqref{cor kernel}. To see this, we start with a proposition
which gives us the explicit structure of the biorthogonal ensemble
in the confluent case.
\begin{prop}
In confluent case $a_j\to a>0$, $b_j\to b> 0$, $j=1,\ldots,n$, the
positions of the non-intersecting squared Bessel paths at time
$t\in(0,T)$ are a biorthogonal ensemble \eqref{pdf different a b}
with functions
\begin{align}
f_{2j-1}(x)&=x^{j-1}p_t^\al(a,x), &&  j=1,\ldots,n_1:=\lceil n/2
\rceil, \label{f_(2j-1)}\\
f_{2j}(x)&=x^{j-1}p_t^{\al+1}(a,x), &&  j=1,\ldots,n_2:=n-n_1, \label{f_(2j-1)2}\\
g_{2j-1}(x)&=x^{j-1}p_{T-t}^\al(x,b), && j=1,\ldots,n_1, \label{g2j1}\\
g_{2j}(x)&=x^{j-1}p_{T-t}^{\al-1}(x,b), && j=1,\ldots,n_2.
\label{g2j}
\end{align}
\end{prop}
\begin{proof}
The functions \eqref{f_(2j-1)} and \eqref{f_(2j-1)2} follow as in
the proof of Proposition 2.1 in \cite{KMW09}.

In the confluent limit $b_j\to b$, the linear space spanned by the
functions $x\mapsto p_{T-t}^\alpha(x,b_j)$, $j=1,\ldots,n$, tends to
the linear space spanned by
\begin{equation}\label{eq:linear_span}
 x\mapsto \frac{\partial^{j-1}}{\partial y^{j-1}}p_{T-t}^\alpha(x,b), \quad j=1,\ldots,n.
\end{equation}
Using the differential relations satisfied by the modified Bessel
functions of the first kind (see \cite{HB92}), we can prove that
\begin{align*}
 \frac{\partial}{\partial y}p_{t}^\alpha(x,y)&=\frac{1}{2t}(p_{t}^{\alpha-1}(x,y)-p_t^\alpha(x,y)),
 \\
  y\frac{\partial}{\partial y}p_{t}^{\alpha-1}(x,y)&=\frac{x}{2t}p_{t}^{\alpha}(x,y)
 -\left(\frac{y}{2t}-(\alpha-1)\right)p_t^{\alpha-1}(x,y).
\end{align*}
Now it is easy to show inductively that the linear span of
\eqref{eq:linear_span} is the same as the linear space spanned by
the functions  \eqref{g2j1} and \eqref{g2j}.
\end{proof}

Next, we introduce two sets of weight functions $w_{1,1}, w_{1,2}$
and $w_{2,1}, w_{2,2}$,
\begin{align*}
w_{1,1}(x)&=x^{\al/2}e^{-\frac{nx}{t}}I_{\al}\left(\frac{2n\sqrt{ax}}{t}\right),
\quad &&w_{1,2}(x)=x^{(\al+1)/2}e^{-\frac{nx}{t}}
I_{\al+1}\left(\frac{2n\sqrt{ax}}{t}\right),
\\
w_{2,1}(x)&=x^{-\al/2}e^{-\frac{nx}{1-t}}
I_{\al}\left(\frac{2n\sqrt{bx}}{1-t}\right), \quad
&&w_{2,2}(x)=x^{-(\al-1)/2}e^{-\frac{nx}{1-t}}I_{\al-1}
\left(\frac{2n\sqrt{bx}}{1-t}\right).
\end{align*}
which are collected in two row vectors
\begin{equation}
\vec{w}_1=(w_{1,1}, w_{1,2}), \qquad \vec{w}_2=(w_{2,1},w_{2,2}).
\end{equation}
Note that by  \eqref{p_t alpha} and \eqref{time scaling}
\begin{equation} \label{fgandwij}
\begin{aligned}
f_{2j-1}(x) & \propto x^{j-1} w_{1,1}(x), & f_{2j}(x) & \propto x^{j-1} w_{1,2}(x), \\
g_{2j-1}(x) & \propto x^{j-1} w_{2,1}(x), & g_{2j}(x) & \propto x^{j-1} w_{2,2}(x),
\end{aligned}
\end{equation}
where $f_j$, $g_j$, $j=1,\ldots,n$ are given in \eqref{f_(2j-1)}--\eqref{g2j}.
The proportionality constants are irrelevant for us.

We then look for a $4 \times 4$ matrix valued function $Y$
satisfying the following RH problem:

\begin{enumerate}[(1)]
  \item $Y$ is defined and analytic in $ \mathbb{C} \setminus \mathbb{R}^{+}$.

  \item For $x >0 $, $Y$ possesses continuous boundary values
  $Y_{+}(x)$ (from the upper half plane) and $Y_{-}(x)$ (from the lower half
  plane), which satisfy
  \begin{equation}\label{Jump for Y}
  Y_+(x)=Y_-(x)
  \begin{pmatrix}
  I_2 & W(x) \\
  0 & I_2
  \end{pmatrix},
  \end{equation}
  where $I_k$ denotes the $k \times k$ identity matrix and $W(x)$ is the rank-one matrix
  \begin{align}\label{weight syt scaling}
  \nonumber W(x) & =\vec{w}_1^T(x) \vec{w}_2(x) \\
  &=e^{-\frac{nx}{(1-t)t}}\begin{pmatrix}
  I_{\al}\left(\frac{2n\sqrt{ax}}{t}\right)I_{\al}\left(\frac{2n\sqrt{bx}}{1-t}\right)
  &
  \sqrt{x}I_{\al}\left(\frac{2n\sqrt{ax}}{t}\right)I_{\al-1}\left(\frac{2n\sqrt{bx}}{1-t}\right)
  \\
  \sqrt{x}I_{\al+1}\left(\frac{2n\sqrt{ax}}{t}\right)I_{\al}\left(\frac{2n\sqrt{bx}}{1-t}\right)
  &
  x I_{\al+1}\left(\frac{2n\sqrt{ax}}{t}\right)I_{\al-1}\left(\frac{2n\sqrt{bx}}{1-t}\right)
  \end{pmatrix}.
  \end{align}

  \item As $z\to \infty$, $z\in\mathbb{C}\setminus \mathbb{R}^{+}$, we have
  \begin{equation}
  Y(z)=(I_4 + O(1/z)) ~\textrm{diag} (z^{n_1}, z^{n_2}, z^{-n_1},
  z^{-n_2}).
  \end{equation}

  \item $Y(z)$ has the following behavior near the origin:
  \begin{equation}\label{zero behavior of Y}
  Y(z)=O\begin{pmatrix}
  1 & 1 & h(z) & h(z) \\
  1 & 1 & h(z) & h(z) \\
  1 & 1 & h(z) & h(z) \\
  1 & 1 & h(z) & h(z) \\
  \end{pmatrix},
  \quad
  Y(z)^{-1}=O\begin{pmatrix}
  h(z) & h(z) & h(z) & h(z) \\
  1 & 1 & 1 & 1 \\
  1 & 1 & 1 & 1 \\
  1 & 1 & 1 & 1 \\
  \end{pmatrix}
  \end{equation}
  as $z\to 0$, $z\in\mathbb{C}\setminus\mathbb{R}^+$, where
  \begin{equation}\label{h(z)}
  h(z)=\begin{cases}
           |z|^{\alpha}, &   \text{ if } -1<\al<0, \\
           \log|z|, & \text{ if } \alpha=0, \\
           1, & \text{ if } \al>0,
         \end{cases}
  \end{equation}
  and the $O$ condition in \eqref{zero behavior of Y} is taken to be
  entrywise.
\end{enumerate}
There exists a unique solution $Y$ to the above RH problem which involves
multiple orthogonal polynomials of mixed type associated with the modified
Bessel weights $\vec{w}_1(x)$, $\vec{w}_2(x)$ and their Cauchy transforms; see
\cite{DK} for details. Due to the singularity of the weight matrix near the
origin, we need the condition \eqref{zero behavior of Y} to ensure the uniqueness of the solution.

It also follows from the results in \cite{DK} that the correlation kernel
\eqref{cor kernel} admits the following representation in terms of the solution of the RH problem:
\begin{equation}\label{kernel representation}
K_{n}(x,y)=\frac{1}{2\pi i(x-y)}\begin{pmatrix}0 &0 & w_{2,1}(y)&
w_{2,2}(y)\end{pmatrix} Y_{+}^{-1}(y)Y_{+}(x)
\begin{pmatrix}
w_{1,1}(x) \\ w_{1,2}(x) \\ 0 \\ 0
\end{pmatrix}.
\end{equation}
The representation \eqref{kernel representation} is based on the Christoffel-Darboux
formula for multiple orthogonal polynomials of mixed type, see also \cite{AlFiMa}.

We will obtain the limit \eqref{Knlimit} from a Deift/Zhou steepest descent analysis
\cite{Deift99book,DKMVZ992,DKMVZ991} for the RH problem for $Y$. It
consists of a series of explicit and invertible transformations
\begin{equation} \label{SDtransformations}
Y\mapsto X \mapsto U \mapsto T \mapsto R,
\end{equation}
which leads to a RH problem for a matrix valued function $R$ which is such that
$R$ tends to the identity matrix as $n\to\infty$. Analyzing the effect of the
transformations \eqref{SDtransformations} we obtain the limit \eqref{Knlimit}.

%\subsection{Outline of the rest of the paper}
The rest of this paper is organized as follows. In
Section~\ref{analysis of equ pro} we analyze the equilibrium problem
and establish Theorem~\ref{theorem:properties of minimizer} and
Proposition~\ref{prop:variation conditions}. In
Section~\ref{section:Riemannsurface} we introduce the Riemann
surface built from the solution to the vector equilibrium problem
and define some auxiliary functions. Section~\ref{sec:steepest
descent} contains the steepest descent analysis of the RH problem
for $Y(z)$ and Section \ref{section:proof:limitdistribution} gives
the proof of the main Theorem~\ref{theorem:limiting mean
distribution}. Throughout this paper, it is also assumed that $n$ is
an even number so that
\begin{equation} \label{n1isn2}
    n_1=n_2=n/2.
    \end{equation}
This assumption is not essential and is only made to simplify the presentation.

%--------------------------------------------------------------

\section{Analysis of the vector equilibrium problem}
\label{analysis of equ pro}

The main purpose of this section is to prove Theorem
\ref{theorem:properties of minimizer} and
Proposition~\ref{prop:variation conditions}, which yields the
existence and main properties of the minimizers $\mu_1$, $\mu_2$ and
$\mu_3$. Our basic tool is the balayage (sweeping out) of a measure.

Let $\nu$ be a finite positive measure on a closed set $K$ with positive
capacity. We denote with $\| \nu \|$ the total mass of $\nu$. A positive measure
$\hat \nu$ on $K$ is called the balayage of $\nu$ if $\|\nu \|= \|\hat \nu \|$ and
\begin{equation}\label{def of balayage}
U^{\nu}(x)=U^{\hat \nu}(x)+C, \qquad \textrm{q.e. $x\in K$},
\end{equation}
for some constant $C$, where ``q.e.'' means quasi-every, i.e., with
the exception of a set with zero capacity, and $U^{\nu}$ is the
logarithmic potential \eqref{logarithmic potential}, see
\cite{SaffTotikBook} for background on potential theory. To
emphasize the dependence on $K$, we also write
\begin{equation}
\hat \nu=\Bal(\nu, K).
\end{equation}
A relation between $\nu$ and $\Bal(\nu, K)$ is established as
follows:
\begin{equation}\label{bal of nu}
\Bal (\nu, K) = \int \Bal(\delta_s, K) d\nu(s),
\end{equation}
where $\delta_z$ denotes the Dirac delta measure at the point $z$.

For the purpose of this paper, we are interested in the case where
\[ K=K_c=(-\infty,-c], \qquad \text{with } c\geq 0. \]
Then $C=0$ in \eqref{def of balayage}, since $K_c$ is unbounded. By
using simple contour integration, it can be shown that, if $s\in\mathbb{R}^+$,
the balayage of $\delta_s$ onto $K_c$ has the density
\begin{equation}\label{bal on Kc}
\frac{d\Bal(\delta_s,K_c)}{dx}=\frac{1}{\pi}
\frac{\sqrt{s+c}}{\sqrt{|x+c|}(s-x)}, \qquad x\in K_c.
\end{equation}
In particular, for $c = 0$,
\begin{equation}\label{bal of delta}
\frac{d\Bal(\delta_s,\mathbb{R}^-)}{dx}=\lim_{c\to
0+}\frac{d\Bal(\delta_s,K_c)}{dx}=\frac{1}{\pi}
\frac{\sqrt{s}}{\sqrt{|x|}(s-x)}, \qquad x\in\mathbb{R}^-.
\end{equation}
Observe that the density \eqref{bal on Kc} is decreasing
as $|x|$ increases, with a decay rate $O(|x|^{-3/2})$ as $|x|\to\infty$. It
will become clear in what follows how the concept of balayage measure is linked
to the equilibrium problem.

%---------------------------------------------------------------------------
\subsection{Equilibrium problems for $\nu_1$ and $\nu_3$}
If $\nu_2$ and $\nu_3$ are given, the equilibrium problem for
$\nu_1$ is to minimize
\begin{equation}\label{eq pro for nu1}
I(\nu)-\int U^{\nu_2}(x)d\nu(x),
\end{equation}
among all measures $\nu$ on $\mathbb{R}^-$ with total mass $1/2$ and satisfying
the constraint $\nu \leq \rho_1$, where $\rho_1$ is the measure on
$\mathbb{R}^-$ with density given by \eqref{sigma a}. Clearly, this problem
only depends on $\nu_2$. The properties of $\nu_1$ are given in the following
lemma.

\begin{lem}\label{prop of nu1} (The measure $\nu_1$)
Suppose $\nu_2$ and $\nu_3$ are fixed so that conditions $(E2)$ and $(E3)$ of
the equilibrium problem are satisfied. Then the measure $\nu_1$ that minimizes
$E(\nu_1,\nu_2,\nu_3)$ subject to the condition $(E1)$ exists. It has the
following properties:
\begin{enumerate}[$($i$)$]
 \item If the constraint $\rho_1$ is not active, then
  the measure $\nu_1$ is given by
  \begin{equation}
  \nu_1=\frac{1}{2}\Bal(\nu_2,\mathbb{R}^-),
  \end{equation}
  or, equivalently, in view of \eqref{bal of nu} and \eqref{bal of
  delta},
  \begin{equation}\label{nu1 without constraint}
  \frac{d\nu_1(x)}{dx}=\frac{1}{2\pi\sqrt{|x|}}
  \int\frac{\sqrt{s}}{s-x}d\nu_2(s), \qquad x\in \mathbb{R}^-.
  \end{equation}

\item If the constraint $\rho_1$ is active, then there exists a
  constant $r_1>0$, such that
  \begin{equation}
  \supp (\rho_1-\nu_1)=(-\infty, -r_1].
  \end{equation}
Moreover, we have
\begin{equation}\label{var con for nu1}
\begin{cases}
2U^{\nu_1}(x)=U^{\nu_2}(x), &   x\in \supp(\rho_1-\nu_1), \\
2U^{\nu_1}(x)<U^{\nu_2}(x), &   x\in \mathbb{R}^-\setminus \supp(\rho_1-\nu_1),
\end{cases}
\end{equation}
and $\rho_1-\nu_1$ vanishes like a square root at $-r_1$.

\item Suppose $\nu_2$ has a compact support. Then here exists a
constant $\tilde K >0$, depending only on $\max \supp(\nu_2)$
 such that
\begin{equation}\label{bounds for nu1}
\frac{d\nu_1(x)}{dx}\leq \frac{\tilde K}{|x|^{3/2}}, \qquad x\in \mathbb{R}^-.
\end{equation}

\end{enumerate}
\end{lem}
\begin{proof}
\begin{enumerate}[(i)]
\item Suppose that $\nu_1=\Bal(\nu_2,\mathbb{R}^-)/2$. Then
  $2U^{\nu_1}(x)=U^{\nu_2}(x)$ for $x\in\mathbb{R}^-$, which
  is the variational condition for \eqref{eq pro for nu1} without
  constraint. This proves part (i).

\item Part (ii) is proved by using iterated balayage algorithm of
\cite{KD99}. The main idea is to construct a sequence of real
numbers $\{r_{1,k}(t)\}$ and a sequence of measures $\{\nu_{1,k}\}$
with limit $r_1(t)$ and $\nu_1$, respectively. Since this argument
is similar to those given in \cite{DuitK2009}, we omit the
proof here.

\item  Suppose $\nu_2$ has a compact support and let
\[ b = \max \supp(\nu_2). \]
By part (ii), the constraint is active on some interval $[-r_1,0)$
(which could be empty). Since $\| \nu_1 \| = 1/2$, we then have that
$r_1 \leq c_1$ where $c_1 > 0$ is the value for which
$\rho_1([-c_1,0]) = 1/2$. By properties of balayage we then have
that the restriction of $\nu_1$ to $K_{c_1}$ is less than the
balayage of $\nu_2/2$ onto $K_{c_1}$. Hence by \eqref{bal of nu} and
\eqref{bal on Kc}
\begin{equation}
\frac{d\nu_1(x)}{dx} \leq \frac{1}{2\pi \sqrt{|x+c_1|}} \int\frac{\sqrt{s+c_1}}{s-x}d\nu_2(s), \qquad
x\in(-\infty,-c_1].
\end{equation}
The function $s \mapsto \frac{\sqrt{s+c_1}}{s-x}$ is increasing for $s \in [-c_1, -x - 2c_1]$.
Thus if $x \leq - 2c_1 - b$, then the maximum on $\supp(\nu_2)$ is taken at its right end point $s=b$,
and therefore
\begin{equation}
\frac{d\nu_1(x)}{dx} \leq \frac{1}{2\pi \sqrt{|x+c_1|}} \frac{\sqrt{b+c_1}}{b-x}, \qquad
x\in(-\infty, - 2c_1 -b].
\end{equation}
Then there is a constant $\tilde K$, which only depends on $b$, such that \eqref{bounds for nu1} holds for $x \leq - 2c_1 -b$.
Since $\nu_1 \leq \rho_1$, we can then adjust the constant if necessary, such that \eqref{bounds for nu1}
holds for every $x < 0$.
\end{enumerate}
\end{proof}

If $\nu_1$ and $\nu_2$ are fixed, the equilibrium problem for $\nu_3$ is to
minimize \eqref{eq pro for nu1} among all measures $\nu$ on $\mathbb{R}^-$ with
total mass $1/2$ and satisfying the constraint $\nu \leq \rho_3$, where
$\rho_3$ is given in \eqref{sigma b}. Hence, the properties of $\nu_3$ can be
derived in a manner similar to the analysis given above. We omit the proof and
state the results in the following lemma:

\begin{lem}\label{prop of nu3} (The measure $\nu_3$)
Suppose $\nu_1$ and $\nu_2$ are fixed so that conditions $(E1)$ and $(E2)$ of
the equilibrium problem are satisfied. Then the measure $\nu_3$ that minimizes
$E(\nu_1,\nu_2,\nu_3)$ subject to the condition $(E3)$ exists. It has the
following properties:
\begin{enumerate}[$($i$)$]
  \item If the constraint $\rho_3$ is not active,
  the measure $\nu_3$ is given by
  \begin{equation}\label{nu3 without constraint}
  \frac{d\nu_3(x)}{dx}=\frac{1}{2\pi\sqrt{|x|}}
  \int\frac{\sqrt{z}}{z-x}d\nu_2(z), \qquad x\in\mathbb{R}^-.
  \end{equation}
  \item If the constraint $\rho_3$ is active, there exists a
  constant $r_3(t)>0$, such that
  \begin{equation}
  \supp (\rho_3-\nu_3)=(-\infty, -r_3(t)].
  \end{equation}
Moreover, we have
\begin{equation}\label{var con for nu3}
\begin{cases}
2U^{\nu_3}(x)=U^{\nu_2}(x), &  x\in \supp(\rho_3-\nu_3), \\
2U^{\nu_3}(x)<U^{\nu_2}(x), &  x\in \mathbb{R}^-\setminus \supp(\rho_3-\nu_3),
\end{cases}
\end{equation}
and $\rho_3-\nu_3$ vanishes like a square root at $-r_3(t)$.

\item Suppose $\nu_2$ has a compact support. Then there exists a
constant $\tilde K >0$, depending only on $\max \supp(\nu_2)$
 such that
\begin{equation}\label{bounds for nu3}
\frac{d\nu_3(x)}{dx}\leq \frac{\tilde K}{|x|^{3/2}}, \qquad x\in\mathbb{R}^-.
\end{equation}
\end{enumerate}
\end{lem}

%------------------------------------------------------------------------
\subsection{Equilibrium problem for $\nu_2$}
If $\nu_1$ and $\nu_3$ satisfying conditions $(E1)$ and $(E3)$ are fixed, the
equilibrium problem for $\nu_2$ is to minimize
\begin{equation}\label{equ pro for nu2}
I(\nu)+\int(V(x)-U^{\nu_1}(x)-U^{\nu_3}(x))d\nu(x)
\end{equation}
among all probability measures $\nu$ on $\mathbb{R}^+$, where
$V$ is given by \eqref{V(x)}. This is a usual equilibrium problem with external field
\begin{equation}\label{external field for nu2}
V(x)-U^{\nu_1}(x)-U^{\nu_3}(x)
\end{equation}
see e.g.\ \cite{Deift99book, SaffTotikBook}.

\begin{lem}\label{prop of nu2} (The measure $\nu_2$)
Suppose $\nu_1$ and $\nu_3$ are fixed so that conditions $(E1)$ and $(E3)$ in
the equilibrium problem are satisfied, in particular we have $\nu_j \leq \rho_j$ for $j=1,3$.
Then the measure $\nu_2$ that minimizes
$E(\nu_1,\nu_2,\nu_3)$ subject to condition $(E2)$ exists and has the following
properties:
\begin{enumerate}[$($i$)$]
  \item The external field \eqref{external field for nu2} is strictly convex
  on $\mathbb R^+$, and there exist two constants $0\leq p<q$ such that
  \begin{equation}
  \supp (\nu_2)=[p, q].
  \end{equation}
  \item If $0 \not\in \supp(\rho_1-\nu_1) \cup \supp(\rho_3 - \nu_3)$ (so that both constraints
  are active in a neighborhood of $0$), then \eqref{external field for nu2}
  is real analytic on $\mathbb R^+$ (including $0$).
% Hence, the density of $\nu_2$ blows up like $O(1/\sqrt{x})$ at the origin and
% vanishes like a square root at $q$.
\end{enumerate}
\end{lem}

\begin{proof}
\begin{enumerate}[(i)]
  \item The second derivative of \eqref{external field for nu2} is, for $x > 0$,
\begin{multline} \label{derivative of external field}
V''(x)-(U^{\nu_1})''(x)-(U^{\nu_3})''(x)
\\
=\frac{\sqrt{a}}{2t}x^{-3/2}+\frac{\sqrt{b}}{2(1-t)}x^{-3/2}
-\int_{-\infty}^0\frac{d\nu_1(s)}{(x-s)^2}
-\int_{-\infty}^0\frac{d\nu_3(s)}{(x-s)^2}.
\end{multline}
Since $\nu_1$ satisfies the constraint $\nu_1\leq\rho_1$ we have for $x > 0$,
\begin{align}\label{inequality for nu1}
\int_{-\infty}^0\frac{d\nu_1(s)}{(x-s)^2}
&<\int_{-\infty}^0\frac{d\rho_1(s)}{(x-s)^2}  =\frac{\sqrt{a}}{\pi
t}\int_{-\infty}^0\frac{ds}{\sqrt{|s|}(x-s)^2}=\frac{\sqrt{a}}{2
t}x^{-3/2},
\end{align}
where we used the explicit expression \eqref{sigma a} for $\rho_1$.
The above inequality is strict since $\nu_1$ cannot be equal to
$\rho_1$ on the whole negative real axis. Similarly, we find
\begin{align}\label{inequality for nu3}
\int_{-\infty}^0\frac{d\nu_3(s)}{(x-s)^2}
<\frac{\sqrt{b}}{2(1-t)}x^{-3/2}, \qquad x>0.
\end{align}
Combining \eqref{derivative of external field}--\eqref{inequality
for nu3}, we find that \eqref{derivative of external field} is
positive, and therefore \eqref{external field for nu2} is strictly convex.
It then follows from
\cite[Theorem IV.1.10]{SaffTotikBook} that the support of $\nu_2$ consists of
one interval, say, $[p,q]$ with $0\leq p< q$.

\item Suppose that $s > 0$ is such that $\nu_1=\rho_1$ and $\nu_3=\rho_3$ on $(-s,0)$.
Thus, if $x\geq 0$,
\begin{align}\label{U nu 1}
-U^{\nu_1}(x)&=\int_{-s}^{0}\log|x-y|d\rho_1(y)+ \int_{-\infty}^{-s} \log|x-y|d\rho_1(y)
\end{align}
where the second term is real analytic on $\mathbb R^+$, and for the first term
we use the explicit expression \eqref{sigma a} for $\rho_1$. Then we obtain
for the derivative of \eqref{U nu 1} (where ``real analytic'' means a real analytic function on $\mathbb R^+$)
\begin{align}\label{derivative of u nu1}
(-U^{\nu_1})'(x)&=\frac{\sqrt{a}}{\pi
t}\int_{-s}^{0}\frac{dy}{(x-y)\sqrt{-y}}+\textrm{``real
analytic''} \nonumber \\
%&=\frac{\sqrt{a}}{\pi t}\left(\int_{-\infty}^{0}-\int_{-\infty}^{-s}\right)
%\frac{dy}{(x-y)\sqrt{-y}}+\textrm{``real
%analytic''} \nonumber \\
&=\frac{\sqrt{a}}{\pi t}\int_{-\infty}^{0}
\frac{dy}{(x-y)\sqrt{-y}}+\textrm{``real
analytic''} \nonumber \\
&=\frac{\sqrt{a}}{t}x^{-1/2}+ \textrm{``real analytic''}.
\end{align}
Then by integrating \eqref{derivative of u nu1}
we find that $ U^{\nu_1}(x) + \frac{2 \sqrt{ax}}{t}$
is real analytic for $x \in \mathbb R^+$ (including at $x=0$).

Similarly, we find that $U^{\nu_3}(x) + \frac{2 \sqrt{bx}}{1-t}$ is real analytic on $\mathbb R^+$
as well. Then $V$ is real analytic because of
\eqref{V(x)} and \eqref{external field for nu2}.
\end{enumerate}
\end{proof}

%---------------------------------------------------------------
\subsection{Proofs of Theorem \ref{theorem:properties of minimizer}
and Proposition~\ref{prop:variation conditions}}
\paragraph{Proof of uniqueness.}
We rewrite the energy functional $E(\nu_1,\nu_2,\nu_3)$ as
\begin{equation}\label{energy functional 2}
E(\nu_1,\nu_2,\nu_3) = \frac{1}{2}I(\nu_2)
+\frac{1}{4}I(\nu_2-2\nu_1)+\frac{1}{4}I(\nu_2-2\nu_3)+\int
V(x)d\nu_2(x).
\end{equation}
It is well-known that if two distinct measures $\nu$ and $\mu$ have finite
logarithmic energy and $\int d\nu =\int d\mu$, then
\begin{equation}\label{I(nu-mu)>0}
I(\nu-\mu) \geq 0.
\end{equation}

Hence, it follows that the energy functional is strictly convex. The minimizer
is therefore uniquely determined by the Euler-Lagrange variational conditions
listed in Proposition~\ref{prop:variation conditions}. \qed

\paragraph{Proof of existence.}
It is easily seen from \eqref{energy functional 2}--\eqref{I(nu-mu)>0} that
\begin{equation}
E(\nu_1,\nu_2,\nu_3) \geq \frac{1}{2}I(\nu_2)+\int V(x)d\nu_2(x).
\end{equation}
Thus the energy functional is bounded from below. As a consequence, we can find
a sequence $(\nu_{1,n},\nu_{2,n},\nu_{3,n})$ such that
\begin{equation}
E(\nu_{1,n},\nu_{2,n},\nu_{3,n}) \leq \frac{1}{n}+ \inf
E(\nu_1,\nu_2,\nu_3),
\end{equation}
where both the sequence and the infimum are taken subject to conditions
$(E1)$--$(E3)$ of the equilibrium problem. In addition, we may assume that
\begin{equation*}
\supp(\nu_{2,n})\subset [p,q],
\end{equation*}
and, in view of \eqref{bounds for nu1}, \eqref{bounds for nu3},
\begin{equation*}
\frac{d\nu_{i,n}(x)}{dx}\leq \frac{\tilde K}{|x|^{3/2}}, \qquad i=1,3
\end{equation*}
with $0\leq p<q$ and $\tilde K>0$ independent of $n$. Therefore, there exists a
weakly convergent subsequence of $(\nu_{1,n},\nu_{2,n},\nu_{3,n})$, with a
certain weak limit $(\nu_1,\nu_2,\nu_3)$. By the fact that the energy
functional $E$ is lower semi-continuous \cite{HK}, the vector of measures
$(\nu_1,\nu_2,\nu_3)$ is then the required minimizer of the energy minimization
problem. \qed

\paragraph{Proofs of Theorem \ref{theorem:properties of minimizer}
and Proposition~\ref{prop:variation conditions}.} We already showed
that there exists a unique minimizer $(\mu_1,\mu_2,\mu_3)$ for the
equilibrium problem. The parts (a)--(c) of Theorem
\ref{theorem:properties of minimizer} and
Proposition~\ref{prop:variation conditions} are now immediate by
Lemmas \ref{prop of nu1}--\ref{prop of nu2}. \qed

%---------------------------------------------------------------

\section{Riemann surface and auxiliary functions}
\label{section:Riemannsurface}
In this section, we shall construct a Riemann
surface $\mathcal{R}$ with the minimizer $(\mu_1,\mu_2,\mu_3)$ of the vector
equilibrium problem and introduce some auxiliary functions for further use.

\subsection{A four-sheeted Riemann surface $\mathcal{R}$}
\label{Riemann surface}

The Riemann surface $\mathcal{R}$ has four sheets $\mathcal{R}_j$,
$j=1,\ldots,4$, which are defined as
\begin{equation}
\begin{aligned}
\mathcal {R}_1&=\mathbb{C}\setminus \supp(\rho_1-\mu_1), \qquad
\mathcal {R}_2=\mathbb{C}\setminus (\supp(\rho_1-\mu_1)\cup
\supp(\mu_2)),
\\
\mathcal {R}_3&=\mathbb{C}\setminus (\supp(\mu_2)\cup
\supp(\rho_3-\mu_3)), \qquad \mathcal {R}_4=\mathbb{C}\setminus
\supp(\rho_3-\mu_3).
\end{aligned}
\end{equation}
These four sheets are connected as follows: $\mathcal {R}_1$ is connected to
$\mathcal {R}_2$ via $ \supp(\rho_1-\mu_1)$, $\mathcal {R}_2$ is connected to
$\mathcal {R}_3$ via $ \supp(\mu_2)$, and $\mathcal {R}_3$ is connected to
$\mathcal {R}_4$ via $ \supp(\rho_3-\mu_3)$. Each connection is in the usual
crosswise manner; see Figure \ref{fig:Riemann surface}.
\begin{figure}[t]
\begin{center}
\begin{overpic}[scale=1]{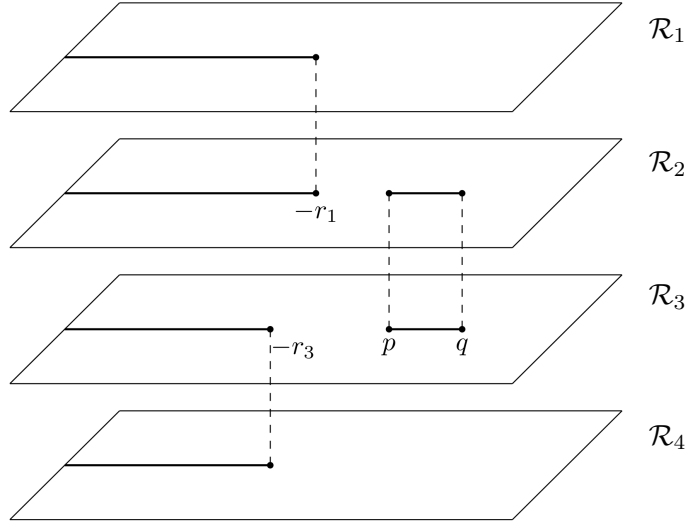}%,viewport=36 31 346 292
\put(100,80){$\mathcal{R}_1$} \put(100,60){$\mathcal{R}_2$}
\put(100,40){$\mathcal{R}_3$} \put(100,19){$\mathcal{R}_4$}
\end{overpic}
\caption{The Riemann surface $\mathcal{R}$.} \label{fig:Riemann
surface}
\end{center}
\end{figure}

The Riemann surface is compactified by adding two points at
infinity: one is common to the sheets $\mathcal{R}_1$ and
$\mathcal{R}_2$, and the other one is common to the sheets
$\mathcal{R}_3$ and $\mathcal{R}_4$. Since the Riemann surface
$\mathcal{R}$ has 4 sheets and 6 simple branch points, it then
follows from Hurwitz's theorem (see e.g. \cite{MRick}) that it has
genus 0. This fact will be helpful in our future construction of the
global parametrix.

\subsection{The $\xi$-functions}
We use $F_j$ to denote the Cauchy transform of the measure $\mu_j$,
i.e.,
\begin{equation}
F_j(z)=\int\frac{1}{z-x}d\mu_j(x), \quad z\in\mathbb{C}\setminus
\supp(\mu_j),\quad j=1,2,3.
\end{equation}
With the above notation, the Sokhotski-Plemelj formulas now read
\begin{align}
F_{j,+}(z)+F_{j,-}(z)&=2 \PV \int \frac{1}{z-x}d\mu_j(x),
\label{Sokhotski-Plemelj 1}\\
F_{j,+}(z)-F_{j,-}(z)&=-2\pi i \frac{d\mu_j}{dz}
\label{Sokhotski-Plemelj 2},
\end{align}
for $z\in \supp(\mu_j)$, where $\PV$ denotes the Cauchy principal value. The
$\xi$-functions are then defined as follows.
\begin{thm}\label{definition of xi function}
The function $\xi: \bigcup_{j=1}^{4} \mathcal{R}_j \rightarrow
\mathbb{C}$ defined by
\begin{equation}\label{xi function}
\xi(z)=\begin{cases}
-F_1(z)+\frac{\sqrt{a}}{t\sqrt{z}}+\frac{1}{t(1-t)}, &  z\in \mathcal{R}_1,\\
F_1(z)-F_2(z)-\frac{\sqrt{a}}{t\sqrt{z}}+\frac{1}{t(1-t)}, & z\in \mathcal{R}_2, \\
F_2(z)-F_3(z)+\frac{\sqrt{b}}{(1-t)\sqrt{z}}, & z\in \mathcal{R}_3, \\
F_3(z)-\frac{\sqrt{b}}{(1-t)\sqrt{z}}, & z\in\mathcal{R}_4,
\end{cases}
\end{equation}
has an extension to a meromorphic function (also denoted by $\xi$) on
$\mathcal{R}$. In \eqref{xi function}, the principal branch of the square root function $\sqrt{z}$ is used, i.e.,
with a branch cut along $(-\infty,0]$.
\end{thm}
\begin{proof}
We first need to show the analyticity of $\xi$ on each individual sheet
$\mathcal{R}_j$. To see this, we note that $\mu_1=\rho_1$ on
$\mathbb{R}^-\setminus \supp(\rho_1-\mu_1)=(-r_1,0)$. It then follows from
\eqref{Sokhotski-Plemelj 2} and the definition \eqref{sigma a} of $\rho_1$ that
\begin{align}
    F_{1,+}(x)-F_{1,-}(x)& =-2 \pi i \frac{d \mu_1(x)}{dx}= -2 \pi i
    \frac{d \rho_1(x)}{dx} \nonumber \\
    &=-2i\frac{\sqrt{a}}{t|x|^{1/2}}
    =\frac{\sqrt{a}}{t} \left(\frac{1}{\sqrt{x}_+}- \frac{1}{\sqrt{x}_-}\right),
\end{align}
for $x\in (-r_1,0)$. By the definition of $\xi$, this implies that $\xi$ has an
analytic continuation to $(-r_1,0)$ on both $\mathcal{R}_1$ and
$\mathcal{R}_2$. Similarly, we have by \eqref{Sokhotski-Plemelj 2} and \eqref{sigma b}
\begin{align}
    F_{3,+}(x)-F_{3,-}(x)=\frac{\sqrt{b}}{1-t} \left(\frac{1}{\sqrt{x}_+}- \frac{1}{\sqrt{x}_-}\right),
\end{align}
for $x \in (-r_2,0)$, which gives the analyticity of $\xi$ on the sheets $\mathcal{R}_3$
and $\mathcal{R}_4$.

It remains to check the analyticity of $\xi$ if one moves from
one sheet to another via a specific cut. This is always the case in
view of the variational conditions for the equilibrium problem. For
convenience, we will only show that $\xi$ is analytic when one
crosses $\supp(\rho_1-\mu_1)$, since the other cases can be verified
similarly.

We start with differentiating $U^{\mu_1}$ on $\supp(\rho_1-\mu_1)$
and obtain
\begin{align}\label{derivative of Umu1 1}
2\frac{d}{dx}U^{\mu_1}(x) = -2 \PV\int \frac{1}{x-s}d\mu_1(s) =
-F_{1,+}(x)-F_{1,-}(x),
\end{align}
where the second equality follows from Sokhotski-Plemelj formula
\eqref{Sokhotski-Plemelj 1}. On the other hand, differentiating the
right-hand side of the variational condition \eqref{var con mu1 1},
we have
\begin{align}\label{derivative of Umu1 2}
2\frac{d}{dx}U^{\mu_1}(x) = \frac{d}{dx}U^{\mu_2}(x) = - \int \frac{1}{x-s}
d\mu_2(s)=-F_2(x).
\end{align}
Combining \eqref{derivative of Umu1 1} and \eqref{derivative of Umu1
2} leads to
\begin{equation}
F_{1,+}(x)+F_{1,-}(x)=F_2(x), \qquad x\in\supp(\rho_1-\mu_1).
\end{equation}
This, together with the fact that $\sqrt{x}_- = - \sqrt{x}_+$ for $x\in
\mathbb{R}^-$, gives the analyticity of $\xi$ when one crosses
$\supp(\rho_1-\mu_1)$ and moves from $\mathcal{R}_1$ to $\mathcal{R}_2$.
\end{proof}

As a consequence of the above theorem, the following result about the
asymptotic behavior of $F_j(z)$ at infinity is easy to prove.
\begin{lem}\label{asy of F_j}
There exist two constants $c_1$ and $c_3$ such that
\begin{align}
F_{1}(z)&=\frac{1}{2z}-\frac{c_1}{z^{3/2}}+O\left(\frac{1}{z^2}\right), \\
F_{2}(z)&=\frac{1}{z}+O\left(\frac{1}{z^2}\right), \\
F_{3}(z)&=\frac{1}{2z}-\frac{c_3}{z^{3/2}}+O\left(\frac{1}{z^2}\right),
\end{align}
as $z\to \infty$.
\end{lem}

We use $\xi_j(z)$ to denote the restriction of $\xi$ to $\mathcal{R}_j$. The
relations between $\xi_j$ and the solution to the equilibrium problem can be
established in the following way.

\begin{prop}
We have that
\begin{align}
d\mu_1(x)&=\frac{1}{2\pi i}(\xi_{1,+}(x)-\xi_{1,-}(x))dx+d\rho_1(x),
\quad &&x\in\mathbb{R}^-, \label{dmu_1}\\
d\mu_2(x)&=\frac{1}{2\pi i}(\xi_{2,+}(x)-\xi_{2,-}(x))dx, \quad
&&x\in\mathbb{R}^+, \label{dmu_2}\\
d\mu_3(x)&=\frac{1}{2\pi i}(\xi_{3,+}(x)-\xi_{3,-}(x))dx+d\rho_3(x), \quad
&&x\in\mathbb{R}^-. \label{dmu_3}
\end{align}
\end{prop}
\begin{proof}
This follows by a combination of \eqref{xi function}, the Sokhotski-Plemelj
formula \eqref{Sokhotski-Plemelj 2}, and the definitions \eqref{sigma a} and \eqref{sigma b}
of $\rho_1$ and $\rho_2$.
\end{proof}

The Riemann surface has two points at infinity and the meromorphic function \eqref{xi function}
takes on two distinct finite values at these points (namely $\frac{1}{t(1-t)}$ and $0$).
Thus $\xi$ is not a constant, and being a meromorphic function on a compact Riemann surface
it must have at least one pole. From \eqref{xi function} it is clear that the only possible poles
are at the values of $z=0$ that correspond to branch points. It thus follows that at least one of $r_1, p$, and $r_3$
is zero, and we can identify the various cases in terms of the structure of $\mathcal R$.

In Case I we have $r_1=r_3 = 0$ and $p >0$. It is only in this case that $\xi$ has two poles. In all other
cases, $\xi$ has exactly one pole, and then $\xi$ is actually a conformal map from $\mathcal R$ onto the
Riemann sphere.

%--------------------------------------------------------------------
\subsection{The spectral curve}
From Theorem \ref{definition of xi function}, it follows that the four
function $\xi_j(z)$, $j=1,2,3,4$ are solutions to an algebraic equation of degree four
\begin{multline}\label{eq:spectral-curve}
    0=P(\xi,z)=(\xi-\xi_1)(\xi-\xi_2)(\xi-\xi_3)(\xi-\xi_4)\\
        =\xi^4+A(z)\xi^3+B(z)\xi^2+C(z)\xi+D(z),
\end{multline}
with coefficients $A(z)$, $B(z)$, $C(z)$, $D(z)$ that can only have
a pole in $z=0$. In fact, because of the behavior of \eqref{xi
function} at $z=0$, we have that $A(z)$ has no pole at $z=0$ (and
therefore is a constant), $B(z)$ and $C(z)$ have at most a simple
pole at $z=0$. Also $D(z)$ has at most a simple pole at $z=0$,
unless we are in Case I, in which case $D(z)$ has a double pole at
$z=0$. Indeed, in this last case we have that all functions $\xi_j$
behave like $z^{-1/2}$ as $z \to 0$, and then $D(z) = \xi_1(z)
\xi_2(z) \xi_3(z) \xi_4(z)$ behaves like $z^{-2}$ as $z \to 0$.

Straightforward calculations using \eqref{xi function} and Lemma \ref{asy of F_j} yield in all cases
\begin{align}
    A(z) & = -\xi_1-\xi_2-\xi_3-\xi_4  = -\frac{2}{t(1-t)}, \label{Az} \\
    B(z) & = \xi_1\xi_2+\xi_1\xi_3+\xi_1\xi_4+\xi_2\xi_3+\xi_2\xi_4+\xi_3\xi_4 \nonumber\\
    & = {\frac {1}{{t}^{2} \left( 1-t \right) ^{2}}}-\frac {b}{ z(1-t)^{2}}-\frac {a}{zt^{2}}+\frac {1}{zt (1-t) }, \label{Bz} \\
    C(z) & =  \frac {2b}{zt(1-t)^{3}}- \frac{1}{zt^{2}(1-t)^{2}}, \label{Cz} \\
    D(z) & =   - \frac{b}{zt^2(1-t)^4} + \frac{c}{z^2 t^2(1-t)^2}, \label{Dz}
    \end{align}
where $c=0$ in all cases but in Case I.

\begin{lem}
The spectral curve is given by \eqref{eq:spectral-curve} with coefficients
given by \eqref{Az}--\eqref{Dz}. The constant $c$ that appears in \eqref{Dz} is equal to
\begin{equation} \label{valueforc}
    c = \begin{cases}  \left( \sqrt{ab}-\frac{1}{2} \right)^2 & \text{ in Case I,} \\
      0 & \text{ in all other cases.}
      \end{cases}
      \end{equation}
In Case I, we have
\begin{equation}
\begin{aligned} \label{sqrtpq}
    \sqrt{p} &= (1-t) \sqrt{a} + t \sqrt{b} - \sqrt{2t(1-t)}  > 0, \\
    \sqrt{q} & = (1-t) \sqrt{a} + t \sqrt{b} + \sqrt{2t(1-t)}.
    \end{aligned}
    \end{equation}
\end{lem}
\begin{proof}
We already saw that the spectral curve is given by \eqref{eq:spectral-curve},  \eqref{Az}--\eqref{Dz}
and that $c=0$ in all cases, except in Case I.

Now suppose we are in Case I, so that $r_1 =  r_3 = 0$ and $p > 0$.
Consider the following functions
\begin{align}  \label{eta1eta2}
    \eta_1(z) & = \begin{cases} z \xi_2(z^2), &  \Re z > 0, \\
           z \xi_1(z^2), & \Re z < 0, \end{cases} \qquad
           \eta_2(z) = \begin{cases} z \xi_1(z^2), & \Re z > 0, \\
           z \xi_2(z^2), & \Re z < 0, \end{cases} \\
   \eta_3(z) & = \begin{cases} z \xi_3(z^2), &  \Re z > 0, \\
           z \xi_4(z^2), & \Re z < 0, \end{cases} \qquad
           \eta_4(z) = \begin{cases} z \xi_4(z^2), & \Re z > 0, \\
           z \xi_3(z^2), & \Re z < 0. \end{cases} \label{eta3eta4}
           \end{align}
Since, we are in Case I, we have that all functions $\eta_j$ are
analytic across the imaginary axis. Also the singularities at $z=0$
are removable because of the extra factors $z$ in \eqref{eta1eta2} and \eqref{eta3eta4}.
It follows that $\eta_1$ and $\eta_3$
are analytic in $\mathbb C \setminus [\sqrt{p}, \sqrt{q}]$, while
$\eta_2$ and $\eta_4$ are analytic in $\mathbb C \setminus [-\sqrt{q}, -\sqrt{p}]$.

In fact the two functions $\eta_1$ and $\eta_3$ give a meromorphic function on
the two-sheeted Riemann surface with cut along $[\sqrt{p}, \sqrt{q}]$, whose
only pole is at infinity on the first sheet.
Since
\begin{align*}
    \eta_1(z) & = \frac{z}{t(1-t)} - \frac{\sqrt{a}}{t} - \frac{1}{2z} + O(z^{-2}) \\
    \eta_3(z) & = \frac{\sqrt{b}}{1-t} + \frac{1}{2z} + O(z^{-2})
    \end{align*}
    as $z \to \infty$, it then follows that $\eta_1(z)$ and $\eta_3(z)$ are the two solutions of
    the quadratic equation
    \begin{equation} \label{eta13equation}
        \eta^2 - \left(\frac{z}{t(1-t} - \frac{\sqrt{a}}{t} + \frac{\sqrt{b}}{1-t}\right) \eta
        + \left( \frac{\sqrt{b}z}{t(1-t)^2} - \frac{\sqrt{ab}}{t(1-t)}  + \frac{1}{2t(1-t)} \right) = 0.
        \end{equation}

Similarly, $\eta_2(z)$ and $\eta_4(z)$ are the two solutions of
    \[ \eta^2 - \left(\frac{z}{t(1-t)} + \frac{\sqrt{a}}{t} - \frac{\sqrt{b}}{1-t}\right) \eta
        + \left( -\frac{\sqrt{b}z}{t(1-t)^2} - \frac{\sqrt{ab}}{t(1-t)}  + \frac{1}{2t(1-t)} \right) = 0. \]

In particular, it follows that
\begin{align*}
    \eta_1(z) \eta_2(z) \eta_3(z) \eta_4(z)
    = - \frac{b z^2}{t^2(1-t)^4} + \frac{(\sqrt{ab} - \frac{1}{2})^2}{t^2(1-t)^2}.
    \end{align*}
Then in view of \eqref{eta1eta2} and \eqref{eta3eta4} we obtain
\begin{align*}
    z^4 D(z^2) =
    z^4 \xi_1(z^2) \xi_2(z^2) \xi_3(z^2) \xi_4(z^2)
    = - \frac{bz^2}{t^2(1-t)^4} + \frac{(\sqrt{ab} - \frac{1}{2})^2}{t^2(1-t)^2}.
    \end{align*}
Comparing this with \eqref{Dz} we indeed find that $c$ is given by \eqref{valueforc} in Case I.

The branch points of \eqref{eta13equation} are equal to $\sqrt{p}$ and $\sqrt{q}$.
Simple calculations then lead to \eqref{sqrtpq}.
\end{proof}

\begin{remark}
We notice that the spectral curve in Case~I is equivalent with the one in
\cite{DelKui1}, while in Cases~II and III it is equivalent with the one in
\cite{DKV} (with $\sqrt{a},\sqrt{b}$ instead of $a,b$). This implies that the
limiting distribution of the non-intersecting squared Bessel paths is obtained
from the one of the corresponding non-intersecting Brownian motions by a simple
squaring transform.

One could extend the above ideas even further by also lifting the vector
equilibrium problem into its square-root form. The lifted measures are then
supported on the real or imaginary axis. This approach can be used to obtain an
equilibrium problem interpretation for the spectral curve in \cite{DKV}, which
allows in turn to obtain an alternative derivation of the results in that
paper. We will not go into the details.
\end{remark}

\begin{prop}
We are in Case I if and only if $ab > 1/4$.
\end{prop}

\begin{proof}
The spectral curve depends continuously on the parameters $a$, $b$ and $t$.
Therefore the change from Case I to one of the other cases can only occur when $ab = 1/4$.

If we are in Case I, then $\sqrt{p}$ is given by \eqref{sqrtpq}
which in particular implies that $(1-t) \sqrt{a} + t \sqrt{b} -
\sqrt{2t(1-t)} > 0$. For $t=1/2$ and $a=b < 1/2$ this inequality is
not satisfied and so for these values of parameters we are not in
Case I. Then it follows that for all $a$ and $b$ with $ab \leq 1/4$
we are not in Case I.

If we are not in Case I, then $c=0$ and the spectral curve
\eqref{eq:spectral-curve} has four real branch points $-r_1$,
$-r_2$, $p$ and $q$. For the specific values $t=1/2$, $a=1$ and
$b=1/2$, we calculated with Maple that there are two non-real branch
points. Thus for those values of parameters we cannot have $c=0$,
and so are in Case I. By continuity we are in Case I for all $a$ and
$b$ with $ab > 1/4$. This completes the proof of the proposition.
\end{proof}

%--------------------------------------------------------------------
\subsection{Phase transitions}\label{subsec:phase transition}

\begin{figure}
\centering
\begin{overpic}[width=.4\textwidth]{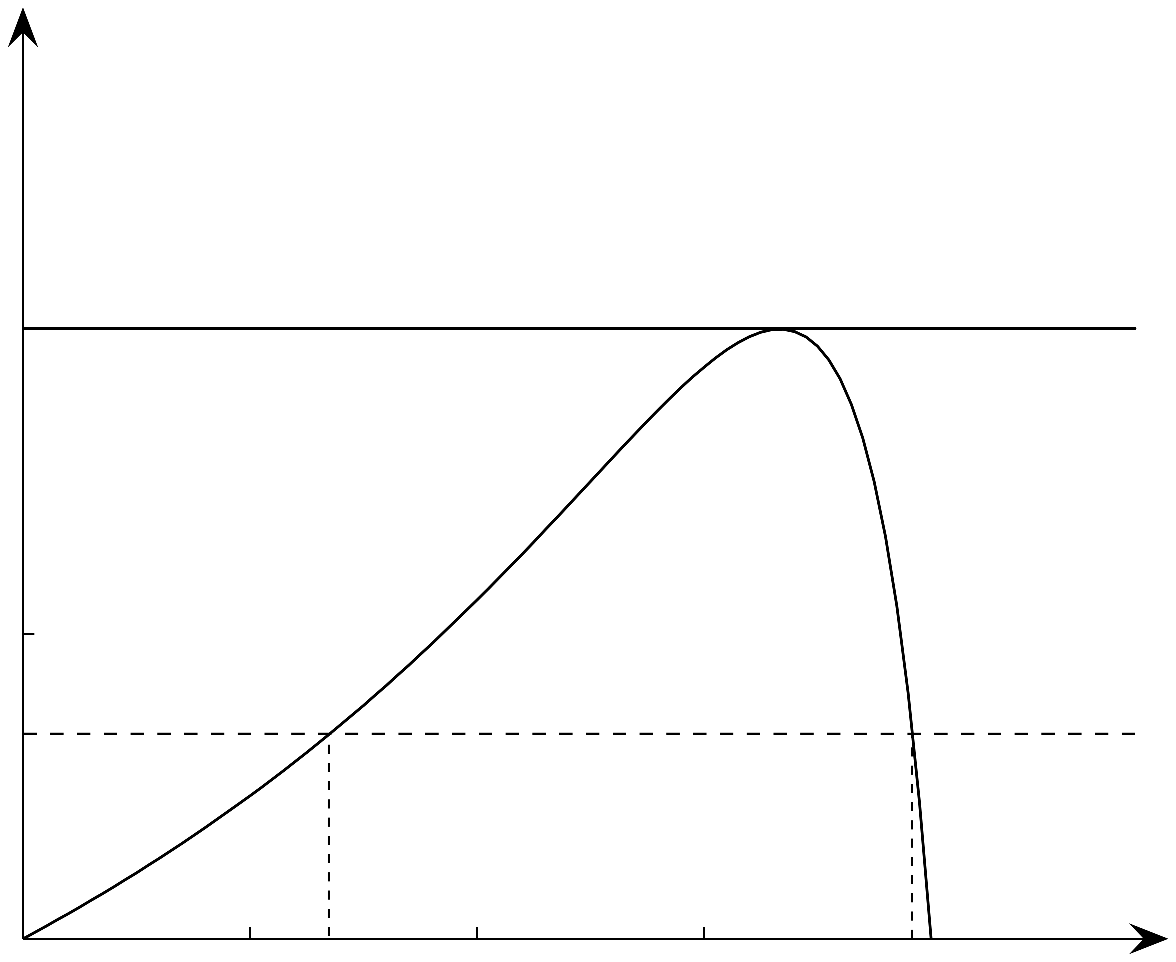}
\small{\put(20,35){Case II} \put(50,35){Case III} \put(76,35){Case
II} \put(45,60){Case I} \put(5,35){$a$} \put(5,21.5){$\frac{1}{3}$}
\put(5,52){$1$} \put(5,4){$0$} \put(32,3.5){$t_1$}
\put(74,3.5){$t_2$} \put(53,3.5){$t$} \put(85,3.5){$1$}}
\end{overpic}
\begin{overpic}[width=.4\textwidth]{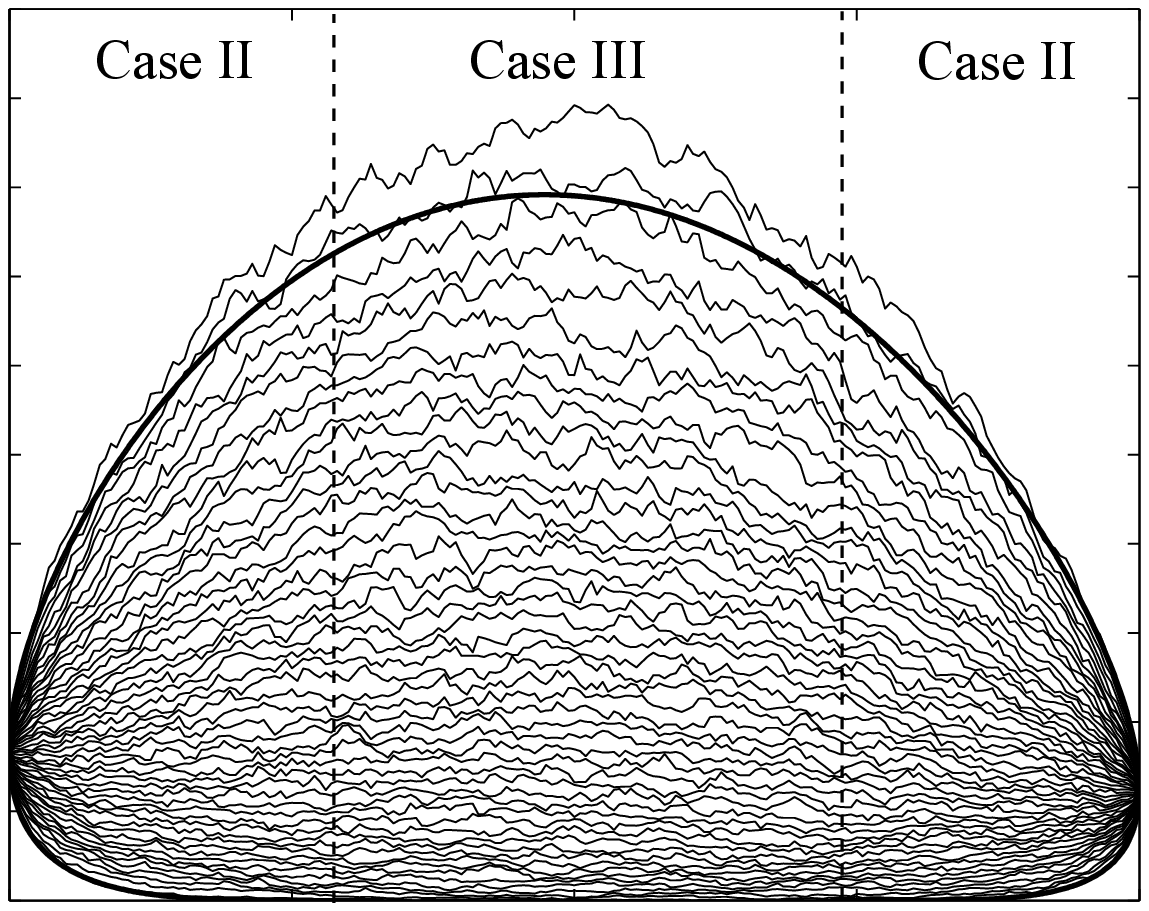}
\small{\put(8.5,3.5){$0$} \put(35,3.5){$t_1$} \put(68,3.5){$t_2$}
\put(54,3.5){$t$} \put(90,3.5){$1$}}
\end{overpic}
\caption{\label{fig:phase_diagram} The left figure shows the phase diagram for
the non-intersecting  squared Bessel paths for $b=1/4$. The solid curve is
 given in \eqref{eq:curve_t}. The right figure shows a simulation of
50 non-intersecting paths for $a=1/3$ which corresponds to the dashed
line in the left figure.}
\end{figure}

Let us assume that the ending point $b$ is fixed. In figure \ref{fig:phase_diagram}
we give a phase diagram divided into three different regions according with the
limiting distribution of the non-intersecting squared Bessel paths as $n\to \infty$.
The horizontal axis of the figure denotes the time $t\in (0,1)$ and the vertical
axis denotes the starting point $a$.

In the case $ab<1/4$, for $t$ small, we are in Case II. The paths start
 at the positive value $a$ at $t=0$ and they initially remain positive.
 The limiting mean density is supported in
 the interval $[p,q]$ with $p>0$ (see Theorems \ref{theorem:properties of minimizer}
 and \ref{theorem:limiting mean distribution}). For this case there is only one active constraint.

We have computed the
discriminant $Disc(z)$ of \eqref{eq:spectral-curve} with coefficients given by \eqref{Az}--\eqref{Dz}
and $c=0$ in \eqref{Dz} with respect to $\xi$ using Maple
and we obtain
\[ Disc(z)=a_4z^4+a_3z^3+a_2z^2+4(4ab-1)(a-2at+at^2+bt^2+t^2-t)^3z \]
for some constants $a_4$, $a_3$ and $a_2$. We do not include the
explicit expressions of $a_4$, $a_3$ and $a_2$ because they are too
long and not too interesting. Therefore there are exactly four zeros
$Disc(z)$ which correspond to the  branch points of the Riemann
surface.

For each $ab<1/4$ there are two critical times $t_{1,2}$ where
$Disc(z)$ has a double zero at $z=0$, namely
\begin{equation}
\label{eq:curve_t} t_{1,2}=\frac {2a+1\pm\sqrt{1-4ab}}{2(a+b+1)}.
\end{equation}
The critical times $t_1$ and $t_2$ are the only two values of $t$
for which the coefficient of $z$ in $Disc(z)$ is equal to zero. The
time $t_1$ corresponds to the time where the first paths touch the
wall. If $t_1<t<t_2$ then we are in Case III. Here the limiting mean
density of the paths is supported on the interval $[0,q]$ (see
Theorems \ref{theorem:properties of minimizer} and
\ref{theorem:limiting mean distribution}). Both constraints are
active in this case. If $t>t_2$ then we are back to the Case II.

The figure on the right of Figure \ref{fig:phase_diagram} shows the transitions
between Case II and Case III for the choice of the parameters
$a=1/3$ and $b=1/4$. This choice corresponds to the dashed line in the left figure. The
critical times $t_1$ and $t_2$ where the transition occurs are given by \eqref{eq:curve_t}.

On the other hand, if $ab>1/4$ then the paths start at the positive value $a$
at $t=0$ and they remain positive. In this case the limiting mean density is
supported in the interval $[p,q]$ with $p>0$. No constraint is active in this
case.

%--------------------------------------------------------------------
%\subsection{The $g$-functions}
\subsection{The $\lam$-functions}

For the measures $\mu_j$, $j=1,2,3$ that minimize the energy
functional \eqref{energy functional}, we first define the associated
$g$-functions as
\begin{equation}\label{def of g_j}
g_j(z)=\int\log(z-s)d\mu_j(s),
\end{equation}
where we take the principal branch of the logarithm, i.e.,
\begin{equation}\label{log:principal:branch}
\log(z-s)=\log|z-s|+i\arg(z-s), \quad \arg(z-s)\in(-\pi,\pi).
\end{equation}
On account of Theorem \ref{theorem:properties of minimizer}, it is
easily seen that $g_1$ and $g_3$ are analytic in
$\mathbb{C}\setminus\mathbb{R}^-$, while $g_2$ is analytic in
$\mathbb{C}\setminus(-\infty,q]$.

Noting that
\begin{equation}\label{derivative of g_j}
g_j'(z)=F_j(z),
\end{equation}
we find the asymptotic behavior of $g_j(z)$ at infinity as given in
the following lemma.
\begin{lem}\label{asy of g-function}
With the constants $c_1$ and $c_3$ given in Lemma \ref{asy of F_j},
we have that
\begin{align}\label{g1 asymptotics}
g_{1}(z)&=\frac{1}{2}\log
z+\frac{2c_1}{\sqrt{z}}+O\left(\frac{1}{z}\right),
\\ \label{g2 asymptotics}
g_{2}(z)&=\log z+O\left(\frac{1}{z}\right), \\
\label{g3 asymptotics} g_{3}(z)&=\frac{1}{2}\log
z+\frac{2c_3}{\sqrt{z}}+O\left(\frac{1}{z}\right),
\end{align}
as $z\to \infty$.
\end{lem}
\begin{proof}
The asymptotics of $g_2$ follows from the fact that $\mu_2$ is a
probability measure with compact support.

The formulas \eqref{g1 asymptotics} and \eqref{g3 asymptotics} for $g_1$ and
$g_3$ can be derived by integrating the asymptotic formulas for $F_1$ and $F_3$
given in Lemma \ref{asy of F_j}. The integration constant is zero, due to the
fact that
\begin{equation*}
\int\log(z-s)d\mu_j(s)
=\log(z)||\mu_j||+\int\log\left(1-\frac{s}{z}\right)d\mu_j(s),
\qquad j=1,3,
\end{equation*}
and $\int\log(1-s/z)d\mu_j(s)\to 0$ as $z\to \infty$.
\end{proof}

%---------------------------------------------------------------------

The $\lam$-functions are then defined as the following
anti-derivatives of the $\xi$-functions:
\begin{align}
\lam_1(z) &= -g_1(z)+\frac{2\sqrt{az}}{t}+\frac{z}{t(1-t)}-l,\label{lambda1:def}\\
\lam_2(z) &= g_1(z)-g_2(z)-\frac{2\sqrt{az}}{t}+\frac{z}{t(1-t)}-l,\label{lambda2:def}\\
\lam_3(z) &= g_2(z)-g_3(z)+\frac{2\sqrt{bz}}{1-t},\label{lambda3:def}\\
\lam_4(z) &= g_3(z)-\frac{2\sqrt{bz}}{1-t},\label{lambda4:def}
\end{align}
where $l\in\er$ is the variational constant in
Proposition~\ref{prop:variation conditions}. Note that $\lam_1$ and
$\lam_4$ are analytic in $\cee\setminus(-\infty,0]$ while $\lam_2$
and $\lam_3$ are analytic in $\cee\setminus(-\infty,q]$. Some
properties of $\lam$-functions are shown in the following two
lemmas.
%These functions, as we shall see later, will be helpful in the second
%transformation of our RH analysis.

\begin{lem}\label{lemma:lambda:var} We have
$$\begin{array}{ll}
\lam_{1,+}(x) = \lam_{1,-}(x),\qquad & x\in (-r_1,0),\\
\lam_{2,+}(x) = \lam_{2,-}(x)-2\pi i,\qquad & x\in (-r_1,p),\\
\lam_{3,+}(x) = \lam_{3,-}(x)+2\pi i,\qquad & x\in (-r_3,p), \\
\lam_{4,+}(x) = \lam_{4,-}(x),\qquad & x\in (-r_3,0),
\end{array}$$
$$ \begin{array}{ll}
\lam_{1,\pm}(x)=\lam_{2,\mp}(x)\mp\pi i,\qquad & x\in \Delta_1,\\
\lam_{2,\pm}(x)=\lam_{3,\mp}(x),\qquad& x\in \Delta_2,\\
\lam_{3,\pm}(x)=\lam_{4,\mp}(x)\mp\pi i,\qquad& x\in \Delta_3,
\end{array}$$
and $$\begin{array}{ll}
\Re(\lam_{1}(x)-\lam_{2}(x))<0,\qquad& x\in (-r_1,0),\\
\Re(\lam_{2}(x)-\lam_{3}(x))>0,\qquad& x\in \er^+\setminus \Delta_2,\\
\Re(\lam_{3}(x)-\lam_{4}(x))<0,\qquad& x\in (-r_3,0). \end{array}$$
\end{lem}

\begin{proof} These are reformulations of the variational conditions of the
equilibrium problem in Proposition~\ref{prop:variation conditions}.
\end{proof}

\begin{lem}\label{lemma:lambda:asy}
The $\lam$-functions have the following asymptotics for $z\to\infty$:
\begin{align*} \lam_1(z) &=
\frac{z}{t(1-t)}+\frac{2\sqrt{az}}{t}-l-\frac{1}{2}\log z-\frac{2c_1}{\sqrt{z}}+O(z^{-1}),\\
\lam_2(z) &=
\frac{z}{t(1-t)}-\frac{2\sqrt{az}}{t}-l-\frac{1}{2}\log z+\frac{2c_1}{\sqrt{z}}+O(z^{-1}),\\
\lam_3(z) &=
\frac{2\sqrt{bz}}{1-t}+\frac{1}{2}\log z-\frac{2c_3}{\sqrt{z}}+O(z^{-1}),\\
\lam_4(z) &=
-\frac{2\sqrt{bz}}{1-t}+\frac{2c_3}{\sqrt{z}}+\frac{1}{2}\log
z+O(z^{-1}).
\end{align*}
\end{lem}

\begin{proof} Obvious from the definitions \eqref{lambda1:def}--\eqref{lambda4:def}
and from \eqref{g1 asymptotics}--\eqref{g3 asymptotics}.
\end{proof}

\subsection{The $\phi$-functions}
We end this section with the introduction of the $\phi$-functions,
which will be used to simplify the jump matrices during our steepest
descent analysis. Recall that $\xi_{j}$ is the restriction of $\xi$
to the sheet $\mathcal{R}_j$ of the Riemann surface $\mathcal{R}$.
We define
\begin{align}
\phi_1(z)&=\frac{1}{2}\int_{-r_1}^{z}(\xi_1(y)-\xi_2(y))dy,
\quad z\in\mathbb{C}\setminus\left((-\infty,-r_1]\cup[p,\infty)\right), \label{phi 1}\\
\phi_2(z)&=\frac{1}{2}\int_{q}^z(\xi_2(y)-\xi_3(y))dy,
\quad z\in\mathbb{C}\setminus(-\infty,q],\label{phi 2}\\
\phi_3(z)&=\frac{1}{2}\int_{-r_3}^{z}(\xi_3(y)-\xi_4(y))dy, \quad
z\in\mathbb{C}\setminus\left((-\infty,-r_3]\cup[p,\infty)\right).
\label{phi 3}
\end{align}
%where the path of integrations all lie entirely in the region $z\in
%\mathbb{C}\setminus (-\infty,q]$, except for the initial points.

%With the above preparations, we are ready to apply the steepest
%descent analysis to the RH problem for $Y$. It consists of a series
%of transformations and will finally lead to the proof of
%Theorem~\ref{theorem:limiting mean distribution}.
%---------------------------------------------------------------------

\section{Steepest descent analysis for $Y(z)$}\label{sec:steepest descent}
In this section, we will perform the steepest descent analysis of
the RH problem for $Y$. It consists of a series of explicit and
invertible transformations. In
Section~\ref{section:proof:limitdistribution}, these transformations
will be used to prove Theorem \ref{theorem:limiting mean
distribution}.

To this end, we mainly follow the theme laid out in \cite{KMW09},
although the argument is somehow more involved since our RH problem
is of size $4\times 4$ whereas the dimension treated in \cite{KMW09}
is $3\times 3$. Furthermore, we have more generic cases to consider
here.

\subsection{First transformation $Y \mapsto X$ }
\label{section:firsttransfo}

It is the aim of this transformation to simplify the block matrix \eqref{weight
syt scaling} appearing in the jump condition \eqref{Jump for Y} for $Y$. The
idea is to use the special properties of modified Bessel functions, in a very
similar way as in \cite{KMW09}. To start with, we set four functions:
\begin{equation}\label{def of y_i}
\begin{aligned}
y_1(z)&=z^{(\al+1)/2}I_{\al+1}(2\sqrt{z}), \quad
&&y_2(z)=z^{(\al+1)/2}K_{\al+1}(2\sqrt{z}), \\
y_3(z)&=z^{-(\al-1)/2}I_{\al-1}(2\sqrt{z}), \quad
&&y_4(z)=z^{-(\al-1)/2}K_{\al-1}(2\sqrt{z}),
\end{aligned}
\end{equation}
where $K_\nu$ is the modified Bessel function of the second kind;
see \cite[Section 9.6]{HB92} for its main properties.
%Note that the
%functions $y_1(z)$ and $y_2(z)$ are the same as those used in
%\cite{KMW09}.
It is readily seen that $y_{3}(z)$ is well defined in
the whole complex plane, while the function $y_{j}(z), j=1,2,4$ is
analytic in the complex plane with a branch cut along the negative
real axis. Indeed, it follows from formulas 9.6.30--9.6.31 in
\cite{HB92} that their jumps on $\mathbb{R}^-$ are given in the
following ways
\begin{equation}\label{jump for y_i}
\begin{aligned}
y_{1,+}(x) &= e^{2i \al \pi}y_{1,-}(x), \\
y_{2,+}(x) &= y_{2,-}(x)+ i \pi e^{i\al \pi}y_{1,-}(x), \\
y_{4,+}(x) &= e^{-2i \al \pi }y_{4,-}(x)+ i \pi e^{ - i \al \pi}
y_{3}(x),
\end{aligned}
\end{equation}
for $x < 0$. In view of the derivative properties of the modified
Bessel functions \cite[9.6.26]{HB92}, we also have
\begin{equation}\label{derivative of y_i}
\begin{aligned}
y_1'(z)&=z^{\al/2}I_{\al}(2\sqrt{z}), \quad &&y_2'(z)=
-z^{\al/2}K_{\al}(2\sqrt{z}) , \\
y_3'(z)&=z^{-\al/2}I_{\al}(2\sqrt{z}), \quad &&y_4'(z)=
-z^{-\al/2}K_{\al}(2\sqrt{z}).
\end{aligned}
\end{equation}

A combination of \eqref{def of y_i}, \eqref{derivative of y_i} and
\eqref{weight syt scaling} implies that we can represent the weight matrix
$W(x)$ in terms of the functions $y_1$, $y_3$ and their derivatives as
\begin{align}
  \nonumber W(x) &=
  e^{-\frac{nx}{t(1-t)}}\begin{pmatrix}x^{\al/2}I_{\al}\left(\frac{2n\sqrt{ax}}{t}\right)\\
  x^{(\al+1)/2}I_{\al+1}\left(\frac{2n\sqrt{ax}}{t}\right)
  \end{pmatrix}
  \begin{pmatrix}
  x^{-\al/2}I_{\al}\left(\frac{2n\sqrt{bx}}{1-t}\right) &
  x^{-(\al-1)/2}I_{\al-1}\left(\frac{2n\sqrt{bx}}{1-t}\right)
  \end{pmatrix} \\
  \label{weight syt in y_i}
  &=
  e^{-\frac{nx}{t(1-t)}}\begin{pmatrix}
   \tau_1^{-\al} y_1'(\tau_1^2 x) \\
   \tau_1^{-\al-1} y_1(\tau_1^2 x)
  \end{pmatrix}
  \begin{pmatrix}
   \tau_3^\al y_3'(\tau_3^2 x)
   &\tau_3^{\al-1} y_3(\tau_3^2 x)
  \end{pmatrix},
  \end{align}
where
\begin{equation}
\tau_1= \tau_{1,n}=\frac{n\sqrt{a}}{t}, \qquad \tau_3=
\tau_{3,n}=\frac{n\sqrt{b}}{1-t}.
\end{equation}
For later use, we also need the following Wronskian relations:
\begin{equation}\label{Wronskian relation}
\begin{aligned}
y_1'(z)y_2(z)-y_1(z)y_2'(z)&=z^\alpha/2, \quad
y_3'(z)y_4(z)-y_3(z)y_4'(z)&=z^{-\alpha}/2,
\end{aligned}
\end{equation}
for $z \in \mathbb{C} \setminus \mathbb{R}^-$; see formula 9.6.15 of
\cite{HB92}.

The first transformation $Y\mapsto X$ is then defined by (compare with
\cite{KMW09})
\begin{align}\label{def:X}
X(z)=C_1Y(z)\diag(A_1(z),A_2(z)),
%\begin{pmatrix}
%A_1(z) & 0 \\
%0 & A_2(z)
%\end{pmatrix},
\end{align}
where
\begin{eqnarray}\label{def:A1}
A_1(z)&=& \tau_1^{-\alpha}z^{-\alpha/2}\begin{pmatrix} -\frac{1}{\pi
i}y_2'(\tau_1^2z) &
y_1'(\tau_1^2z) \\
-\frac{1}{\pi i}\tau_1^{-1}y_2(\tau_1^2z) & \tau_1^{-1}y_1(\tau_1^2z)
\end{pmatrix},
\\ \label{def:A2}A_2(z)&=& 2\tau_3^{\alpha}z^{\alpha/2}\begin{pmatrix} y_4(\tau_3^2z) &
-\pi i y_3(\tau_3^2z) \\
-\tau_3 y_4'(\tau_3^2z) & \pi i \tau_3 y_3'(\tau_3^2z)
\end{pmatrix},
\end{eqnarray}
and with $C_1$ the constant $4\times 4$ matrix
\begin{equation*}
C_1 = \diag\left(\sqrt{2\pi\tau_1}
\begin{pmatrix}i&0\\ \frac{4(\al+1)^2-1}{16\tau_1} & 1\end{pmatrix},
\frac{1}{\sqrt{2\pi\tau_3}}\begin{pmatrix}0&1\\
i & -i\frac{4(\al-1)^2-1}{16\tau_3}\end{pmatrix}\right).
%
%C_1= \begin{pmatrix}
%        i & 0 & 0 & 0 \\
%        \frac{4(\al+1)^2-1}{16\tau_1} & 1 & 0 & 0\\
%        0 & 0 & 0 & 1 \\ 0 & 0 & i & -i\frac{4(\al-1)^2-1}{16\tau_3}
%        \end{pmatrix}
%\begin{pmatrix}
%(2\pi\tau_1)^{1/2} & 0 & 0 & 0 \\
%0 & (2\pi\tau_1)^{1/2} & 0 & 0 \\
%0 & 0 & (2\pi\tau_3)^{-1/2} & 0 \\
%0 & 0 & 0 & (2\pi\tau_3)^{-1/2}
%\end{pmatrix}.
\end{equation*}
%if $n$ is even, and
%\begin{equation*}C_1= \begin{pmatrix}
%        1 & i\frac{4\al^2-1}{16\tau_3} & 0 & 0 \\
%        0 & 1 & 0 & 0\\
%        0 & 0 & 1 & 0 \\
%        0 & 0 & -\frac{4\al^2-1}{16\tau_1} & 1
%        \end{pmatrix}
%\begin{pmatrix}
%(2\pi\tau_3)^{1/2} & 0 & 0 & 0 \\
%0 & -i(2\pi\tau_3)^{1/2} & 0 & 0 \\
%0 & 0 & (2\pi\tau_1)^{-1/2} & 0 \\
%0 & 0 & 0 & i(2\pi\tau_1)^{-1/2}
%\end{pmatrix},
%\end{equation*}
%if $n$ is odd.
Note that, on account of \eqref{Wronskian relation}, we have
\begin{equation}
\det A_1=-\frac{1}{2\pi\tau_1} i, \qquad \det A_2 = 2\pi\tau_3 i.
\end{equation}
Hence, it is easily seen that $\det{X}=1$. Now, we have
\begin{prop}\label{prop:RHP for X}
The matrix valued function $X(z)$ defined by \eqref{def:X} is the unique
solution of the following RH problem.
\begin{enumerate}[(1)]
  \item ${X}(z)$ is analytic in
  $\mathbb{C}\setminus\mathbb{R}$.

  \item For $x\in\mathbb{R}$,  ${X}(x)$ satisfies the
following jump condition
  \begin{equation}\label{jump for X}
  X_+(x)=X_-(x)
  \begin{cases}
  I+e^{-\frac{nx}{t(1-t)}}E_{23}, & \text{ if } x>0, \\
  \diag(e^{-i\pi\al},e^{i\pi\al},e^{-i\pi\al},e^{i\pi\al})-E_{21}-E_{43}, & \text{ if }x<0,
  \end{cases}
  \end{equation}
where we denote by $E_{ij}$ the $4 \times 4$ elementary matrix whose
entries are all 0, except for the $(i,j)$-th entry, which is $1$.
  \item As $z\to \infty$, $z\in \mathbb{C}\setminus\mathbb{R}$, we
  have
  \begin{align*}\label{asy of X}
  X(z)=&\left(I+O\!\left(\frac{1}{z}\right)\right)
  \diag\left(z^{-1/4},z^{1/4},z^{-1/4},z^{1/4}\right)
  %\begin{pmatrix}
%  z^{-1/4} & 0 & 0 & 0 \\
%  0 & z^{1/4} & 0 & 0 \\
%  0 & 0 & z^{1/4} & 0 \\
%  0 & 0 & 0 & z^{-1/4}
%  \end{pmatrix}
\frac{1}{\sqrt{2}} \diag\left(\begin{pmatrix} 1&i\\i&1
\end{pmatrix},\begin{pmatrix} 1&i\\i&1
\end{pmatrix}\right)
  \nonumber \\
  %&\times
  %\diag\left(z^{\al/2}, z^{-\al/2}, z^{\al/2}, z^{-\al/2}\right)
  % \nonumber \\
  &\times \diag\left(
  z^{n/2}e^{-2n\sqrt{az}/t}, z^{n/2}e^{2n\sqrt{az}/t}, z^{-n/2}e^{-2n\sqrt{bz}/(1-t)},
  z^{-n/2}e^{2n\sqrt{bz}/(1-t)}
  \right),
  \end{align*}
  uniformly valid for $z$ bounded away from the negative real axis.

  \item $X(z)$ has the following behavior near the origin:
  %\begin{align}\label{zero behavior of X:1}
  \begin{equation}\label{zero behavior of X} \begin{array}{ll}
  X(z)\diag(|z|^{\al/2},|z|^{-\al/2},|z|^{\al/2},|z|^{-\al/2}))=O(1),&
  %X(z)=O\begin{pmatrix}
%  |z|^{-\al/2} & |z|^{\al/2} & |z|^{-\al/2} &|z|^{\al/2} \\
%  |z|^{-\al/2} & |z|^{\al/2} & |z|^{-\al/2} &|z|^{\al/2} \\
%  |z|^{-\al/2} & |z|^{\al/2} & |z|^{-\al/2} &|z|^{\al/2} \\
%  |z|^{-\al/2} & |z|^{\al/2} & |z|^{-\al/2} &|z|^{\al/2} \\
%  \end{pmatrix},&
%   %~~X^{-1}(z)=O\begin{pmatrix}
%  |z|^{\al/2} & |z|^{\al/2} & |z|^{\al/2} & |z|^{\al/2}\\
%   |z|^{-\al/2} & |z|^{-\al/2} & |z|^{-\al/2} & |z|^{-\al/2} \\
%  |z|^{\al/2} & |z|^{\al/2} & |z|^{\al/2} & |z|^{\al/2} \\
%   |z|^{-\al/2} & |z|^{-\al/2} & |z|^{-\al/2} & |z|^{-\al/2}\\
%  \end{pmatrix},
\qquad \textrm{if $\alpha>0$},\\
  X(z)\diag((\log |z|)^{-1},1,(\log |z|)^{-1},1)=O(1),&
  %O\begin{pmatrix}
%  \log |z| & 1 & \log |z| & 1 \\
%  \log |z| & 1 & \log |z| & 1 \\
%  \log |z| & 1 & \log |z| & 1 \\
%  \log |z| & 1 & \log |z| & 1 \\
%  \end{pmatrix},
%  % ~~X^{-1}(z)=O\begin{pmatrix}
%  1 & 1 & 1 & 1 \\
%   \log |z| & \log |z| & \log |z| & \log |z|\\
%  1 & 1 & 1 & 1 \\
%   \log |z| & \log |z| & \log |z| & \log |z|\\
%  \end{pmatrix},
  \qquad \textrm{if $\alpha=0$},
  \\
  X(z)=O(z^{\al/2}),~~X^{-1}(z)=O(z^{\al/2}),&\qquad \textrm{if $-1<\al<0$.}
  \end{array}\end{equation}
  %where $C_{\al}$, $\al\geq 0$ are certain constant matrices of size $4\times 4$.
\end{enumerate}
\end{prop}

\begin{proof}
The proof follows \cite{KMW09}. It is easily seen from
\eqref{def:X}--\eqref{def:A2} and \eqref{Jump for Y} that
\begin{multline}\label{formal jump for tilde X}
X^{-1}_{-}(x)X_{+}(x) =\begin{pmatrix}
A_{1,-}^{-1}(x) & 0 \\
0 & A_{2,-}^{-1}(x)
\end{pmatrix}Y^{-1}_{-}(x)Y_{+}(x)
\begin{pmatrix}
A_{1,+}(x) & 0 \\
0 & A_{2,+}(x)
\end{pmatrix}
\\
=\left\{
    \begin{array}{ll}
     \begin{pmatrix}
I_2 & A_{1,-}^{-1}(x)W(x)A_{2,+}(x) \\
0 & I_2
\end{pmatrix}, & \text{if } x>0, \\
       \begin{pmatrix}
A_{1,-}^{-1}(x)A_{1,+}(x) &  0 \\
0 & A_{2,-}^{-1}(x)A_{2,+}(x)
\end{pmatrix}, & \text{if }x<0.
    \end{array}
  \right.
\end{multline}
In view of \eqref{weight syt in y_i} and \eqref{Wronskian relation},
we obtain from a direct calculation that
\begin{align}
&A_{1,-}^{-1}(x)W(x)A_{2,+}(x) \nonumber \\
&=e^{-\frac{nx}{(1-t)t}}A_{1,-}^{-1}(x)
  \begin{pmatrix}
   \tau_1^{-\al} y_1'(\tau_1^2 x) \\
   \tau_1^{-\al-1} y_1(\tau_1^2 x)
  \end{pmatrix}
  \begin{pmatrix}
   \tau_3^\al y_3'(\tau_3^2 x)
   &\tau_3^{\al-1} y_3(\tau_3^2 x)
  \end{pmatrix}A_{2,+}(x) \nonumber \\
&=e^{-\frac{nx}{(1-t)t}}\begin{pmatrix}
   0 \\
   x^{\alpha/2}
  \end{pmatrix}
  \begin{pmatrix}
   x^{-\alpha/2}
   &0
  \end{pmatrix}=e^{-\frac{nx}{(1-t)t}}\begin{pmatrix}
   0 & 0 \\
   1 & 0
  \end{pmatrix},
\end{align}
for $x>0$. Also, by \eqref{jump for y_i} and \eqref{derivative of
y_i}, it can be shown that
\begin{equation}\label{jump on neg R for tilde X}
A_{1,-}^{-1}(x)A_{1,+}(x) =\begin{pmatrix}
   e^{-i\pi\al}& 0 \\
   -1 & e^{i\pi\al}
  \end{pmatrix}
~~~\textrm{and}~~~ A_{2,-}^{-1}(x)A_{2,+}(x) =\begin{pmatrix}
   e^{-i\pi\al} & 0 \\
   -1 & e^{i\pi\al}
  \end{pmatrix},
\end{equation}
for $x<0$. The jump condition \eqref{jump for X} now follows from a
combination of \eqref{formal jump for tilde X}--\eqref{jump on neg R
for tilde X}.

The large $z$ behavior of $X(z)$ shown in item $(3)$, follows very similarly as
in \cite{KMW09}. The relevant results are
\begin{align}\label{asy of A1 2}
A_1(z) &=\frac{1}{\sqrt{2\pi\tau_1}}\left[
\begin{pmatrix}
-i & 0 \\
i\frac{4(\alpha+1)^2-1}{16\tau_1}  & 1
\end{pmatrix}
+O(z^{-1})\right]\frac{z^{-\sigma_3/4}}{\sqrt{2}}\begin{pmatrix} 1 & i \\
i & 1
\end{pmatrix} e^{-2\tau_1\sqrt{z}\sigma_3},
\end{align}
and
\begin{align}\label{asy of A2}
A_2(z) =\sqrt{2\pi\tau_3}\left[
\begin{pmatrix}
\frac{4(\alpha-1)^2-1}{16\tau_3} & -i \\
1 & 0
\end{pmatrix}
+O(z^{-1})\right]\frac{z^{-\sigma_3/4}}{\sqrt{2}}\begin{pmatrix} 1 & i \\
i & 1
\end{pmatrix} e^{-2\tau_3\sqrt{z}\sigma_3},
\end{align}
as $z\to \infty$, uniformly for $z$ bounded away from the negative real axis.
Substituting the asymptotic formulas \eqref{asy of A1 2} and \eqref{asy of A2}
of $A_1(z)$ and $A_2(z)$ into \eqref{def:X}, the asymptotic behavior of $X(z)$
at infinity is immediate.

Finally, the known behavior of the modified Bessel functions near zero given by
formulas 9.6.7--9.6.9 in \cite{HB92} yields
\begin{align*}
y_1(z)&\sim \frac{1}{\Gamma(\alpha+2)}z^{\alpha+1}, &&\qquad
y_1'(z)\sim \frac{1}{\Gamma(\alpha+1)}z^{\alpha}, \\
y_2(z)&\sim\frac{1}{2}\Gamma(\alpha+1),
&&\qquad y_2'(z)\sim\left\{
\begin{array}{ll}
-\frac{1}{2}\Gamma(\alpha), &  \alpha >0, \\
\frac{1}{2}\log(z), & \alpha=0, \\
-\frac{1}{2}\Gamma(-\alpha)z^{\alpha}, &  \alpha<0,
\end{array}
\right.
\end{align*}
and
\begin{align*}
y_3(z)&\sim \frac{1}{\Gamma(\alpha)}, &&\qquad
y_3'(z)\sim \frac{1}{\Gamma(\alpha+1)}, \\
y_4(z)&\sim\left\{
\begin{array}{ll}
\frac{1}{2}\Gamma(\alpha-1)z^{1-\alpha}, & \alpha >1, \\
-\frac{1}{2}\log(z), & \alpha=1, \\
\frac{1}{2}\Gamma(1-\alpha), & \alpha<1,
\end{array}
\right. &&\qquad y_4'(z)\sim\left\{
\begin{array}{ll}
-\frac{1}{2}\Gamma(\alpha)z^{-\alpha}, & \alpha >0, \\
\frac{1}{2}\log(z), & \alpha=0, \\
-\frac{1}{2}\Gamma(-\alpha), & \alpha<0.
\end{array}
\right.
\end{align*}
The behavior at zero of $X(z)$ in item (4) then follows from a straightforward
calculation.

This completes the proof of Proposition \ref{prop:RHP for X}.
\end{proof}

\subsection{Second transformation $X \mapsto U$}
\label{section:secondtransfo}

The second transformation $X \mapsto U$ is defined by
\begin{align}\label{def:U}
U(z)=C_2X(z)\diag\left(e^{n\left(\lam_1(z)-\frac{z}{t(1-t)}\right)},
e^{n\left(\lam_2(z)-\frac{z}{t(1-t)}\right)},e^{n\lam_3(z)},e^{n\lam_4(z)}\right),
\end{align}
where $C_2$ is the constant matrix
\begin{equation}\label{C2}
C_2= (I +2inc_1E_{21}+2inc_3 E_{34})\diag (e^{nl},e^{nl},1,1),
\end{equation}
with $c_1$ and $c_3$ as in Lemma \ref{asy of F_j} and with $l$ the variational
constant in Proposition~\ref{prop:variation conditions}.

In view of the jump relations \eqref{jump for X} for $X$ and the definition
\eqref{def:U}, we see that the jump relations for $U$ are given by
\begin{equation}\label{jump for U:a}
  U_+=U_-
  \left\{
  \begin{array}{ll}
  \Lam_-^{-1}\left(I+E_{23}\right)\Lam_+, & \text{on } \mathbb{R}^+, \\
  \Lam_-^{-1}\left(\diag(e^{-i\pi\al},e^{i\pi\al},e^{-i\pi\al},e^{i\pi\al})
-E_{21}-E_{43}\right)\Lam_+, & \text{on } \mathbb{R}^-,
  \end{array}
  \right.
  \end{equation}
  where we define the diagonal matrix
  \begin{equation}\label{def:Lambda}
  \Lam(z)=\diag(e^{n\lam_1(z)},e^{n\lam_2(z)},e^{n\lam_3(z)},e^{n\lam_4(z)}).
\end{equation}

%\begin{equation}\label{jump for U:a1}
%\begin{pmatrix}
%          1 & 0 & 0 & 0 \\
%          0 & e^{n(\lam_{2,+}(x)-\lam_{2,-}(x))} &
%          e^{n(\lam_{3,+}(x)-\lam_{2,-}(x))} & 0 \\
%          0 & 0 & e^{n(\lam_{3,+}(x)-\lam_{3,-}(x))} & 0 \\
%          0 & 0 & 0 & 1 \\
%\end{pmatrix},
%\end{equation}
%for $x>0$, and by
%\begin{equation}\label{jump for U:a2}
%\begin{pmatrix}
%e^{-i\pi\al}e^{n\left(\lam_{1,+}(x)-\lam_{1,-}(x)\right)} & 0 & 0 & 0 \\
%-e^{n\left(\lam_{1,+}(x)-\lam_{2,-}(x)\right)} &
%e^{i\pi\al}e^{n\left(\lam_{2,+}(x)-\lam_{2,-}(x)\right)}& 0 & 0 \\
%0 & 0 & e^{-i\pi\al}e^{n(\lam_{3,+}(x)-\lam_{3,-}(x))} & 0 \\
%0 & 0 & -e^{n\left(\lam_{3,+}(x)-\lam_{4,-}(x)\right)} &
%e^{i\pi\al}e^{n\left(\lam_{4,+}(x)-\lam_{4,-}(x)\right)}
%\end{pmatrix},
%\end{equation}
%for $x<0$.
The jump matrix for $U$ can be reformulated with the help of the
$\phi$-functions in \eqref{phi 1}--\eqref{phi 3} as stated in the
following proposition.
\begin{prop}\label{prop:RHP for U}
The matrix valued function $U(z)$ defined by \eqref{def:U} is the unique
solution of the following RH problem.

\begin{enumerate}[(1)]
  \item $U(z)$ is analytic in
  $\mathbb{C}\setminus\mathbb{R}$.

  \item For $x\in\mathbb{R}$, we have that
  $
  U_+=U_-J_{U},
  $
  where
  \begin{align}
  J_{U}&=
         %\begin{pmatrix}
%          1 & 0 & 0 & 0 \\
%          0 & e^{2n\phi_{2,+}(x)} &
%          1 & 0 \\
%          0 & 0 & e^{2n\phi_{2,-}(x)} & 0 \\
%          0 & 0 & 0 & 1
%         \end{pmatrix}
   \diag\left(1,\begin{pmatrix}e^{2n\phi_{2,+}}
   & 1\\ 0 & e^{2n\phi_{2,-}}\end{pmatrix},1\right)
         \quad \textrm{on $\Delta_2$},
         \label{jump on delta_2}\\
  J_{U}&=I+e^{-2n\phi_{2,+}}E_{23}
         \quad \textrm{on $\mathbb{R}^+ \setminus (0 \cup
         \overline{\Delta_2})$},\\
%  J_{U}(x)&=\label{jump on delta_3}
%  \diag\left(\begin{pmatrix}
%  e^{-i\pi\al}e^{2n\phi_{1,+}(x)} & 0\\
%  -1 & e^{i\pi\al}e^{2n\phi_{1,-}(x)}
%  \end{pmatrix},\begin{pmatrix}
%  e^{-i\pi\al}e^{2n\phi_{3,+}(x)} & 0  \\
%  -1 & e^{i\pi\al}e^{2n\phi_{3,-}(x)}
%  \end{pmatrix}\right),
\label{jump on delta_3} J_{U}&=\diag \left(
          \begin{pmatrix}
          e^{-i\pi\al}e^{2n\phi_{1,+}} & 0 \\
          -1 & e^{i\pi\al}e^{2n\phi_{1,-}}
          \end{pmatrix},
          \begin{pmatrix}
           e^{-i\pi\al}e^{2n\phi_{3,+}} & 0  \\
           -1 & e^{i\pi\al}e^{2n\phi_{3,-}}
           \end{pmatrix}\right)
~ \textrm{on $\Delta_3$},
%J_{U}&=\begin{pmatrix}\label{jump on delta_3}
%          e^{-i\pi\al}e^{2n\phi_{1,+}} & 0 & 0 & 0 \\
%          -1 & e^{i\pi\al}e^{2n\phi_{1,-}} & 0 & 0 \\
%          0 & 0 & e^{-i\pi\al}e^{2n\phi_{3,+}} & 0  \\
%          0 & 0 & -1 & e^{i\pi\al}e^{2n\phi_{3,-}}
%         \end{pmatrix}
%\quad \textrm{on $\Delta_3$},
  \\
\label{for case II and III}  J_{U}&=\left(\begin{pmatrix}
          e^{-i\pi\al}e^{2n\phi_{1,+}} & 0 \\
          -1 & e^{i\pi\al}e^{2n\phi_{1,-}} \end{pmatrix}
          \begin{pmatrix} e^{-i\pi\al} & 0  \\
           -e^{2n\phi_{3}} & e^{i\pi\al}
           \end{pmatrix}\right)
         ~\textrm{on $\Delta_1 \setminus
         \overline{\Delta_3}$ },
  %J_{U}&=\begin{pmatrix}\label{for case II and III}
%          e^{-i\pi\al}e^{2n\phi_{1,+}} & 0 & 0 & 0 \\
%          -1 & e^{i\pi\al}e^{2n\phi_{1,-}} & 0 & 0 \\
%          0 & 0 & e^{-i\pi\al} & 0  \\
%          0 & 0 & -e^{2n\phi_{3}} & e^{i\pi\al}
%         \end{pmatrix}\quad \textrm{on $\Delta_1 \setminus
%         \overline{\Delta_3}$ },
         \\
         \label{for case III}
  J_{U}&=\diag(e^{-i\pi\al},e^{i\pi\al},e^{-i\pi\al},e^{i\pi\al})-e^{2n\phi_{1}}E_{21}
         -e^{2n\phi_{3}}E_{43}
%\begin{pmatrix}
%          1 & 0 & 0 & 0 \\
%          -|x|^{\alpha}e^{2n\phi_{1}(x)} & 1 & 0 & 0 \\
%          0 & 0 & 1 & 0  \\
%          0 & 0 & |x|^{\alpha}e^{2n\phi_{3}(x)} & 1
%         \end{pmatrix}
\quad \textrm{on $\mathbb{R}^- \setminus
         \overline{\Delta_1}$},
  \end{align}
with $\Delta_j$, $j=1,2,3$ being defined in \eqref{def of Delta} and
$\overline{\Delta_j}$ denotes the closure of $\Delta_j$.

  \item As $z\to \infty$, $z\in \mathbb{C}\setminus\mathbb{R}$, we
  have that
  \begin{align}\label{asy of U}
 U(z)=&~\left(I+O\!\left(\frac{1}{z}\right)\right)
  \diag\left(z^{-1/4},z^{1/4},z^{-1/4},z^{1/4}\right)
  %\begin{pmatrix}
%  z^{-1/4} & 0 & 0 & 0 \\
%  0 & z^{1/4} & 0 & 0 \\
%  0 & 0 & z^{1/4} & 0 \\
%  0 & 0 & 0 & z^{-1/4}
%  \end{pmatrix}
\frac{1}{\sqrt{2}} \diag\left(\begin{pmatrix} 1&i\\i&1
\end{pmatrix},\begin{pmatrix} 1&i\\i&1
\end{pmatrix}\right)
%\begin{pmatrix}
%  1 & i & 0 & 0 \\
%  i & 1 & 0 & 0 \\
%  0 & 0 & 1 & i \\
%  0 & 0 & i & 1
%  \end{pmatrix}
  %\nonumber \\
  %&\times
  %\diag\left(z^{\al/2},z^{-\al/2},z^{\al/2},z^{-\al/2}\right),
  %\begin{pmatrix}
%  z^{\al/2} & 0 & 0 & 0 \\
%  0 & z^{-\al/2} & 0 & 0 \\
%  0 & 0 & z^{\al/2} & 0 \\
%  0 & 0 & 0 & z^{-\al/2}
%  \end{pmatrix}
  \end{align}
  uniformly valid for $z$ bounded away from the negative real axis.

  \item $U(z)$ has the same behavior as $X(z)$ near the origin; see
  \eqref{zero behavior of X}.
\end{enumerate}
\end{prop}
Before proceeding to the proof of Proposition \ref{prop:RHP for U}, we remind
the reader that, due to our assumption \eqref{assumption on r}, the jump
condition \eqref{for case II and III} only appears in Cases II and III, while
\eqref{for case III} only appears in Case III.

\begin{proof}
The jump conditions in item (2) are obtained in a straightforward
way from Lemma~\ref{lemma:lambda:var} and the definitions of
$\phi$-functions in \eqref{phi 1}--\eqref{phi 3}. For convenience,
we give a proof of \eqref{jump on delta_2}, the other jumps can be
proved in similar ways.

If $x\in\Delta_2$, it follows from \eqref{jump for U:a} and
Lemma~\ref{lemma:lambda:var} that
\begin{equation}
J_U(x)=\diag\left(1,\begin{pmatrix}e^{n(\lambda_{2,+}-\lambda_{2,-})(x)}
   & 1\\ 0 &
   e^{n(\lambda_{3,+}-\lambda_{3,-})(x)}\end{pmatrix},1\right).
\end{equation}
On account of our definition of $\lambda_{2,3}$ in
\eqref{lambda2:def} and \eqref{lambda3:def}, it suffices to show
that
\begin{equation}\label{g2 phi2 1}
g_{2,+}(x)-g_{2,-}(x)=-2\phi_{2,+}(x)=2\phi_{2,-}(x),
\end{equation}
for $x\in\Delta_2$. To see this, we obtain from \eqref{def of g_j}
and \eqref{dmu_2} that
\begin{align}\label{eq:diff of g2}
g_{2,+}(x)-g_{2,-}(x)=2 \pi i \int_x^{q}d\mu_2(s)
=\int_{q}^{x}(\xi_{2,-}(s)-\xi_{2,+}(s))ds.
\end{align}
The relation \eqref{g2 phi2 1} now is immediate, with the aid of
\eqref{eq:diff of g2} and the fact that $\xi_{2,-}(s)=\xi_{3,+}(s)$
provided $s\in \Delta_2$.

To prove the asymptotic behavior of $U$ near infinity, we note that
the asymptotics of $X$ and the large $z$ behavior of the
$\lam$-functions given in Lemma \ref{lemma:lambda:asy} yield
\begin{align}\label{XGL}
  &X(z)\diag\left(e^{n\left(\lam_1(z)-\frac{z}{t(1-t)}\right)},
e^{n\left(\lam_2(z)-\frac{z}{t(1-t)}\right)},e^{n\lam_3(z)},e^{n\lam_4(z)}\right)\nonumber \\
  &=\left(I+O\!\left(\frac{1}{z}\right)\right)\diag\left(z^{-1/4},z^{1/4},z^{-1/4},z^{1/4}\right)
  %\begin{pmatrix}
%  z^{-1/4} & 0 & 0 & 0 \\
%  0 & z^{1/4} & 0 & 0 \\
%  0 & 0 & z^{1/4} & 0 \\
%  0 & 0 & 0 & z^{-1/4}
%  \end{pmatrix}
\frac{1}{\sqrt{2}} \diag\left(\begin{pmatrix} 1&i\\i&1
\end{pmatrix},\begin{pmatrix} 1&i\\i&1
\end{pmatrix}\right)C(z),
%  \begin{pmatrix}
%  1 & i & 0 & 0 \\
%  i & 1 & 0 & 0 \\
%  0 & 0 & 1 & i \\
%  0 & 0 & i & 1
%  \end{pmatrix}
% \nonumber \\
%  &~~~~\times
  %\begin{pmatrix}
%  z^{\al/2} & 0 & 0 & 0 \\
%  0 & z^{-\al/2} & 0 & 0 \\
%  0 & 0 & z^{\al/2} & 0 \\
%  0 & 0 & 0 & z^{-\al/2}
%  \end{pmatrix}
%  \diag\left(z^{\al/2},z^{-\al/2},z^{\al/2},z^{-\al/2}\right)C(z),
  \end{align}
as $z\to\infty$, where
\begin{equation*}
  C(z)=\diag\left(
  e^{-nl}\left(1-\frac{2nc_1}{z^{1/2}}\right),
  e^{-nl}\left(1+\frac{2nc_1}{z^{1/2}}\right),
  1-\frac{2nc_3}{z^{1/2}},
  1+\frac{2nc_3}{z^{1/2}}
  \right)+O\!\left(\frac{1}{z}\right),
\end{equation*}
where the matrix denoted with $O\!\left(\frac{1}{z}\right)$ is diagonal as
well.

Now we can move $C(z)$ in \eqref{XGL} to the left as in the proof of
Proposition \ref{prop:RHP for X}. The result is that \eqref{XGL} is equal to
\begin{align*}
 C_2^{-1}&\left(I+O\!\left(\frac{1}{z}\right)\right)
  \diag\left(z^{-1/4},z^{1/4},z^{-1/4},z^{1/4}\right)
  %\begin{pmatrix}
%  z^{-1/4} & 0 & 0 & 0 \\
%  0 & z^{1/4} & 0 & 0 \\
%  0 & 0 & z^{1/4} & 0 \\
%  0 & 0 & 0 & z^{-1/4}
%  \end{pmatrix}
\frac{1}{\sqrt{2}} \diag\left(\begin{pmatrix} 1&i\\i&1
\end{pmatrix},\begin{pmatrix} 1&i\\i&1
\end{pmatrix}\right).
%\begin{pmatrix}
%  1 & i & 0 & 0 \\
%  i & 1 & 0 & 0 \\
%  0 & 0 & 1 & i \\
%  0 & 0 & i & 1
%  \end{pmatrix}.
 % \nonumber \\
 % &\times
 % \diag\left(z^{\al/2},z^{-\al/2},z^{\al/2},z^{-\al/2}\right).
  %\begin{pmatrix}
%  z^{\al/2} & 0 & 0 & 0 \\
%  0 & z^{-\al/2} & 0 & 0 \\
%  0 & 0 & z^{\al/2} & 0 \\
%  0 & 0 & 0 & z^{-\al/2}
%  \end{pmatrix}
  \end{align*}
Then \eqref{asy of U} is immediate by the definition of $U$ in \eqref{def:U}.

The behavior of $U$ near the origin follows from the corresponding behavior of
$X$, and the fact that the $\lam$-functions are bounded at the origin.
\end{proof}

%-----------------------------------------------------------------------------
\subsection{Third transformation $U\mapsto T$}
\label{section:thirdtransfo}

Our next transformation involves factorizations of jump matrices and the
so-called lens opening, which is a key ingredient of the steepest descent
analysis. The main goal of this step is to convert the highly oscillatory jump
matrices into a more convenient form (e.g., constant matrices) on the original
contour, thereby introducing extra jump matrices with exponentially small
off-diagonal entries on the new contours.

The existence of lenses is guaranteed by the next lemma.

\begin{lem}\label{lemma:locallenses} (Lenses around $\Delta_j$.)
There exist Jordan curves $\Delta_2^{+}$, $\Delta_2^{-}$ connecting $p$ to $q$
in the upper and lower half plane, respectively, such that $\re ~\phi_2(z)<0$
for every non-real $z$ in the region enclosed by $\Delta_2^{+}$,
$\Delta_2^{-}$. This region is called the \emph{lens around~$\Delta_2$}.

Similarly, for $j=1,3$ there exist Jordan curves $\Delta_j^{+}$, $\Delta_j^{-}$
connecting $-\infty$ to $-r_j$ in the upper and lower half plane, respectively,
such that $\re ~\phi_j(z)>0$ for every non-real $z$ in the region enclosed by
$\Delta_j^{+}$, $\Delta_j^{-}$, $j=1,3$. This region is called the \emph{lens
around~$\Delta_j$}. If we are in Case~I or $r_3=r_1$ in Case~III, then we may
(and do) assume that $\Delta_3^{\pm}= \Delta_1^{\pm}$.
\end{lem}

\begin{proof}
We first prove the claim for $\Delta_2$. Recalling that
$\xi_{2,+}(x)=\xi_{3,-}(x)$ for $x\in\Delta_2$, it follows from
\eqref{dmu_2} and the definition \eqref{phi 2} of $\phi_2(x)$ that
\begin{equation}
\phi_{2,+}(x)=-\phi_{2,-}(x)=\pi i\int_{q}^{x}d\mu_2, \qquad x\in\Delta_2.
\end{equation}
Hence, $\phi_{2,\pm}$ is purely imaginary on $\Delta_2$. Furthermore, by
differentiating the above formula with respect to $x$ on both sides, we obtain
\begin{equation}
\frac{d}{dx}\im ~\phi_{2,+}(x)>0 \quad \textrm{and}\quad \frac{d}{dx}\im
~\phi_{2,-}(x)<0.
\end{equation}
Therefore, an appeal to the Cauchy-Riemann equations yields the existence of a
region around $\Delta_2$, such that $\re ~\phi_2(z)<0$ for every
$z\notin\{p,q\}$ in that region.

To prove the claim for $\Delta_1$, note that $\xi_{1,+}(x)=\xi_{2,-}(x)$ for
$x\in\Delta_1$. Then we obtain from \eqref{phi 1} and \eqref{dmu_1} that
\begin{equation}
\phi_{1,+}(x)=-\phi_{1,-}(x)=\pi i\int_{-r_1}^{x}d(\mu_1-\rho_1), \qquad
x\in\Delta_1.
\end{equation}
In view of the constraint condition $\mu_1\leq\rho_1$, it is immediate that
\begin{equation}
\frac{d}{dx}\im ~\phi_{1,+}(x)<0 \quad \textrm{and}\quad \frac{d}{dx}\im
~\phi_{1,-}(x)>0.
\end{equation}
Again, by the Cauchy-Riemann equations, our claim holds. The claim
for $\Delta_3$ can be proved similarly.
%Finally, in Case~I we have
%$\Re(\phi_1)=\Re(\phi_3)$ so we can take $\Delta_3^{\pm}=
%\Delta_1^{\pm}$.
\end{proof}

\begin{figure}[t]
\centering
\begin{overpic}[scale=0.8]{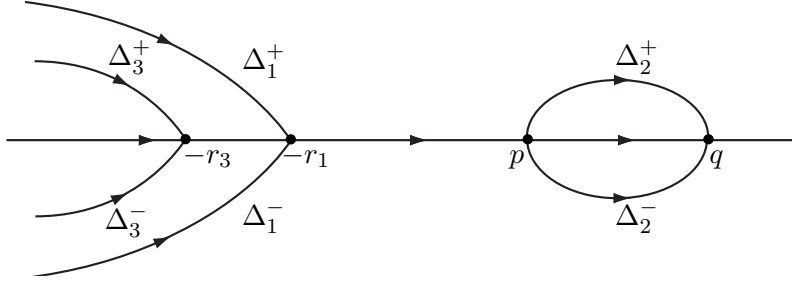}
\put(22.5,15){$-r_3$} \put(35,15){$-r_1$} \put(64,15){$p$} \put(89.5,15){$q$}
\put(12.9,27.5){$\Delta_3^+$} \put(12.5,6.5){$\Delta_3^-$}
\put(30,27){$\Delta_1^+$} \put(30,7){$\Delta_1^-$}
\put(77.5,27.5){$\Delta_2^+$} \put(77.5,7){$\Delta_2^-$}
\end{overpic}
\caption{Lenses around $\Delta_j$, $j=1,2,3$. In Case~I or $r_3=r_1$
in Case~III, we have $\Delta_{1}^{\pm}=\Delta_{3}^{\pm}$.
}\label{fig:LensopeningII}
\end{figure}

An illustration of Lemma~\ref{lemma:locallenses} is shown in
Figure~\ref{fig:LensopeningII}.

%Figure~\ref{fig:local lensopen Case I}.
%\begin{figure}[t]
%%\begin{center}
%%\includegraphics[scale=0.7,viewport=2 8 379 148]{LensopeningI.eps}
%%\caption{Lens opening around $\Delta_1$ and $\Delta_2$ for Case I.}
%%\label{fig:local lensopen Case I}
%%\end{center}
%\centering
%\begin{overpic}[scale=0.7]{LensopeningI.eps}
%\put(20,1){$\Delta_1^{-}$} \put(20,32.5){$\Delta_1^{+}$}
%\put(37,14){0} \put(64,14){$p$} \put(89.5,14){$q$}
%\put(77.5,28){$\Delta_2^{+}$} \put(77.5,5){$\Delta_2^{-}$}
%\end{overpic}
%\caption{Lens opening around $\Delta_1$ and $\Delta_2$ for Case I.}
%\label{fig:local lensopen Case I}
%\end{figure}

The $2\times 2$ middle block of the jump matrix \eqref{jump on delta_2} on
$\Delta_2$ admits the factorization
\begin{align}\label{factorization for phi2}
\begin{pmatrix}
e^{2n\phi_{2,+}} & 1 \\
0 & e^{2n\phi_{2,-}}
\end{pmatrix}
=&\begin{pmatrix}
1 & 0 \\
e^{2n\phi_{2,-}} & 1
\end{pmatrix}
\begin{pmatrix}
0 & 1 \\
-1 & 0
\end{pmatrix} \begin{pmatrix}
1 & 0 \\
e^{2n\phi_{2,+}} & 1
\end{pmatrix}.
\end{align}
Similarly, for the $2\times 2$ upper left and lower right blocks of the jump
matrix \eqref{jump on delta_3} on $\Delta_1$ and $\Delta_3$ we have the
factorizations
\begin{align}\label{factorization for phi13}
\begin{pmatrix}
e^{-i\pi\al}e^{2n\phi_{j,+}} & 0 \\
-1 & e^{i\pi\al}e^{2n\phi_{j,-}}
\end{pmatrix}
=&\begin{pmatrix}
1 & -e^{-i\pi\alpha}e^{-2n\phi_{j,-}}\\
0 & 1
\end{pmatrix}
\begin{pmatrix}
0 & 1 \\
-1 & 0
\end{pmatrix} \nonumber \\
&\times
\begin{pmatrix}
1 & -e^{i\pi\alpha}e^{-2n\phi_{j,+}}\\
0 & 1
\end{pmatrix},\qquad j=1,3.
\end{align}

Based on the factorizations \eqref{factorization for
phi2}--\eqref{factorization for phi13}, we define
\begin{equation}\label{def:T}
T(z)=U(z)\left(I\mp e^{2n\phi_2(z)}E_{32}\right),
\end{equation}
for $z$ in the domain bounded by $\Delta_2^{\pm}$ and $\Delta_2$;
\begin{equation}
T(z)=U(z)\left(I\pm e^{\pm i\pi\al}e^{-2n\phi_{1}(z)}E_{12}\right),
%\begin{pmatrix}
%          1 & \pm e^{\pm\al \pi i}z^{-\al}e^{-2n\phi_{1}(z)} & 0 & 0 \\
%          0 & 1 & 0 & 0\\
%          0 & 0 & 1 & 0  \\
%          0 & 0 & 0 & 1
%\end{pmatrix}
\end{equation}
for $z$ in the domain bounded by $\Delta_1^{\pm}$, $\Delta_3^{\pm}$ and
$\Delta_1$;
\begin{equation}
T(z)=U(z)\left(I\pm e^{\pm i\pi\al}e^{-2n\phi_{1}(z)}E_{12}\pm e^{\pm
i\pi\al}e^{-2n\phi_{3}(z)}E_{34}\right),
%\begin{pmatrix}
%          1 & 0 & 0 & 0 \\
%          0 & 1 & 0 & 0\\
%          0 & 0 & 1 & \mp e^{\pm\al \pi i}z^{-\al}e^{-2n\phi_{3}(z)}  \\
%          0 & 0 & 0 & 1
%\end{pmatrix}
\end{equation}
for $z$ in the domain bounded by $\Delta_3^{\pm}$ and $\Delta_3$, and we let
\begin{equation}
T(z)=U(z),
\end{equation}
for $z$ outside of the lenses.

It then follows from straightforward calculations that $T(z)$ is characterized
by the following RH problem:
\begin{prop}\label{prop:RHP for T}
The matrix valued function $T(z)$ is the unique solution of the following RH
problem.
\begin{enumerate}[(1)]
  \item $T$ is analytic in
  $\mathbb{C}\setminus(\mathbb{R}\cup\Delta_{1}^{\pm}
  \cup\Delta_{2}^{\pm}\cup\Delta_{3}^{\pm})$.

  \item For each of the oriented contours shown in Figure \ref{fig:LensopeningII},
  $T$ has a jump $T_+=T_-J_{T}$,
  where
  \begin{align}
 \nonumber J_{T}&=
  \diag\left(1,\begin{pmatrix}0&1\\-1&0\end{pmatrix},1\right),
          %\begin{pmatrix}
%          1 & 0 & 0 & 0 \\
%          0 & 0 & 1 & 0 \\
%          0 & -1 & 0 & 0 \\
%          0 & 0 & 0 & 1
%         \end{pmatrix}
         ~ \textrm{on $\Delta_2$},
         %\label{jump on delta_2 T Case II and III}
  \\
  \nonumber J_{T}&=I+e^{-2n\phi_{2,+}}E_{23},
  %\begin{pmatrix}
%          1 & 0 & 0 & 0 \\
%          0 & 1 &
%          x^{\al}e^{-2n\phi_{2,+}(x)} & 0 \\
%          0 & 0 & 1 & 0 \\
%          0 & 0 & 0 & 1
%         \end{pmatrix}
         ~ \textrm{on $\mathbb{R}^+ \setminus(0 \cup \overline{\Delta_2})$},
 %\end{align}
% \begin{align}
\\
 \nonumber J_{T}&=\diag(e^{-i\pi\al},e^{i\pi\al},e^{-i\pi\al},e^{i\pi\al})-e^{2n\phi_{1}}E_{21}-
 e^{2n\phi_{3}}E_{43},
          %\begin{pmatrix}
%          1 & 0 & 0 & 0 \\
%          -|x|^{\alpha}e^{2n\phi_{1}(x)} & 1 & 0 & 0 \\
%          0 & 0 & 1 & 0  \\
%          0 & 0 & |x|^{\alpha}e^{2n\phi_{3}(x)} & 1
%         \end{pmatrix}
        ~ \textrm{on $(-r_1,0)$ $($Case III$)$},
 \\
 \nonumber J_{T}&=
 \diag\left(\begin{pmatrix}0&1\\-1&0\end{pmatrix},
 \begin{pmatrix}e^{-i\pi\al} & 0\\ -e^{2n\phi_{3}} & e^{i\pi\al}\end{pmatrix}
 \right),
 %\begin{pmatrix}
%          0 & 1 & 0 & 0 \\
%          -1 & 0 & 0 & 0 \\
%          0 & 0 & e^{-i\pi\al} & 0  \\
%          0 & 0 & -e^{2n\phi_{3}(x)} & e^{i\pi\al}
%         \end{pmatrix}
         ~ \textrm{on $\Delta_1 \setminus
         \overline{\Delta_3}$ $($Cases II and III$)$},
\\
  \nonumber J_{T}&= %\label{Tjump on delta_3 Cases II and III}
  \diag\left(\begin{pmatrix}0&1\\-1&0\end{pmatrix},
  \begin{pmatrix}0&1\\-1&0\end{pmatrix}\right),
         % \begin{pmatrix}
%          0 & 1 & 0 & 0 \\
%          -1 & 0 & 0 & 0 \\
%          0 & 0 & 0 & 1 \\
%          0 & 0 & -1 & 0
%         \end{pmatrix}
         ~ \textrm{on $\Delta_3$},
\\
 \nonumber J_{T}&=I+e^{2n\phi_2}E_{32}, ~ \textrm{on $\Delta_{2}^{\pm}$},
 \\
  \nonumber J_{T}&=I-e^{\pm i \al \pi }e^{-2n\phi_{1}}E_{12},
  %\begin{pmatrix}
%          1 & -e^{\pm\al \pi i}z^{-\alpha}e^{-2n\phi_{1}(z)} & 0 & 0 \\
%          0 & 1 & 0 & 0 \\
%          0 & 0 & 1 & e^{\pm\al \pi i}z^{-\alpha}e^{-2n\phi_{3}(z)}  \\
%          0 & 0 & 0 & 1
%         \end{pmatrix}
~ \textrm{on $\Delta_{1}^{\pm}$},
\\
\nonumber J_{T}&=I-e^{\pm i \al \pi }e^{-2n\phi_{3}}E_{34},
  %\begin{pmatrix}
%          1 & -e^{\pm\al \pi i}z^{-\alpha}e^{-2n\phi_{1}(z)} & 0 & 0 \\
%          0 & 1 & 0 & 0 \\
%          0 & 0 & 1 & e^{\pm\al \pi i}z^{-\alpha}e^{-2n\phi_{3}(z)}  \\
%          0 & 0 & 0 & 1
%         \end{pmatrix}
~ \textrm{on $\Delta_{3}^{\pm}$},
\\
 \nonumber J_{T}&=I-e^{\pm i \al \pi }e^{-2n\phi_{1}}E_{12}
 -e^{\pm i \al \pi }e^{-2n\phi_{3}}E_{34},
  %\begin{pmatrix}
%          1 & -e^{\pm\al \pi i}z^{-\alpha}e^{-2n\phi_{1}(z)} & 0 & 0 \\
%          0 & 1 & 0 & 0 \\
%          0 & 0 & 1 & e^{\pm\al \pi i}z^{-\alpha}e^{-2n\phi_{3}(z)}  \\
%          0 & 0 & 0 & 1
%         \end{pmatrix}
~ \textrm{on $\Delta_{1}^{\pm}=\Delta_{3}^{\pm}$ $($Case~I or
$r_3=r_1$ in Case~III$)$}.
  \end{align}

  \item %$T(z)$ has the same behavior for $z\to\infty$ as $U(z)$, see \eqref{asy of U}.
  As $z\to \infty$, we have
 \begin{align}\nonumber %\label{asy of T}
 T(z)=&~\left(I+O\!\left(\frac{1}{z}\right)\right)
  \diag\left(z^{-1/4},z^{1/4},z^{-1/4},z^{1/4}\right)
  %\begin{pmatrix}
%  z^{-1/4} & 0 & 0 & 0 \\
%  0 & z^{1/4} & 0 & 0 \\
%  0 & 0 & z^{1/4} & 0 \\
%  0 & 0 & 0 & z^{-1/4}
%  \end{pmatrix}
\frac{1}{\sqrt{2}}
%\begin{pmatrix}
%  1 & i & 0 & 0 \\
%  i & 1 & 0 & 0 \\
%  0 & 0 & 1 & i \\
%  0 & 0 & i & 1
%  \end{pmatrix},
\diag\left(\begin{pmatrix}1&i\\i&1\end{pmatrix},\begin{pmatrix}1&i\\i&1\end{pmatrix}\right),
  %\nonumber \\
  %&\times
  %\diag\left(z^{\al/2},z^{-\al/2},z^{\al/2},z^{-\al/2}\right),
  %\begin{pmatrix}
%  z^{\al/2} & 0 & 0 & 0 \\
%  0 & z^{-\al/2} & 0 & 0 \\
%  0 & 0 & z^{\al/2} & 0 \\
%  0 & 0 & 0 & z^{-\al/2}
%  \end{pmatrix}
  \end{align}
  uniformly for $z\in\mathbb{C}\setminus\mathbb{R}$.

  \item $T(z)$ has the same behavior near the origin as $U(z)$ (and $X(z)$), see \eqref{zero behavior of
  X}, provided that $z\to 0$ outside the lenses that end in
  $0$. [The behavior inside the lenses is different but will not be needed for us.]
  \item T is bounded at the finite endpoints of $\Delta_j$, $j=1,2,3$
  other than the origin.
\end{enumerate}
\end{prop}

\begin{proof}
All the properties follow from straightforward calculations.

To show that the asymptotic behavior of $T$ given in item (3) holds uniformly
up to the negative real axis, one needs to trace back the transformations
$Y\mapsto X \mapsto U\mapsto T$. It turns out the entries of $T$ are actually
linear combinations of the modified Bessel functions, whose asymptotic
expansions are uniformly valid up to the negative real axis. We omit the
details here.
%
%Finally, we mention that the uniqueness of $T$ is a result of
%\cite[Lemma 4.1]{KMVV2004}.
\end{proof}

%---------------------------------------------------------------------------------------
\subsection{Global parametrix} \label{section:globalparamatrix}

By the above constructions, it is easily seen from Lemmas
\ref{lemma:lambda:asy} and \ref{lemma:locallenses} that the jump
matrices $J_T$ in the RH problem for $T(z)$ all tend to the identity
matrix exponentially fast as $n\to\infty$, except for the jump
matrices on $\Delta_j$, $j=1,2,3$. Hence, we may approximate $T$ by
a solution $N_\alpha$ of the following $4\times4$ model RH problem:
\begin{enumerate}[(1)]
  \item $N_\alpha$ is analytic in
  $\mathbb{C}\setminus(\overline{\Delta_{1}}\cup\overline{\Delta_{2}}
  \cup\overline{\Delta_{3}})$.

  \item $N_\alpha$ has continuous boundary values on
  $\Delta_{j}$, $j=1,2,3$ and satisfies
  $N_{\alpha,+}=N_{\alpha,-}J_{N_\alpha}$, where
  \begin{equation*}
  J_{N_{\alpha}}=\left\{
                   \begin{array}{ll}
                    \diag\left(\begin{pmatrix}0&1\\-1&0\end{pmatrix},\begin{pmatrix}0&1\\-1&0\end{pmatrix}\right),
                     & \text{on }\Delta_3, \\
                     \diag\left(\begin{pmatrix}0&1\\-1&0\end{pmatrix},e^{-i\pi\al},e^{i\pi\al}\right),
                     & \text{on } \Delta_1 \setminus \overline{\Delta_3}  \text{ (Cases II and III),} \\
                     \diag(e^{-i\pi\al},e^{i\pi\al},e^{-i\pi\al},e^{i\pi\al}), & \text{on  } (-r_1,0)
                     \text{ (Case III),} \\
                     \diag\left(1,\begin{pmatrix}0&1\\-1&0\end{pmatrix},1\right), & \text{on } \Delta_2.
                   \end{array}
                 \right.
  \end{equation*}
  %\begin{align*}%\label{jump for Nalpha 1}
%  N_{\alpha,+}(x)&=N_{\alpha,-}(x)
%  \diag\left(\begin{pmatrix}0&1\\-1&0\end{pmatrix},\begin{pmatrix}0&1\\-1&0\end{pmatrix}\right),
%  %        \begin{pmatrix}
%%          0 & 1 & 0 & 0 \\
%%          -1 & 0 & 0 & 0 \\
%%          0 & 0 & 0 & 1 \\
%%          0 & 0 & -1 & 0
%%         \end{pmatrix},
%         \quad\textrm{$x \in \Delta_3$},
%         \\
%    N_{\alpha,+}(x)&=N_{\alpha,-}(x)
%     \diag\left(\begin{pmatrix}0&1\\-1&0\end{pmatrix},e^{-i\pi\al},e^{i\pi\al}\right),
%  %        \begin{pmatrix}
%%          0 & 1 & 0 & 0 \\
%%          -1 & 0 & 0 & 0 \\
%%          0 & 0 & e^{-i\pi\al} & 0 \\
%%          0 & 0 & 0 & e^{i\pi\al}
%%         \end{pmatrix},
%         \quad\textrm{$x \in \Delta_1 \setminus \overline{\Delta_3}
%         $},
%         \\
%         N_{\alpha,+}(x)&=N_{\alpha,-}(x)\diag(e^{-i\pi\al},e^{i\pi\al},e^{-i\pi\al},e^{i\pi\al}),
%         \quad\textrm{$x \in (-r_1,0)
%         $},
%         \\
%         %\label{jump for Nalpha 3}
%  N_{\alpha,+}(x)&=N_{\alpha,-}(x)
%  \diag\left(1,\begin{pmatrix}0&1\\-1&0\end{pmatrix},1\right),
%          %\begin{pmatrix}
%%          1 & 0 & 0 & 0 \\
%%          0 & 0 & 1 & 0 \\
%%          0 & -1 & 0 & 0 \\
%%          0 & 0 & 0 & 1
%%         \end{pmatrix},
%         \quad \textrm{ $x \in \Delta_2$}.
%  \end{align*}

  \item As $z\to \infty$, $z\in \mathbb{C}\setminus \er^-$,
 \begin{align}
 \nonumber
 %\label{asy of N alpha case II}
 N_{\alpha}(z)=&~\left(I+O\!\left(\frac{1}{z}\right)\right)
  \diag\left(z^{-1/4},z^{1/4},z^{-1/4},z^{1/4}\right)
  %\begin{pmatrix}
%  z^{-1/4} & 0 & 0 & 0 \\
%  0 & z^{1/4} & 0 & 0 \\
%  0 & 0 & z^{1/4} & 0 \\
%  0 & 0 & 0 & z^{-1/4}
%  \end{pmatrix}
\frac{1}{\sqrt{2}}
\diag\left(\begin{pmatrix}1&i\\i&1\end{pmatrix},\begin{pmatrix}1&i\\i&1\end{pmatrix}\right).
%\begin{pmatrix}
%  1&i & 0 & 0 \\
%  i&1 & 0 & 0 \\
%  0 & 0 & 1&i \\
%  0 & 0 & i&1
%  \end{pmatrix}.
%  \nonumber \\
%  &\times
%  \diag\left(z^{\al/2},z^{-\al/2},z^{\al/2},z^{-\al/2}\right).
  %\begin{pmatrix}
%  z^{\al/2} & 0 & 0 & 0 \\
%  0 & z^{-\al/2} & 0 & 0 \\
%  0 & 0 & z^{\al/2} & 0 \\
%  0 & 0 & 0 & z^{-\al/2}
%  \end{pmatrix}
  \end{align}
\end{enumerate}

We construct the solution to this RH problem with the help of
meromorphic differentials, following the idea in \cite{KMo}. We will
give the construction for Case~III; the modifications for Cases~I
and II will be briefly commented at the end of this section.

Consider the Riemann surface in Figure~\ref{fig:Riemann surface}.
Denote with $p$, $q$ (sheets 2 and 3), $-r_1$ (sheets 1 and 2),
$-r_3$ (sheets 3 and 4) the finite branch points. Also denote with
$\infty_{1,2}$ the point at infinity that is common to the first and
second sheet, with $\infty_{3,4}$ the point at infinity that is
common to the third and fourth sheet, and with $0_1$ and $0_4$ the
origin on the first and fourth sheet respectively. Recall that we
are working in Case~III, so $p=0$ and $\min(r_1,r_3)>0$.

Let $\omega_i$, $i=1,2,3,4$ be four meromorphic differentials (of
the third kind) on the Riemann surface with simple poles at the
above listed points (and nowhere else), and with residues at these
points given by the following table:
\begin{equation}\label{meromorphicdiff:1}
\begin{array}{c|cccccccc} x & p & q & -r_1 & -r_3 & \infty_{1,2} & \infty_{3,4}
& 0_1 & 0_4
\\ \hline \Res(\omega_1,x) &
-1/2 & -1/2 &-1/2 &-1/2 &1/2 & 3/2 & -\alpha/2 & \alpha/2
\\
\hline \Res(\omega_2,x) & -1/2 & -1/2 &-1/2 &-1/2 &-1/2 & 5/2 &
-\alpha/2 & \alpha/2
\\
\hline \Res(\omega_3,x) & -1/2 & -1/2 &-1/2 &-1/2 &3/2 & 1/2 &
-\alpha/2 & \alpha/2
\\
\hline \Res(\omega_4,x) & -1/2 & -1/2 &-1/2 &-1/2 &5/2 & -1/2 &
-\alpha/2 & \alpha/2
\end{array}
\end{equation}
The differences among $\omega_i$ are that their residues at
$\infty_{1,2}$ and $\infty_{3,4}$ are all different. Such
meromorphic differentials exist because the sum of the residues is
zero. The fact that our Riemann surface has genus $0$ implies that
each meromorphic differential is unique; cf.~\cite{FK,MRick}.

We then build the first row of $N_\alpha$ with the aid of meromorphic
differential $\omega_1$ introduced in \eqref{meromorphicdiff:1}. Let $p_1$ be
an arbitrary reference point on the first sheet of the Riemann surface, chosen
on the positive real line such that $p_1> q$. Define four functions
$$u_j(z) = \int_{p_1}^{z_j} \omega_1,\qquad j=1,\ldots,4,$$ where we denote with
$z_j$ the point $z$ lying on the $j$th sheet. The integration path
from $p_1$ to $z_j$ obeys the following rules: It can only move from
one sheet to another by going from the lower side of $\Delta_j$ on
sheet $j$ to the upper side of $\Delta_j$ on sheet $j+1$. Moreover,
the path is not allowed to go through a pole of $\omega_1$, and it
must not intersect the intervals $[-r_1,0]$ on the first and second
sheets, nor the intervals $[-r_3,0]$ on the third and fourth sheets,
except possibly at the endpoint of the path.

\begin{figure}[t]
\centering
\begin{overpic}[width=.65\textwidth]{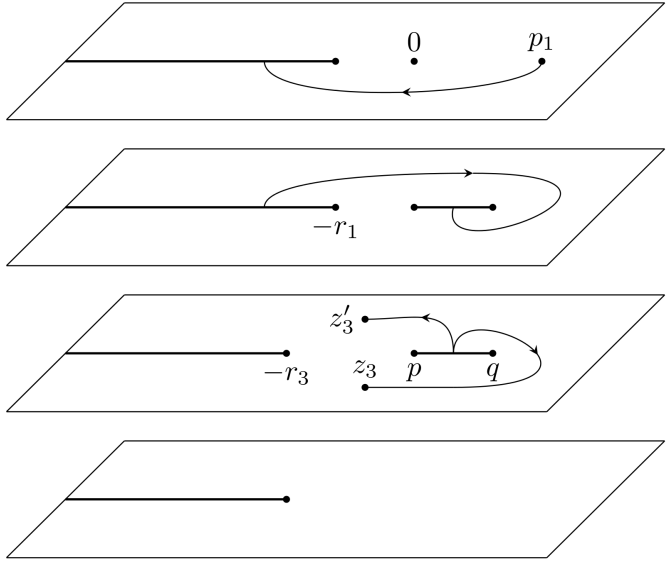}
\end{overpic}
\caption{\label{fig:integration_paths}Integration path for
$u_3(z)$.}
\end{figure}

An illustration of the integration path for $u_3(z)$ is shown in
Figure \ref{fig:integration_paths}.

With the above conventions, the functions $u_j(z)$ satisfy the
following jump properties
\begin{equation}\label{meromorphicdiff:2}
\begin{aligned}
u_{1,+} &= u_{1,-}+2\pi i\Res(\omega_1,0_1)=u_{1,-}-\alpha\pi i, &
\textrm{on $(-r_1,0)$},\\
u_{2,+} &= u_{2,-}-2\pi i\left(\Res(\omega_1,0_1)
+\Res(\omega_1,\infty_{1,2})+\Res(\omega_1,-r_1)\right)\\
&=u_{2,-}+\alpha\pi i, & \textrm{on $(-r_1,0)$},\\
u_{3,+} &= u_{3,-}-2\pi
i\left(\Res(\omega_1,0_4)+\Res(\omega_1,\infty_{3,4})+\Res(\omega_1,-r_3)\right)\\
&=u_{3,-}-2\pi i\left(1+\alpha/2\right), & \textrm{on $(-r_3,0)$},\\
u_{4,+} &= u_{4,-}+2\pi i\Res(\omega_1,0_4)=u_{4,-}+\alpha\pi i, &
\textrm{on $(-r_3,0)$}.
\end{aligned}
\end{equation}
Each of these jumps can be easily verified by deforming the
integration path of $u_{j,+} - u_{j,-}$ to a closed loop and
checking which poles are inside (or outside) the loop. Similarly, we
have
\begin{equation}\label{meromorphicdiff:3}
\begin{aligned}
u_{2,+} &= u_{1,-}, \quad
u_{2,-} = u_{1,+}+2\pi i\Res(\omega_1,\infty_{1,2}) = u_{1,+}+\pi i, \quad &&\textrm{on $\Delta_1$},\\
u_{3,+} &= u_{2,-},\quad
u_{3,-} = u_{2,+}-2\pi i\Res(\omega_1,q)=u_{2,+}+\pi i, \quad &&\textrm{on $\Delta_2$},\\
u_{4,+} &= u_{3,-},\quad u_{4,-} = u_{3,+}+2\pi
i\Res(\omega_1,\infty_{3,4})= u_{3,+}+3\pi i, \quad &&\textrm{on
$\Delta_3$}.
\end{aligned}
\end{equation}

%\begin{equation}\label{meromorphicdiff:3}
%\begin{array}{ll}
%u_{2,+} = u_{1,-} & \textrm{on $(-\infty,-r_1)$},\\
%u_{2,-} = u_{1,+}+2\pi i\Res(\omega,\infty_{1,2}), & \textrm{on $(-\infty,-r_1)$},\\
%u_{3,+} = u_{2,-} & \textrm{on $(p,q)$},\\
%u_{3,-} = u_{2,+}-2\pi i\Res(\omega,q), & \textrm{on $(p,q)$},\\
%u_{4,+} = u_{3,-} & \textrm{on $(-\infty,-r_3)$},\\
%u_{4,-} = u_{3,+}+2\pi i\Res(\omega,\infty_{3,4}), & \textrm{on
%$(-\infty,-r_3)$}.
%\end{array}
%\end{equation}
Now set \begin{equation}\label{def:vj} v_j(z) = \exp(u_j(z)),\qquad
j=1,\ldots,4,
\end{equation}
and consider the row vector
$\vecv(z):=(v_1(z),v_2(z),v_3(z),v_4(z))$. By
\eqref{meromorphicdiff:1}--\eqref{meromorphicdiff:3} we find that
\begin{equation}
\vecv_+=\vecv_-J_{N_\alpha}, \qquad \textrm{on
$\Delta_{1}\cup\Delta_{2}\cup\Delta_{3}$,}
\end{equation}
in Case~III.
%\begin{align*}%\label{jump for vecv}
%  \vecv_+(x)&=\vecv_-(x)
%  \diag\left(\begin{pmatrix}0&1\\-1&0\end{pmatrix},\begin{pmatrix}0&1\\-1&0\end{pmatrix}\right),
%   \quad\textrm{$x \in \Delta_3$},
%         \\
%    \vecv_+(x)&=\vecv_-(x)
%     \diag\left(\begin{pmatrix}0&1\\-1&0\end{pmatrix},e^{-i\pi\al},e^{i\pi\al}\right),
%         \quad\textrm{$x \in \Delta_1 \setminus \overline{\Delta_3}
%         $},
%         \\
%         \vecv_+(x)&=\vecv_-(x)\diag(e^{-i\pi\al},e^{i\pi\al},e^{-i\pi\al},e^{i\pi\al}),
%         \quad\textrm{$x \in (-r_1,0)
%         $},
%         \\
%         %\label{jump for Nalpha 3}
%  \vecv_+(x)&=\vecv_-(x)
%  \diag\left(1,\begin{pmatrix}0&1\\-1&0\end{pmatrix},1\right),
%         \quad \textrm{ $x \in \Delta_2$}.
%  \end{align*}
Thus $\vecv(z)$ satisfies the required jumps for the row vectors of
$N_{\alpha}$.

Next we discuss the asymptotics of $\vecv(z)$ for $z\to\infty$.
Taking into account that $\infty_{1,2}$ is a branch point on the
Riemann surface, its local coordinate on the Riemann surface can be
chosen as $w(z)=1/\sqrt{z}$. Therefore
\begin{align*} u_j(z) &= \Res(\omega_1,\infty_{1,2})\log w(z) + \gamma_j
+O(w(z)) \\
&= -\frac{1}{4}\log z + \gamma_j +O(z^{-1/2}), \qquad z\to\infty,
%u_j(z) = -\frac{1}{2}\Res(\omega,\infty_{1,2})\log z + \gamma_2 +o(1),& \qquad
%j=3,4,
\end{align*}
for $j=1,2$, where $\gamma_1$ and $\gamma_2$ are certain constants.
From the relation $u_{2,+}=u_{1,-}$ on $(-\infty,-r_1)$ we find that
$\gamma_2=\gamma_1+\Res(\omega_1,\infty_{1,2})\pi i=\gamma_1+\pi
i/2$. Similarly we also have
\begin{align*} u_j(z) = -\frac{3}{4}\log z + \gamma_j
+O(z^{-1/2}),& \qquad z\to\infty,
%u_j(z) = -\frac{1}{2}\Res(\omega,\infty_{1,2})\log z + \gamma_2 +o(1),& \qquad
%j=3,4,
\end{align*}
for $j=3,4$, where $\gamma_3$ and $\gamma_4$ are certain constants.
Combining all of this and using \eqref{meromorphicdiff:1} and
\eqref{def:vj}, we find that
$$ \vecv(z) = \left(e^{\gamma_1}z^{-1/4},ie^{\gamma_1}z^{-1/4},0,0 \right)+O(z^{-3/4}),\qquad z\to\infty.
$$
The behavior of $\vecv(z)$ at the origin can be found from similar
arguments as above and is given by
$$ \vecv(z) = O\left(z^{-\alpha/2},z^{-1/4},z^{-1/4},z^{\alpha/2} \right),\qquad z\to 0.
$$
Finally, we also find that
\begin{align}
\vecv(z)=O((|z-\kappa|)^{-1/4}), \qquad \textrm{as $z \to \kappa$},
\end{align}
if $\kappa$ is a branch point other than the origin.

Summarizing, we see that $\frac{e^{-\gamma_1}}{\sqrt{2}}\vecv(z)$
satisfies all the constraints for the first row of $N_{\alpha}$. The
other meromorphic differentials $\omega_i$, $i=2,3,4$ in Table
\ref{meromorphicdiff:1} can be used to build the $i$-th row of
$N_{\alpha}$ in a similar way, we omit the details here.

Our construction of $N_\alpha$ leads to
\begin{align}\label{Nalpha near branchpts}
N_\alpha(z)=O((|z-\kappa|)^{-1/4}), \qquad \textrm{as $z \to
\kappa$},
\end{align}
if $\kappa$ is a branch point other than the origin, and
\begin{equation}\label{Nalpha near the origin:III}
N_\alpha(z) \diag(z^{\alpha/2},z^{1/4},z^{1/4},z^{-\alpha/2}) = O(1),\quad
\textrm{as $z \to 0$}\quad (\textrm{Case~III}).
\end{equation}
Similar constructions can be given in Cases~I and II. For instance,
we may build the first row of $N_\al$ in terms of a meromorphic
differential $\omega$ with prescribed simple poles and residues as
shown in the following tables:
\begin{equation}\label{meromorphicdiff:Case I}
\begin{array}{c|cccccccc} x & p & q & 0_{1,2} & 0_{3,4} & \infty_{1,2} & \infty_{3,4}
\\ \hline \Res(\omega,x) &
-1/2 & -1/2 &-1/2 &-1/2 &1/2 & 3/2
\end{array},
\end{equation}
for Case~I, and
\begin{equation}\label{meromorphicdiff:Case II}
\begin{array}{c|cccccccc} x & p & q & 0_{1,2} & -r_3 & \infty_{1,2} &
\infty_{3,4} & 0_3 & 0_4
\\ \hline \Res(\omega,x) &
-1/2 & -1/2 &-1/2 &-1/2 &1/2 & 3/2 & -\al/2 & \al/2
\end{array},
\end{equation}
for Case~II. Such constructions will again give us \eqref{Nalpha
near branchpts}, but the behavior near the origin is different:
\begin{align}\label{Nalpha near the origin:IandII}
N_\alpha(z) = O(z^{-1/4}),\quad \textrm{as $z \to 0$}\quad (\textrm{Case~I}),\\
N_\alpha(z)\diag(z^{1/4},z^{1/4},z^{\alpha/2},z^{-\alpha/2}) = O(1),\quad
\textrm{as $z \to 0$}\quad (\textrm{Case~II}).
\end{align}

In what follows, we will construct local parametrices near each
branch point.

%--------------------------------------------------------------------------
\subsection{Parametrix near the nonzero branch points}
\label{section:AiryParametrix}

Near each of the branch points in $\{p,q,-r_1,-r_3\}\setminus\{0\}$,
we need to build a local parametrix $P^{\Airy}$ using the $2\times
2$ RH problem for Airy functions. This construction is very
well-known and we omit the details.

%------------------------------------------------------------------
\subsection{Parametrix near the branch point 0 (hard edge)}
\label{section:Besselparametrix}

The local parametrix at the origin is built by means of modified Bessel
functions. Such a construction was given before for a $2\times 2$ RH problem in
\cite{KMVV2004} and in a $3\times 3$ setting in \cite{KMW09,LW08}. Our
construction will be similar to the one in \cite{KMW09}. For $-1<\alpha<0$
there are some extra complications that were solved in an ad hoc way in
\cite{KMW09}. We will be facing similar problems and solve them in a more
conceptual way.

%------------------------------------------------------------------
\subsubsection{RH problem for modified Bessel functions}
\label{subsection:Hankel}

First we recall the construction of \cite{KMVV2004}. Define the $2\times 2$
matrix valued function $\Psi^{\Bessel}(\zeta)$ by
\begin{equation*}%\label{psi bessel}
\Psi^{\Bessel}(\zeta)=
\begin{pmatrix}
I_\al(2\zeta^{1/2}) & \frac{i}{\pi}K_\al(2\zeta^{1/2}) \\
2\pi i\zeta^{1/2}I'_\al(2\zeta^{1/2}) &
-2\zeta^{1/2}K_\al'(2\zeta^{1/2})
\end{pmatrix},
\end{equation*}
for $|\arg \zeta|<2\pi/3$,
\begin{equation*}
\Psi^{\Bessel}(\zeta)=\begin{pmatrix}
\frac{1}{2}H_\al^{(1)}(2(-\zeta)^{1/2}) &
\frac{1}{2}H_\al^{(2)}(2(-\zeta)^{1/2}) \\
\pi\zeta^{1/2}\left(H_\al^{(1)}\right)'(2(-\zeta)^{1/2}) &
\pi\zeta^{1/2}\left(H_\al^{(2)}\right)'(2(-\zeta)^{1/2})
\end{pmatrix}e^{\frac{i \pi \al  }{2}\sigma_3},
\end{equation*}
for $2\pi/3<\arg \zeta<\pi$, and
\begin{equation*}%\label{psi bessel 3}
\Psi^{\Bessel}(\zeta)=\begin{pmatrix}
\frac{1}{2}H_\al^{(2)}(2(-\zeta)^{1/2}) &
-\frac{1}{2}H_\al^{(1)}(2(-\zeta)^{1/2}) \\
-\pi\zeta^{1/2}\left(H_\al^{(2)}\right)'(2(-\zeta)^{1/2}) &
\pi\zeta^{1/2}\left(H_\al^{(1)}\right)'(2(-\zeta)^{1/2})
\end{pmatrix}e^{-\frac{i \pi \al }{2}\sigma_3},
\end{equation*}
for $-\pi<\arg \zeta<-2\pi/3$, where $I_\al$ and $K_\al$ are
modified Bessel functions and $H_\al^{(i)}$, $i=1,2$ is the Hankel
function; cf.~\cite[Chapter 9]{HB92}.

Denote with $\gamma_j$, $j=1,2,3$ the complex rays
$\{\zeta\in\cee\mid\arg\zeta=(j+1)\pi/3\}$. According to
\cite{KMVV2004}, $\Psi^{\Bessel}$ satisfies the following RH
problem:
\begin{itemize}
\item[(1)] $\Psi^{\Bessel}$ is analytic in $\cee\setminus \bigcup_{j=1}^3
\gamma_j$.
\item[(2)] With the rays $\gamma_j$, $j=1,2,3$ all oriented towards the origin,
$\Psi^{\Bessel}$  has the jumps \begin{equation*} \Psi^{\Bessel}_+ =
\Psi^{\Bessel}_- \times\left\{\begin{array}{ll}\begin{pmatrix} 1 & 0 \\
e^{i\pi\al} & 1
\end{pmatrix}, &\text{ on } \gamma_1, \\
\begin{pmatrix} 0 & 1 \\ -1 & 0
\end{pmatrix}, &\text{ on } \gamma_2, \\
\begin{pmatrix} 1 & 0 \\ e^{-i\pi\al} & 1
\end{pmatrix}, &\text{ on } \gamma_3.
\end{array}\right.
\end{equation*}
\item[(3)] Uniformly for $\zeta\to\infty$ we have
\begin{equation}\label{Psi Bessel:asy} \Psi^{\Bessel}(\zeta)
= (2\pi\zeta^{1/2})^{-\sigma_3/2}
\left(\frac{1}{\sqrt{2}}\begin{pmatrix}1 & i \\ i &
1\end{pmatrix}+O(\zeta^{-1/2})\right)e^{2\zeta^{1/2}\sigma_3}.
\end{equation}
\item[(4)] As $\zeta\to 0$ in $|\arg\zeta|<2\pi/3$ we have
\begin{equation}\label{Hankel:zero2}
\Psi^{\Bessel}(\zeta) = \left\{\begin{array}{ll}
O\begin{pmatrix}\zeta^{\al/2} & \zeta^{-\al/2}\\
\zeta^{\al/2} & \zeta^{-\al/2}\end{pmatrix}, &\text{if } \al>0, \\
O\begin{pmatrix}
1 & \log|\zeta| \\
1 & \log|\zeta|
\end{pmatrix}, &\text{if } \al=0,\\
O(\zeta^{\al/2}), &\text{if } \al<0.
\end{array}\right.
\end{equation}
\end{itemize}

%------------------------------------------------------------------
\subsubsection{Construction for $\alpha\geq 0$}
\label{subsection:hardedge:>0}

In the construction of the local parametrix at the origin we will
restrict ourselves to Case~II, which is in some sense the most
general and difficult of the three cases. The modifications for
Cases~I and III will be briefly commented at the end of this
section.

We denote by $B_\delta$ a small fixed disk with radius $\delta>0$
centered at the origin, such that it does not contain any other
branch points and consider all the jump matrices $J_T$ of $T$
restricted in $B_\delta$; cf.~item (2) in Proposition \ref{prop:RHP
for T}. We note that the $(2,3)$ entry of $J_T$ on $(0,\delta)$ is
$e^{-2n\phi_{2,+}}$. Since $\Re\phi_{2,+}=\Re
(\lambda_2-\lambda_3)>c>0$ on $(0,\delta)$ (cf.~Lemma
\ref{lemma:lambda:var}), this entry is exponentially small as
$n\to\infty$. Similarly the $(4,3)$ entry of $J_T$ on $(-\delta,0)$
is $-e^{2n\phi_{3}}$ which is also exponentially small.

By neglecting the exponentially small entries in the jump matrices
for $T(z)$ in $B_\delta$, we are led to the following RH problem for
a $4\times 4$ matrix valued function $Q(z)$:

\begin{figure}[t]
\centering
\begin{overpic}[scale=0.8]{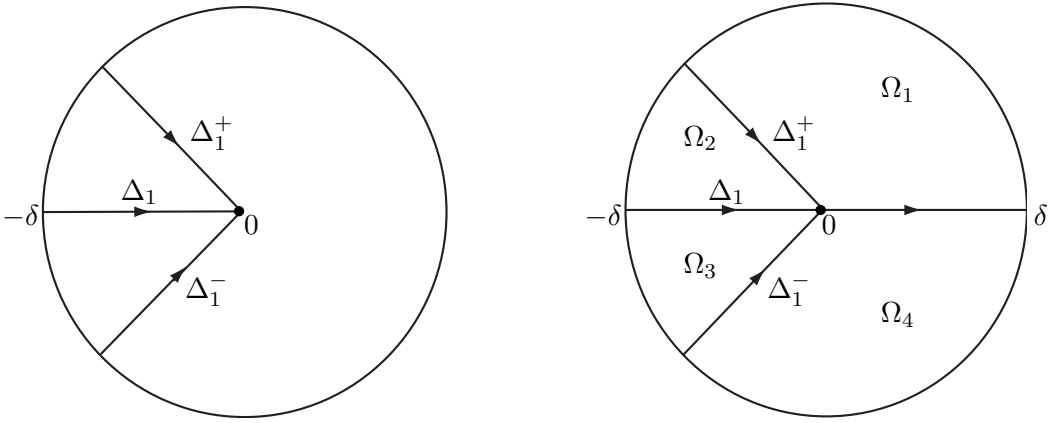}
\put(-4,20){$-\delta$} \put(20.5,19){$0$} \put(8,22.5){$\Delta_1$}
\put(14.5,12.5){$\Delta_1^-$} \put(15,28.5){$\Delta_1^{+}$}
\put(55,20){$-\delta$} \put(100.5,20){$\delta$} \put(79,19){$0$}
\put(67.5,22.5){$\Delta_1$} \put(73.5,12.5){$\Delta_1^-$}
\put(74,28.5){$\Delta_1^{+}$} \put(85,33){$\Omega_1$}
\put(85,10){$\Omega_4$} \put(65,15){$\Omega_3$}
\put(65,28){$\Omega_2$}
\end{overpic}
\caption{Contours for the local parametrix around $0$ in Case~II,
for $\alpha\geq 0$ (left picture) and for $-1<\alpha<0$ (right
picture).} \label{fig:contournear 0 case II}
\end{figure}

\begin{enumerate}[(1)]
  \item $Q$ is analytic in
  $B_\delta \setminus(\Delta_{1}\cup\Delta_{1}^{\pm})$.

  \item On each of the oriented contours in the left picture of
   Figure \ref{fig:contournear 0 case II},
  $Q$ has a jump $Q_+=Q_-J_{Q}$,
  where
\begin{equation*}
J_{Q}=\left\{
        \begin{array}{ll}
          \diag\left(\begin{pmatrix}0&1\\-1& 0\end{pmatrix},e^{-i\pi\alpha},e^{i\pi\alpha}\right),
          & \text{on } \Delta_1\cap B_\delta=(-\delta,0), \\
          I-e^{\pm\al \pi i}e^{-2n\phi_{1}}E_{12}, & \text{on } \Delta_{1}^{\pm}\cap B_\delta.
        \end{array}
      \right.
\end{equation*}
  %\begin{align*}
%  J_{Q}(x)&=
%  \diag\left(\begin{pmatrix}0&1\\-1& 0\end{pmatrix},e^{-i\pi\alpha},e^{i\pi\alpha}\right),
%  \quad\textrm{ },
%  \\
%  J_{Q}(z)&=I-e^{\pm\al \pi i}e^{-2n\phi_{1}(z)}E_{12},
%  \quad \textrm{ $z\in\Delta_{1}^{\pm}\cap B_\delta$}.
%  \end{align*}
  \item As $z\to 0$ outside the lens enclosed by $\Delta_1^{\pm}$ we have
  \begin{equation}\label{zero behavior of Q:>0} \begin{array}{ll}
  Q(z)\diag(|z|^{\al/2},|z|^{-\al/2},|z|^{\al/2},|z|^{-\al/2})=O(1),&
\qquad \textrm{if $\alpha>0$},\\
  Q(z)\diag((\log |z|)^{-1},1,(\log |z|)^{-1},1)=O(1),&
  \qquad \textrm{if $\alpha=0$}.
  %\\
  %Q(z)=O(z^{\al/2}),~~Q^{-1}(z)=O(z^{\al/2}),&\qquad \textrm{if $-1<\al<0$,}
  \end{array}\end{equation}
  %where $C_{\al}$, $\al\geq 0$ are certain constant matrices of size $4\times
  %4$.
  %$Q(z)$ has the same behavior near the origin as $X(z)$, see \eqref{zero behavior of X},
  \item As $n \to \infty$,
 \begin{align}\label{matching condition Q}
Q(z)=N_\alpha(z)\left(I+O\!\left(\frac{1}{n}\right)\right),
  \end{align}
  uniformly for $z\in\partial B_\delta \setminus(\Delta_1\cup
  \Delta_1^{\pm})$, where $N_{\alpha}$ is the global parametrix built in
  Section~\ref{section:globalparamatrix}.

\end{enumerate}

Before solving this RH problem, let us consider the function
\begin{equation}
f_1(z):=(\phi_1(z))^2.
\end{equation}
Since the density of $\rho_1-\mu_1$ blows up like $O(|x|^{-1/2})$ as
$x \to 0$, we obtain from \eqref{dmu_1} and the definition of
$\phi_1(z)$ in \eqref{phi 1} that $f_1(z)$ is analytic in $B_\delta$
and gives a conformal mapping from a neighborhood of the origin onto
itself, such that $f_1(x)$ is real and positive for
$x\in(0,\delta)$. For our purpose, we may deform the contours
$\Delta_1^{\pm}$ near $0$ such that $f_1$ maps $\Delta_1^{\pm}\cap
B_\delta$ to the rays with angles $2\pi/3$ and $-2\pi/3$,
respectively.

We now define
\begin{equation}\label{Q first part Case II}
Q(z)=E(z)\diag\left(
\sigma_1\Psi^{\Bessel}(n^2f_1(z))\sigma_1,z^{-\alpha/2},z^{\alpha/2}\right)
\diag(e^{n\phi_1(z)},-e^{-n\phi_1(z)}, 1,1),
\end{equation}
where $\sigma_1=\begin{pmatrix} 0 & 1 \\
1 & 0
\end{pmatrix}$, and where the prefactor $E(z)$ is analytic in $B_\delta$ and is chosen to satisfy
the matching condition on $\partial B_\delta$, see below. With this
definition, the items $(1)$, $(2)$ and $(3)$ in the RH problem for
$Q(z)$ are satisfied, by virtue of items $(1)$, $(2)$ and $(4)$ in
the RH problem for $\Psi^{\Bessel}(z)$.

To achieve the matching condition in item $(4)$ of the RH problem
for $Q(z)$, we take $E(z)$ in \eqref{Q first part Case II} as
\begin{align}\label{E case II}
E(z) &=N_\al(z)\diag(1,-1,1,1)
\diag\left(\frac{1}{\sqrt{2}}\begin{pmatrix}1 & -i
\\
-i & 1\end{pmatrix},z^{\alpha/2},z^{-\alpha/2}\right) \nonumber \\
&~~~\times\diag((2\pi n)^{-1/2}f_1(z)^{-1/4},(2\pi
n)^{1/2}f_1(z)^{1/4},1,1).
\end{align}
Obviously $E(z)$ is analytic for $z\in B_{\delta}\setminus\er^-$.
Moreover, one checks that $E(z)$ is also analytic across
$(-\delta,0)$. Finally, since $f_1(z)^{1/4}=O(z^{1/4})$ as $z\to 0$
and using \eqref{Nalpha near the origin:III} we find that
\begin{equation}
E(z)=O(z^{-1/2}), \qquad z\to 0,
\end{equation}
so $E(z)$ cannot have a pole at zero. We conclude that $E(z)$ is
analytic in the disk $B_{\delta}$. By virtue of \eqref{Psi
Bessel:asy}, the matching condition \eqref{matching condition Q} in
the RH problem for $Q$ is satisfied.

In summary, we have established the following proposition.

\begin{prop}\label{sol to alpha geq 0}
The matrix valued function $Q$ defined by \eqref{Q first part Case
II} and \eqref{E case II} satisfies the conditions (1)--(4) of the
RH problem for $Q$.
\end{prop}

In view of the fact that $\det N_\alpha(z)=1$ and $\det
\Psi^{\Bessel} =1$ (see \cite{KMVV2004}), we see that
\begin{equation}
\det Q(z) \equiv 1, \qquad z\in B_\delta.
\end{equation}
If we take $Q$ as the local parametrix for $T$, the final
transformation would be defined as
\begin{equation}
R(z)=T(z)Q(z)^{-1}, \qquad z\in B_\delta.
\end{equation}
Then $R$ would be analytic in $B_\delta\setminus (0,\infty)$ with
the following jump on $(0,\delta)$,
\begin{align}\label{jump for R on 0 to delta}
R_-(x)^{-1}R_+(x)&= Q(x)T_-(x)^{-1}T_+(x)Q(x)^{-1} \nonumber \\
&=Q(x)(I+e^{-2n\phi_{2,+}(x)}E_{23})Q(x)^{-1} \nonumber \\
&=I+e^{-2n\phi_{2,+}(x)}Q(x)E_{23}Q(x)^{-1}.
\end{align}
Similarly, the jump on $(-\delta,0)$ would be
\begin{align}\label{jump for R on -delta to 0}
R_-(x)^{-1}R_+(x)&=I-e^{-i\pi\alpha}e^{2n\phi_{3}(x)}Q_+(x)E_{43}Q_+(x)^{-1}.
\end{align}

\begin{lem}\label{QEQinverse}
For $\alpha\geq 0$, the matrix $Q(x)E_{23}Q(x)^{-1}$ is bounded as
$x\to 0$, $x>0$. Similarly, $Q_+(x)E_{43}Q_+(x)^{-1}$ is bounded as
$x\to 0$, $x<0$.
\end{lem}
\begin{proof}
We prove the first statement only; the second follows similarly.
Observe that
\begin{equation}\label{hardedge:boundedjump}
Q(x)E_{23}Q(x)^{-1}=Q(x)\begin{pmatrix} 0 & 1 & 0 & 0
\end{pmatrix}^T
\begin{pmatrix}
0 & 0 & 1 & 0
\end{pmatrix}Q(x)^{-1}.
\end{equation}
If $\alpha>0$, it follows from \eqref{zero behavior of Q:>0} and
$\det Q(z)\equiv 1$ that both $Q(x)\begin{pmatrix} 0 & 1 & 0 & 0
\end{pmatrix}^T$ and $\begin{pmatrix}
0 & 0 & 1 & 0
\end{pmatrix}Q(x)^{-1}$ behave like $O(z^{\alpha/2})$, and the lemma follows.

If $\alpha=0$, we use \eqref{Hankel:zero2} and $\det
\Psi^{\Bessel}(\zeta)=1$ to get
\begin{equation*}
\left(\Psi^{\Bessel}\right)^{-1}(\zeta)= \begin{pmatrix} O(1) & O(1)\\
O(\log|\zeta|) & O(\log|\zeta|)
\end{pmatrix}
\quad \textrm{as $\zeta\to 0$}.
\end{equation*}
Inserting this formula in \eqref{Q first part Case II}, we have
\begin{equation}\label{zero behavior of Q inverse}
Q^{-1}(x)=
\diag\left( \begin{pmatrix} O(1) & O(1)\\
O(\log x) & O(\log x)
\end{pmatrix},
O(1),O(1)\right)
%\begin{pmatrix}
%O(1) & O(1) & 0 & 0 \\
%O(\log x) & O(\log x) & 0 & 0 \\
%0 & 0 & O(1) & O(1) \\
%0 & 0 & O(\log x) & O(\log x)
%\end{pmatrix}
E^{-1}(x),
\end{equation}
as $x \to 0$ and $x>0$, where $E^{-1}(x)$ is bounded near the
origin. Since $Q(x)(0~1~0~0)^{T}$ is bounded by \eqref{zero behavior
of Q:>0} and $(0~0~1~0)Q(x)^{-1}$ is bounded by \eqref{zero behavior
of Q inverse}, the lemma follows for $\alpha=0$ as well.
\end{proof}
The above lemma, together with \eqref{jump for R on 0 to delta} and
the fact that $\re~\phi_{2,+}(x)>c>0$ for some constant $c$ on
$(0,\delta)$, implies that, if $\alpha \geq 0$, the jump matrix for
$R$ is exponentially close to the identity matrix as $n \to \infty$,
uniformly for $x\in(0,\delta)$. A similar statement holds for the
jump on $(-\delta,0)$. We therefore take the parametrix $P=Q$ in
case $\alpha\geq 0$.

%----------------------------------------------------------------
\subsubsection{Construction for $-1<\alpha< 0$}
\label{subsection:hardedge:<0}

If $-1<\alpha<0$, then the $(2,3)$ entry of the jump matrix $J_T$ on
$(0,\delta)$ and the $(4,3)$ entry of $J_T$ on $(-\delta,0)$ cannot
be simplify neglected. Hence, we need to build a $4\times 4$ matrix
valued function $P$ such that
\begin{enumerate}[(1)]
  \item $P$ is analytic in
  $B_\delta \setminus(\er\cup\Delta_{1}^{\pm})$.

  \item For each of the oriented contours shown in the right picture of
   Figure \ref{fig:contournear 0 case II},
  $P$ has a jump $P_+=P_-J_{P}$,
  where
  \begin{equation*}
  J_P=\left\{
        \begin{array}{ll}
          \diag\left(\begin{pmatrix}0&1\\-1&0\end{pmatrix},
         \begin{pmatrix}e^{-i\pi\al} & 0\\ -e^{2n\phi_{3}} & e^{i\pi\al}\end{pmatrix}\right),
      & \text{on } \Delta_1\cap B_\delta=(-\delta,0), \\
          I-e^{\pm\al \pi i}e^{-2n\phi_{1}}E_{12}, & \text{on } \Delta_{1}^{\pm}\cap B_\delta, \\
          I+e^{-2n\phi_{2,+}}E_{23}, & \text{on } (0,\delta).
        \end{array}
      \right.
  \end{equation*}
  %\begin{align*}
%  J_{P}(x)&=
%  \diag\left(\begin{pmatrix}0&1\\-1&0\end{pmatrix},
% \begin{pmatrix}e^{-i\pi\al} & 0\\ -e^{2n\phi_{3}(x)} & e^{i\pi\al}\end{pmatrix}\right),
%  \quad x\in(-\delta,0)=\Delta_1\cap B_\delta,
%  \\
%  J_{P}(z)&=I-e^{\pm\al \pi i}e^{-2n\phi_{1}(z)}E_{12},
%  \quad  z\in\Delta_{1}^{\pm}\cap B_\delta,
%  \\
%  J_{P}(x)&=I+e^{-2n\phi_{2,+}(x)}E_{23},\quad x\in(0,\delta).
%  \end{align*}

  \item
  $P(z)$ behaves near the origin like:
  \begin{equation*}
  P(z)=O(z^{\al/2}),
  \qquad P^{-1}(z)=O(z^{\al/2}).
  \end{equation*}

  \item As $n \to \infty$,
 \begin{align}\label{matching condition P}
  P(z)=N_\alpha(z)\left(I+O\!\left(\frac{1}{n}\right)\right),
  \end{align}
  uniformly for $z\in\partial B_\delta \setminus(\Delta_1\cup
  \Delta_1^{\pm})$, where $N_{\alpha}$ is the global parametrix built in
  Section~\ref{section:globalparamatrix}.
\end{enumerate}

We use the matrix valued function $Q$ given by \eqref{Q first part
Case II} and \eqref{E case II}, which works as the parametrix for
the case $\alpha\geq 0$, to construct $P$. More precisely, recall
the sectors $\Omega_j$, $j=1,\ldots,4$, in the right picture of
Figure~\ref{fig:contournear 0 case II}. By setting
\begin{equation}\label{def:s}
S(z)=\Lam^{-1}(z)S_j(z)\Lam(z), \qquad z\in\Omega_j,
\end{equation}
where $\Lam$ is defined in \eqref{def:Lambda} and $S_j$,
$j=1,\ldots,4$ is a constant matrix that depends on the sector
$\Omega_j$, we construct $P$ in the form
\begin{equation}\label{def of P case II}
P(z)=Q(z)S(z).
\end{equation}
The construction of these four matrices $S_j$ is different in the
three Cases I--III; we will again explain it for Case~II.

Let us give names to the jump matrix $J_P$ on the four jump contours
around zero as follows. $J_{P,1}$ is the jump matrix of $P$ on
$(0,\delta)$, $J_{P,2}$ is the jump matrix on $\Delta_1^+$,
$J_{P,3}$ on $(-\delta,0)$ and $J_{P,4}$ on $\Delta_1^-$. Similarly
we define the jump matrices $J_{Q,k}$, $k=1,\ldots,4$. Note that
$J_{Q,2}=J_{P,2}$, $J_{Q,4}=J_{P,4}$, $J_{Q,1}$ differs from
$J_{P,1}$ in the $(2,3)$ entry, and $J_{Q,3}$ differs from $J_{P,3}$
in the $(4,3)$ entry,

To obtain the correct jumps for $P$, the matrices $S_j$ in
\eqref{def of P case II} should be such that
\begin{align}
\nonumber \Lam J_{P,1}^{-1}\Lam^{-1} = S_1^{-1}(\Lam J_{Q,1}^{-1}\Lam^{-1}) S_4, & \\
\label{Sk:relations}\Lam J_{P,k}\Lam^{-1} = S_k^{-1}(\Lam
J_{Q,k}\Lam^{-1}) S_{k-1},& \qquad \textrm{for }k=2,3,4.
\end{align}
Multiplying these equations into a telescoping product form yields
the following relation for $S_1$:
\begin{equation}\label{eq:S1} \Lam(J_{P,1}^{-1}J_{P,4}J_{P,3}J_{P,2})\Lam^{-1} =
S_1^{-1}\Lam\left( J_{Q,1}^{-1}J_{Q,4}J_{Q,3}J_{Q,2}\right)\Lam^{-1}
S_1.
\end{equation}
Now $\Lam(J_{P,1}^{-1}J_{P,4}J_{P,3}J_{P,2})\Lam^{-1}$ and
$\Lam(J_{Q,1}^{-1}J_{Q,4}J_{Q,3}J_{Q,2})\Lam^{-1}$ are both constant
matrices. A calculation shows that they have the spectral
decompositions
\begin{align}\label{spectral:decomp:P} \Lam (J_{P,1}^{-1}J_{P,4}J_{P,3}J_{P,2})\Lam^{-1} =
\begin{pmatrix}
e^{-i\pi\alpha} &0&0&0\\
-1 & e^{i\pi\alpha} & -e^{-i\pi\alpha} & 0\\
0 & 0 & e^{-i\pi\alpha} & 0 \\
0 & 0 & -1 & e^{i\pi\alpha}
\end{pmatrix} =: V_P D V_P^{-1},
\\ \label{spectral:decomp:Q} \Lam (J_{Q,1}^{-1}J_{Q,4}J_{Q,3}J_{Q,2})\Lam^{-1} =
\diag(e^{-i\pi\alpha},e^{i\pi\alpha},e^{-i\pi\alpha},e^{i\pi\alpha})-E_{21}
=: V_Q D V_Q^{-1},
\end{align}
with common eigenvalue matrix
\begin{equation}\label{eigenvalue:matrix:D} D =
\diag(e^{-i\pi\alpha},e^{i\pi\alpha},e^{-i\pi\alpha},e^{i\pi\alpha}),\end{equation}
and where the eigenvector matrices $V_P$ and $V_Q$ can be chosen as
$$ V_P =
I+\frac{1}{2i\sin(\pi\al)}(E_{21}+e^{-i\pi\alpha}E_{23}+E_{43}),\qquad
V_Q=I+\frac{1}{2i\sin(\pi\al)}E_{21}.
$$
On account of the spectral decompositions
\eqref{spectral:decomp:P}--\eqref{spectral:decomp:Q}, the equation
\eqref{eq:S1} can be rewritten as
$D(V_Q^{-1}S_1V_P)=(V_Q^{-1}S_1V_P)D$, which allows the solution
$S_1 = V_Q V_P^{-1}$. So we can take
\begin{equation}\label{eq:S1 expli}
S_1 = V_Q V_P^{-1} =
I-\frac{1}{2i\sin(\pi\al)}(e^{-i\pi\alpha}E_{23}+E_{43}).
\end{equation}
The constant matrices $S_2$, $S_3$ and $S_4$ are then obtained
recursively from $S_1$ and \eqref{Sk:relations}, which now read
\begin{align}
S_2&=I+\frac{1}{2i\sin(\pi\al)}\left(E_{13}-e^{-i\al\pi}E_{23}-E_{43}\right),
\\
S_3&=I-\frac{1}{2i\sin(\pi\al)}(E_{13}+e^{i\pi\al}E_{23}+E_{43}),
\\
S_4&=I-\frac{1}{2i\sin(\pi\al)}(e^{i\pi\al}E_{23}+E_{43}).\label{eq:S4}
\end{align}

With $S_k$, $k=1,\ldots,4$, given in \eqref{eq:S1
expli}--\eqref{eq:S4}, we conclude that $P$ defined in \eqref{def of
P case II} indeed satisfies the items (1)--(4) in the RH problem for
$P(z)$ stated above.

The above description was for Case~II. For Cases~I and III similar
constructions can be given. The key point is that in each case, the
cyclic products of jump matrices in the left hand sides of
\eqref{spectral:decomp:P} and \eqref{spectral:decomp:Q},
respectively, allow spectral decompositions with common eigenvalue
matrix \eqref{eigenvalue:matrix:D}. We leave these constructions to
interested readers.

Combining the results in this subsection and Section
\ref{subsection:hardedge:>0}, we take $P=QS$ as the parametrix in
$B_\delta$ for Case II, where $S=I$ if $\alpha\geq 0$, and $S$ is
given by \eqref{def:s}, if $-1<\alpha<0$.

%-------------------------------------------------------------------
\subsection{Final transformation} \label{section:finaltransfo}
\begin{figure}[t]
\centering
\begin{overpic}[scale=0.7]{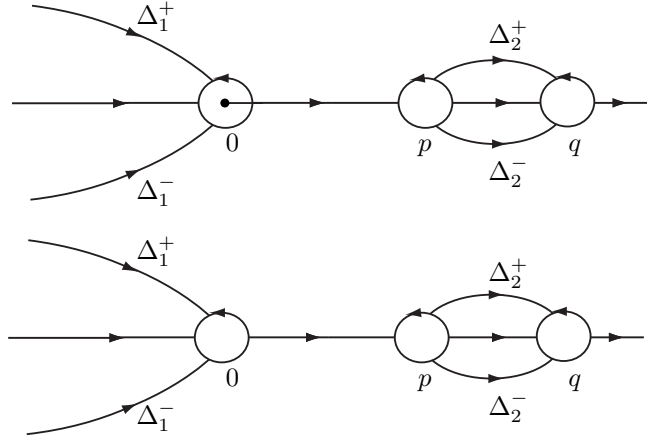}
\small{\put(34,45){$0$} \put(64,45){$p$} \put(87.5,45){$q$}
\put(20,65){$\Delta_1^+$} \put(20,37.5){$\Delta_1^{-}$}
\put(75,62){$\Delta_2^{+}$} \put(75,40.5){$\Delta_2^{-}$}
\put(20,28){$\Delta_1^+$} \put(20,2){$\Delta_1^{-}$} \put(34,8){$0$}
\put(64,8){$p$} \put(87.5,8){$q$} \put(75,24.5){$\Delta_2^{+}$}
\put(75,3.5){$\Delta_2^{-}$}}
\end{overpic}
\caption{Jump contours $\Sigma_{R}$ for the RH problem for $R$ in
Case I, for $\alpha \geq 0$ (top) and for $-1<\alpha<0$ (bottom).}
\label{fig:contoursforR CaseI}
\end{figure}

Using the local parametrices $P$, $P^{\Airy}$ and the global
parametrix $N_\alpha$, we define the final transformation as follows
\begin{equation}
R(z)=\left\{
       \begin{array}{ll}
         T(z)(P^{\Airy}(z))^{-1}, & \text{in the disks around } \{p,q,-r_1,-r_3\}\setminus\{0\}, \\
         T(z)(P(z))^{-1}, & \text{in the disk } B_\delta
         \text{ around the origin,} \\
        T(z)(N_\alpha(z))^{-1}, & \text{elsewhere.}
       \end{array}
     \right.
\end{equation}

\begin{figure}[t]
\centering
\begin{overpic}[scale=0.75]{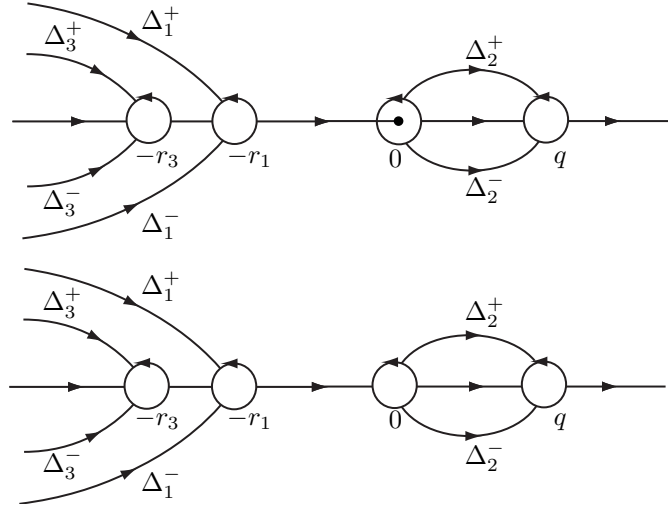}
\put(19,52.5){\small{$-r_3$}} \put(33,52.5){\small{$-r_1$}}
\put(57.5,52){\small{$0$}} \put(82.5,52.5){\small{$q$}}
\put(19,12.5){\small{$-r_3$}} \put(33,12.5){\small{$-r_1$}}
\put(57.5,12){\small{$0$}} \put(82.5,12.5){\small{$q$}}
\put(5.2,70.7){\small{$\Delta_3^{+}$}}
\put(5,45.5){\small{$\Delta_3^{-}$}}
\put(5,30){\small{$\Delta_3^{+}$}}
\put(5,5.4){\small{$\Delta_3^{-}$}}
\put(20,73){\small{$\Delta_1^{+}$}}
\put(20,42){\small{$\Delta_1^{-}$}}
\put(20,32.5){\small{$\Delta_1^{+}$}}
\put(20,2.5){\small{$\Delta_1^{-}$}}
\put(69,68.5){\small{$\Delta_2^{+}$}}
\put(69,47.5){\small{$\Delta_2^{-}$}}
\put(69,28.5){\small{$\Delta_2^{+}$}}
\put(69,6.5){\small{$\Delta_2^{-}$}}
\end{overpic}
\caption{Jump contours $\Sigma_{R}$ for the RH problem for $R$ in
Case III, for $\alpha \geq 0$ (top) and for $-1<\alpha<0$
(bottom).}\label{fig:contoursforR CaseIII}
\end{figure}

From our construction of the parametrices, it follows that $R$
satisfies the following RH problem:
\begin{enumerate}[(1)]
\item $R$ is analytic in $\mathbb{C} \setminus \Sigma_R$, where $\Sigma_R$ depends on
$\alpha$ and is different for all three cases. For Cases I and III,
an illustration is shown in Figures \ref{fig:contoursforR CaseI} and
\ref{fig:contoursforR CaseIII}, respectively.

\item $R$ has jumps $R_+=R_-J_R$ on $\Sigma_R$ that satisfy
\begin{equation}
J_R(z)=I+O(1/n),
\end{equation}
uniformly for $z$ on the boundaries of the disks;
\begin{equation}
J_R(x)=I+O(x^{\alpha} e^{-cn}),
\end{equation}
on $(0,\delta)$ (Case I), or on $(-\delta, 0)\cup(0,\delta)$ (Case
II), or on $(-\delta, 0)$ (Case III), for some constant $c>0$, if
$\alpha \geq 0$;
\begin{equation}
J_R(z)=I+O(e^{-cn|z|}),
\end{equation}
on the other parts of $\Sigma_R$, for some constant $c>0$.

\item $R(z)=I+O(1/z)$ as $z\to \infty$.

\end{enumerate}

Then, as in \cite{DKMVZ992,DKMVZ991,KMW09}, we may conclude that
\begin{equation}\label{large n behavior of R}
R(z)=I+O\left(\frac{1}{n(|z|+1)}\right),
\end{equation}
as $n\to\infty$, uniformly for $z$ in the complex plane outside of
$\Sigma_R$.

%------------------------------------------------------------------------------
\section{Proof of Theorem \ref{theorem:limiting mean distribution}}
\label{section:proof:limitdistribution}

With the above preparations, we are ready to establish Theorem
\ref{theorem:limiting mean distribution}. For this purpose, we need
to represent the correlation kernel $K_n(x,y)$ in terms of $T$ by
following the sequence of transformations $Y \mapsto X \mapsto U
\mapsto T$ in the RH analysis.

Let $x,y \in \Delta_2$. Recalling the representation of $K_n(x,y)$
in \eqref{kernel representation}, we first find from \eqref{def:X}
that
\begin{equation*}
K_n(x,y)=\frac{1}{2\pi i(x-y)}
\begin{pmatrix}
0 & 0 & y^{-\alpha/2} e^{-\frac{ny}{1-t}} & 0
\end{pmatrix}X_{+}^{-1}(y)X_{+}(x)
\begin{pmatrix}
0 & x^{\alpha/2} e^{-\frac{nx}{t}} & 0 & 0
\end{pmatrix}^T.
\end{equation*}
In view of \eqref{def:U}, this becomes
\begin{align*}
K_n(x,y)&=\frac{x^{\alpha/2}y^{-\alpha/2}}{2\pi
i(x-y)}\begin{pmatrix} 0 & 0 & e^{n\lam_{3,-}(y)} & 0
\end{pmatrix} U_{+}^{-1}(y)U_{+}(x)
\begin{pmatrix}
0 & e^{-n\lam_{2,+}(x)}  & 0 & 0
\end{pmatrix}^T.
\end{align*}
By \eqref{def:T}, it follows that
\begin{multline}\label{kernel in terms of T}
K_n(x,y)=\frac{x^{\alpha/2}y^{-\alpha/2}}{2\pi
i(x-y)}\begin{pmatrix} 0 & -e^{n\lam_{2,-}(y)} & e^{n\lam_{3,-}(y)}
& 0
\end{pmatrix}
%\nonumber \\
%&~~~~\times
T_{+}^{-1}(y)T_{+}(x)
\\ \times\begin{pmatrix}
0 & e^{-n\lam_{2,+}(x)}  & e^{-n\lam_{3,+}(x)} & 0
\end{pmatrix}^T.
\end{multline}

Now, we obtain from \eqref{large n behavior of R} and standard
arguments (e.g. \cite[Section 9]{BK04}) that
\begin{equation}\label{Tinverse time T}
T_{+}^{-1}(y)T_{+}(x)=I+O(x-y), \qquad \textrm{as $x\to y$,}
\end{equation}
uniformly in $n$. Also note that $\lam_{3,\pm}=\lam_{2,\mp}$. Thus,
by taking $y\to x$, it follows from \eqref{Tinverse time T} and
L'H\^{o}pital's rule that
\begin{equation}
K_n(x,x)=\frac{n}{2\pi
i}(\xi_{2,+}(x)-\xi_{2,-}(x))+O(1)=n\frac{d\mu_2}{dx}(x)+O(1),
\end{equation}
where the last equality follows from \eqref{dmu_2}, and so
\begin{equation}
\lim_{n\to \infty} \frac{1}{n}K_n(x,x)= \frac{d\mu_2}{dx}(x), \quad
x\in \Delta_2.
\end{equation}
If $x\in\mathbb{R}^+\setminus \Delta_2$, a similar argument as given
above yields
\begin{equation}
\lim_{n\to \infty} \frac{1}{n}K_n(x,x)= 0.
\end{equation}
This completes the proof of Theorem \ref{theorem:limiting mean
distribution}.

\section*{Acknowledgement}
Steven Delvaux is a Postdoctoral Fellow of the Fund for Scientific Research - Flanders (Belgium).

Arno Kuijlaars is supported by K.U.~Leuven research grant OT/08/33,
FWO-Flanders projects G.0427.09 and G.0641.11, by the
Belgian Interuniversity Attraction Pole P06/02,
and by grant MTM2008-06689-C02-01 of the Spanish Ministry
of Science and Innovation.

Pablo Rom\'an is supported by K.U.~Leuven research grant OT/08/33.

Lun Zhang is supported by the
Belgian Interuniversity Attraction Pole P06/02.

%------------------------------------------------------------------

\end{document}